\documentclass{jfm}
\usepackage{tikz}
\usepackage[pdftex]{pict2e}
\usepackage{graphicx}
\usepackage{epstopdf,epsfig}
\usepackage{xcolor}
\usepackage{epstopdf}
\PrependGraphicsExtensions{.tif, .tiff}
\usepackage{newtxtext}
\usepackage{newtxmath}

\usepackage{amsmath} 
\usepackage{multirow}
\usepackage{placeins}
\usepackage{romannum}
\usepackage{tabularx}
\usepackage{float} 
\usepackage{natbib}
\usepackage{hyperref}
\usepackage{xcolor}
\usepackage[dvipsnames]{xcolor}
\usepackage{setspace}
\hypersetup{
    colorlinks = true,
    urlcolor   = blue,
    citecolor  = blue,
}

\usepackage[ruled,vlined]{algorithm2e}
\SetKw{KwBy}{by}

\newcommand{\RomanNumeralCaps}[1]

\title{\vspace{-2.0cm} Particle Force-Based Continuum Model for Multicomponent Size Segregating Mixtures
}

\author{Soniya Kumawat \and Anurag Tripathi \corresp{\email{anuragt@iitk.ac.in}}}

\affiliation{Department of Chemical Engineering, Indian Institute of Technology Kanpur 208016, India}

\begin{document}
\pagenumbering{arabic}
\maketitle
\begin{abstract}
We investigate size difference driven segregation in dense granular flows of multicomponent mixtures down a periodic chute using continuum model and Discrete Element Method (DEM) simulations. A previously developed particle force-based segregation model for binary mixtures is systematically extended to mixtures comprising three or more particle species differing in size. The generalized model accounts for inter-species interactions by computing the net force on each component in the presence of all others, without relying on empirical percolation velocity. This segregation model is coupled with a mixture rheology model and incorporated into the species transport and momentum balance equations to develop a continuum model that predicts the spatial and temporal evolution of species concentration and velocity fields. The continuum model predictions are found to be in agreement with DEM simulation data for ternary and quaternary mixtures over a wide range of mixture compositions and chute inclinations at moderate size ratios for well-mixed and small-near-base configurations. For larger size ratios, the one dimensional model predictions capture the qualitative segregation trend while showing relatively larger quantitative differences from DEM data. For an initial configuration, having large particles near base and small particles near the free surface, a Rayleigh-Taylor like instability at early times is observed. Due to the presence of this instability, two dimensional evolution of the species concentration fields is present for initial part of the flow. Predictions of such features requires the extension of the one dimensional continuum model to two dimensions.
\end{abstract}

\section{Introduction}
\label{sec:intro}
Size-\textcolor{black}{difference} driven segregation in multicomponent granular mixtures is a prevalent phenomenon across a wide range of geophysical flows and industrial applications.
This phenomenon poses practical challenges in industries from pharmaceuticals to mining and plays a critical role in shaping natural processes such as debris flows, avalanches, and silo discharge patterns. 
\textcolor{black}{Considerable} experimental~\citep{gray1997pattern,HSIAU2002,Jain2005,bhattacharya2014chute,asachi2018experimental,liao2020behavior,pillitteri2020size,BRANDAO20201} and computational~\citep{pereira2011insights,chand2012discrete,bhattacharya2014chute,combarros2014segregation,ayeni2015discrete,Pereira2017,Qiao2021} works have reported segregation in binary mixtures. 

Theoretical modeling of segregation in binary granular mixtures has evolved through several approaches. The kinetic theory-based formulation~\citep{jenkins1985kinetic,savage1988particle,khakhar1999mixing,kumaran2008dense,kumaran2015kinetic,larcher2013segregation,larcher2015evolution,jenkins2020segregation} \textcolor{black}{ are the most detailed and theoretically grounded approaches, but relatively less applicable to real life dense flow situations.} \textcolor{black}{Empirical} approaches attempt \textcolor{black}{to model segregation by} fitting the simulation or experimental data of percolation velocity with other flow properties~\citep{fan2014modelling,jones2018asymmetric,Duan2021,singhandhennan2024continuum}. More recently, particle force-based methods~\citep{jing2020rising,tripathi2021size,yennemadi2023drag} have also been used to predict size segregation \textcolor{black}{that appear to be promising.}

\textcolor{black}{In contrast to the large number of studies available for bidisperse system, only a limited number of studies have investigated the size segregation in multicomponent granular mixtures, despite its practical relevance in both industrial and geophysical contexts.} \citet{gray_ancey_2011} extended their earlier advection–diffusion–segregation framework~\citep{grayandchugunov2006particle} from binary to multicomponent mixtures by formulating coupled transport equations for each species. However, their model does not account for the effect of shear rate on the segregation velocity. \citet{schlick2016} accounted for this effect and extended the linear empirical segregation model developed by \citet{fan2014modelling} (for binary mixtures) to model segregation in polydisperse granular mixtures. The model demonstrated the ability to qualitatively capture segregation in polydisperse flows but relies on prior knowledge of the velocity fields. \citet{deng2018continuum} subsequently employed this extended model to predict the time evolution of segregation in bounded heap flows and rotating cylinder configurations. \textcolor{black}{However, the quadratic empirical models that appear to improvise over the linear empirical model~\citep{jones2018asymmetric,Duan2021} are limited to binary mixtures and cannot be readily extended to polydisperse mixtures.}

\citet{trewhela2021experimental} recently proposed the empirical segregation model based on single intruder experiments conducted for binary size mixtures. \citet{barker2021OpenFoam} incorporated this segregation model along with the granular mixture rheology~\citep{barker2017wellposedrheology} to solve the species transport equation and momentum balance equations in OpenFOAM. Their three-dimensional continuum model successfully captured qualitative behavior of segregation in various flow configurations, including chute flows and rotating drum systems. \textcolor{black}{\citet{maguire2024particle} recently extended the continuum framework to predict segregation dynamics in a triangular rotating drum and further generalized it to ternary mixtures, demonstrating qualitative agreement with experimental observations.}
Building upon their earlier work, \citet{trewhela2024segregation} employed the segregation model in a one-dimensional continuum model to predict time-dependent segregation behavior in plane shear flows.
In another empirical study,
\citet{liuandhennan2023coupled} developed a coupled model combining shear-strain-rate-gradient-driven size segregation with a non-local rheology to predict segregation in vertical and annular shear flows. This model was later extended by \citet{singhandhennan2024continuum} to include pressure-gradient-driven segregation, resulting in predictions that closely match DEM simulations for chute and plane shear flows.

Theoretical studies \textcolor{black}{based} on \textcolor{black}{particle level} force-based approaches, particularly for addressing size segregation~\citep{guillard2016scaling,jing2020rising,jing2021unified,tripathi2021size,duan2022segregation,yennemadi2023drag} \textcolor{black}{have also gained popularity in last decade}. 
In these studies, the segregation velocity is determined by considering the forces acting on \textcolor{black}{a} single large intruder in a mixture of small particles.  
A similar particle force-based approach for density segregation \textcolor{black}{has been successful in predicting} the steady ~\citep{tripathi2013density,sahu_kumawat_agrawal_tripathi_2023} as well as transient~\citep{kumawat2025transient} evolution of segregation for binary and multicomponent mixtures. 

Motivated by the success of particle force-based approach in case of density segregation, in this work, we generalize the particle force-based segregation model \textcolor{black}{proposed by}~\citep{tripathi2021size,kumawat2025sizetransient} to multicomponent size mixtures. \textcolor{black}{We} study the time-dependent behavior of flow and segregation of \textcolor{black}{polydisperse} granular mixtures and explore the applicability of this particle force-based model to accurately predict the evolution of species concentration and velocity. We solve the momentum balance equations by utilizing $\mu-I$ rheology for \textcolor{black}{granular} mixtures along with the segregation-diffusion equation \textcolor{black}{by intercoupling the segregation model with rheology}.
We perform \textcolor{black}{a large number of} DEM simulations for ternary as well as quaternary mixtures differing in sizes. \textcolor{black}{We consider a variety of} mixture compositions and size ratios flowing over different inclination angles. The results presented in this work show that transient segregation in a periodic chute can be very well captured by our $1D$ continuum model for ternary and quaternary mixtures for moderate size ratios for small-near-base and well-mixed initial configurations.

The \textcolor{black}{organization} of the paper is as follows. Section~\S\ref{chp4_sec:theory} presents the generalized particle force-based segregation model for multicomponent mixtures, along with the continuum model that solves the time-dependent momentum balance and species transport equations \textcolor{black}{utilizing} the segregation model and granular mixture rheology. We report the comparison of continuum model predictions with the DEM simulations data for ternary and quaternary mixtures in Section \S \ref{sec:results_discussion} \textcolor{black}{followed by results for large sizes in Section~\S \ref{sec:solidsfraction_fl}}. Finally, Section~\S\ref{chp4_sec:conclusion} summarizes the key findings of this work.

\section{Governing equations for continuum segregation model}
\label{chp4_sec:theory}
\textcolor{black}{Segregation behavior in multicomponent granular systems is considerably more complex than in binary mixtures. During free surface flow of such mixtures, the larger particle species typically rise to the free surface, while the smallest species consistently tend to concentrate in regions away from the free surface. For intermediate-sized species, the concentration of particles depends on their competing interactions with smaller and larger particles. In order to predict segregation in multicomponent mixtures, we need the segregation flux expression for small \textcolor{black}{and} intermediate species as well. 
However, previous studies based on the particle force-based approach~\citep{tripathi2021size,yennemadi2023drag} primarily focus on computing forces acting on a single large intruder particle in the dilute limit. These approaches do not address the behavior of small or intermediate species in binary mixtures. Therefore, before extending the model to multicomponent systems, we \textcolor{black}{first discuss} segregation in binary mixtures \textcolor{black}{to obtain the expression} of segregation flux of small species in binary mixture \textcolor{black}{using particle force-based approach}.}

\subsection{\textcolor{black}{Size} segregation model \textcolor{black}{for} binary mixtures}
\begin{figure}
    \centering
    \includegraphics[width=0.3\linewidth]{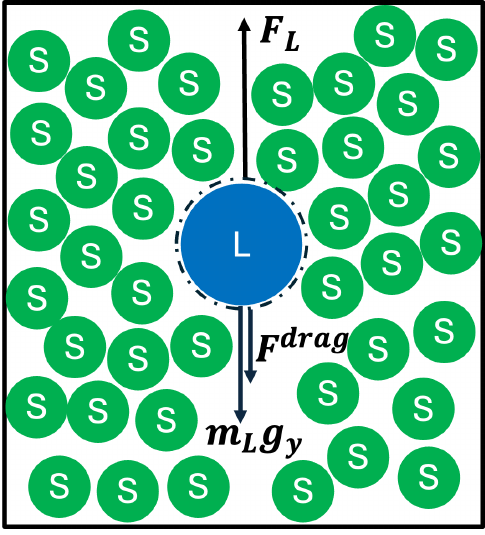}
    \caption{Forces on \textcolor{black}{a rising} single large intruder in binary mixture in \textcolor{black}{an assembly} of small particles
    }
    \label{fig:Single_binary}
\end{figure}
Let us consider a binary mixture of small and large size particles having identical densities. \textcolor{black}{Following \cite{tripathi2013density} and \cite{tripathi2021size}, we use an effective medium approach and} write the force balance on a single large intruder particle rising in a mixture of small particles \textcolor{black}{(shown in the Figure~\ref{fig:Single_binary})} flowing over an inclined plane at an inclination angle $\theta$ as
\begin{equation}
    F^{drag}=  F_{L} - m_L g_y,
\end{equation}
where $m_L$ is the mass of the large intruder. $F_{L}$ is the total upward force (due to flow induced buoyancy and lift) \textcolor{black}{and} the \textcolor{black}{downward acting} drag force is given as $ F^{drag} = c \pi \eta d_L v_L$ \citep{tripathi2013density,yennemadi2023drag}. Here, $c$ is the Stokes drag coefficient, $d_L$ and $v_L$ are the diameter and velocity of the large intruder particle, respectively. By simplifying the above equation, we obtain the segregation velocity of a large intruder as 
\begin{equation}
     v_{L} = \frac{(F_{L} - m_L g_y)}{c_L \pi\eta d_L}.
     \label{eq:seg_vel_large}
\end{equation}
\textcolor{black}{For} moderately large size particles, \textcolor{black}{$v_L > 0$, i.e., they} move upward ($F_L > m_L g_y$). \cite{tripathi2021size} \textcolor{black}{account for this fact by expressing} the total upward force $F_L$ on a large intruder in an assembly of small particles as 
\begin{equation}
    F_{L}=(1 + \alpha_{L})m_L g_y,
\label{eq:ch4_large_force_binary_tripathi}
\end{equation}
\textcolor{black}{where, $\alpha_{L}$ is the dimensionless measure of net upward force on the large size particles. \textcolor{black}{The value of $\alpha_L$ is particle size dependent which is estimated from DEM simulations to be smaller than unity.} For sufficiently dilute mixtures \textcolor{black}{containing only a} few larger \textcolor{black}{size} particles i.e., if the concentration of large particles $f_L << 1$, the presence of other large particles does not alter their segregation velocity. \textcolor{black}{In other words, $\alpha_{L}$ is independent of species concentration in a dilute mixture containing only a few large particles in an assembly of small size particles, i.e., $\alpha_{L} (f_L \to 0) = \alpha_{LS}^{0}$ }. Using this, we obtain the total upward force on a large particle in a \textcolor{black}{\textit{dilute mixture}} as}
\begin{equation}
    F_{LS}=(1 + \alpha_{LS}^{0})m_L g_y.
    \label{eq:ch4_large_force_binary}
\end{equation}

The force on a large particle in an assembly of identical size large particles (i.e., in the limit of no small particles) can be written \textcolor{black}{in a similar fashion} as $F_{LL} = (1 + \alpha_{LL}^{0}) m_L g_y$. In such mixture with $f_L \to 1$, there are no net upward motion of large particles and the upward force equals to the gravitational force, \textcolor{black}{giving}, $\alpha_{LL}^{0} = 0$. For a non-dilute binary mixture of large and small particles, however, the net upward motion of the large particle seems to depend upon the local concentrations of large and small species~\citep{tripathi2021size,trewhela2021experimental}.  
\textcolor{black}{We assume that} the total upward force acting on a large particle (due to the combined effect of large and small particles in its vicinity) \textcolor{black}{is given by} the volume-averaged contribution of upward force from its interactions in these limiting cases, i.e., 
\begin{equation}
    F_{L} = f_S F_{LS} + f_L F_{LL}, 
    \label{sum_binary_FL}
\end{equation}
where $f_L$ and $f_S$ are concentration of large and small particles in mixture, respectively.
Equation~\ref{eq:ch4_large_force_binary_tripathi}, written for a single large size intruder, can be generalized for a binary mixture in the same form i.e., as $F_{L} = (1 + \alpha_{L}) m_L g_y$ to account for the local species concentration dependent total upward force, where $\alpha_L$ depends on both $f_L$ and $f_S$ i.e., $\alpha_L = \alpha_L(f_L,f_S)$. Equating this expression with that of equation~\ref{sum_binary_FL} we get 
$\alpha_L =f_L \alpha_{LL}^{0}  + f_S \alpha_{LS}^{0}  $. In other words, the dimensionless net upward force $\alpha_L$ on the large particle in a binary mixture is the volume-averaged value of the limiting values $\alpha_{LL}^{0}$ and $\alpha_{LS}^{0}$. Since $\alpha_{LL}^0 = 0$, the dimensionless measure of the net upward force on a large particle, $\alpha_L$, reduces to the contribution from interactions with small particles and can be expressed as $\alpha_L = f_S \alpha_{LS}^{0}$. Using this relation, the net upward force acting on a large particle of a binary mixture in the vertical direction becomes $F_{L} = (1 + f_S \alpha_{LS}^{0}) m_L g_y$. Substituting this expression for $F_{L}$ into equation~\ref{eq:seg_vel_large} yields the segregation velocity of the large particles as
 \begin{equation}
     v_L  = \frac{m_L g_y}{c_L\pi\eta d_L}  \alpha_{LS}^{0} f_S .
     \label{eq:ch4_large_segvel_binary}
 \end{equation}
\subsubsection{Segregation flux of large species}
The segregation flux of the large particles is defined as $J^{seg}_L = v_L f_L$. Upon substituting the segregation velocity from equation~\ref{eq:ch4_large_segvel_binary}, the resulting expression for the segregation flux becomes $J^{seg}_L = v_L f_L  = \frac{m_L g_y}{c_L\pi\eta d_L}  \alpha_{LS}^{0} f_S f_L.$ This leads to a symmetric flux expression which is consistent with the formulation proposed by~\cite{grayandchugunov2006particle} and~\cite{fan2014modelling}. However, researchers have experimentally~\citep{golick2009mixing,van2015underlying} shown that a small particle percolates downwards at a faster speed in an assembly of large particles compared to the rising velocity of a large particle in an assembly of small particles. Due to this reason, they proposed asymmetric segregation flux expressions.
\cite{tripathi2021size} accounted for this asymmetry in the segregation flux \textcolor{black}{by an} additional linear dependence on the concentration of large species ($f_L$) in bi-disperse mixture given as
\begin{equation}
     J^{seg}_L = \frac{m_L g_y}{c_L\pi\eta d_L}  \alpha_{LS}^{0} (1 + k_{LS} ~f_L) f_S f_L  .
     \label{eq:large_flux}
\end{equation}
The asymmetry parameter $k_{LS}$ is found to be nearly identical to the size ratio \textcolor{black}{$r_{LS}$}, i.e., $k_{LS}  = r_{LS} = d_L/d_S$, where $d_L$ and $d_S$ are the diameters of large and small particles, respectively. 
\textcolor{black}{The term $(1 + k_{LS} ~f_L) f_S$ in equation~\ref{eq:large_flux} is the correction in the dilute limit net upward force $\alpha_{LS}^{0}~m_L g_y$ due to the presence of both small and large size particles. Thus $\alpha_{LS}$ for large species with respect to small species in binary mixture is given as} 
\begin{equation}
    \alpha_{LS} = \alpha_{LS}^{0}  (1 + k_{LS} ~f_L).
    \label{eq:alpha_LS_assymetry_dl}
\end{equation}
\textcolor{black}{Using this, the expression of segregation flux on large particles can be written as  
\begin{equation}
     J^{seg}_L = \frac{m_L g_y}{c_L\pi\eta d_L}  \alpha_{LS} f_S f_L.
     \label{eq:large_flux_expression}
\end{equation}
We note that equation~\ref{eq:alpha_LS_assymetry_dl} can be generalized as}
\begin{equation}
    \alpha_{Li} = \alpha_{Li}^{0}  (1 + k_{Li} ~f_L),
    \label{eq:alpha_Li_assymetry}
\end{equation}
\textcolor{black}{to any other mixture of large species particles with $i^{th}$ species particles provided species $i$ is smaller in size than the large species particle. Following~\citep{tripathi2021size,kumawat2025sizetransient}, the parameter $k_{Li}$ is assumed to be equal to the size ratio, $k_{Li} = d_L/d_i$. In the dilute limit of large particles, $f_L \to 0$, $\alpha_{Li}$ approaches to $\alpha_{Li}^0$. As before $\alpha_{Li}^{0}$ represents the dimensionless upward force on large particles in the dilute limit in an assembly of $i^{th}$ species particles.}

\subsubsection{Segregation flux of small species}
\label{sec:small_seg_flux}
We next consider a single small intruder in an assembly of large particles. Following an approach similar to that for large particles (equation~\ref{eq:large_flux_expression}), we obtain the segregation flux of small particles in a non-dilute binary mixture as
\textcolor{black}{
\begin{equation}
    J^{seg}_{S} = \frac{ m_S g_y}{c_S \pi \eta d_S} \alpha_{SL} f_L f_S, 
    \label{eq:ch4_segflux_small}
\end{equation}
}
We employ a species segregation flux balance in binary mixtures to determine the expression of $\alpha_{SL}$ for small size particles. \textcolor{black}{In the case} of \textcolor{black}{segregation in} a unidirectional periodic chute \textcolor{black}{flow of small and large species,} the species mass balance equation for large and small species in a binary mixture are as follows (\cite{kumawat2025sizetransient})
\begin{equation}
   \frac{\partial f_L}{\partial t} =  - \frac{\partial}{\partial y } ( J^{seg}_{L} + J^{diff}_{L})
   \label{eq:large_con_seg_diff_eq}
\end{equation}
and
\begin{equation}
   \frac{\partial f_S}{\partial t} =  - \frac{\partial}{\partial y } ( J^{seg}_{S} + J^{diff}_{S}),
   \label{eq:small_con_seg_diff_eq}
\end{equation}
respectively. The expression of the diffusion flux is given as $J_{i} = - D \frac{\partial f_i}{\partial y}$, where $D$ is the diffusion coefficient. Adding equations~\ref{eq:large_con_seg_diff_eq} and~\ref{eq:small_con_seg_diff_eq} and using the expressions for segregation flux of large (Equation~\ref{eq:large_flux_expression}) and small (Equation~\ref{eq:ch4_segflux_small}) species along with the diffusion flux, we obtain 
\begin{equation}
   \frac{\partial (f_L + f_S)}{\partial t} =   - \frac{ m_S g_y }{c_S \pi \eta d_S} \alpha_{SL} f_L f_S  - \frac{ m_L g_y }{c_L \pi \eta d_L} \alpha_{LS} f_S f_L - D \frac{\partial (f_L + f_S)}{\partial y}
\label{eq:large_small_con_seg_diff_eq}
\end{equation}
For a binary mixture, we use the relation $f_L + f_S = 1$ in the above equation~\eqref{eq:large_small_con_seg_diff_eq} and get
\textcolor{black}{
\begin{equation}
    \frac{ m_S g_y }{c_S \pi \eta d_S} \alpha_{SL} f_L f_S  + \frac{ m_L g_y }{c_L \pi \eta d_L} \alpha_{LS} f_S f_L  = 0.
\label{eq:ch4_sum_small_large}
\end{equation}
}
\textcolor{black}{
Simplifying the above expression, we obtain
\begin{equation}
    \alpha_{SL} = -  \frac{c_S}{c_L}  r_{LS}^2 \alpha_{LS},
\end{equation}} 
$r_{LS} = d_L/d_S$ \textcolor{black}{being} the ratio of large to small particle size. Using the expression of $\alpha_{LS} = \alpha_{LS}^0 (1 + k_{LS} f_L)$, we get
\textcolor{black}{
\begin{equation}
   \alpha_{SL}= -  \alpha_{LS}^{0} \frac{c_S}{c_L}  r_{LS}^2 (1 + k_{LS}~ f_L).
   \label{eq:alpha_SL_from_fluxbalance}
\end{equation}
}
\textcolor{black}{By substituting $f_L = 1 - f_S$ and rearranging, we obtain the following expression for $\alpha_{SL}$:
\begin{equation}
    \alpha_{SL} = -  \alpha_{LS}^{0} \frac{c_S}{c_L}  r_{LS}^2 ~(1 + k_{LS}) ~ \bigg(1 - \frac{k_{LS}}{1 + k_{LS}} f_S \bigg).
    \label{eq:small_alpha_sl}
\end{equation}
Writing $\alpha_{SL}^{0} = - \alpha_{LS}^{0} \frac{c_S}{c_L}  r_{LS}^2 ~(1 + k_{LS})$ and $k_{SL} = - \frac{k_{LS}}{1 + k_{LS}}$, equation~\ref{eq:small_alpha_sl} can be represented as 
\begin{subequations}
    \begin{align}
    \alpha_{SL} = \alpha_{SL}^{0}  (1 + k_{SL} f_S),~~~~~~~~~~ \\
 \text{where}~~~~   \alpha_{SL}^{0} = - \alpha_{LS}^{0} \frac{c_S}{c_L}  r_{LS}^2 ~(1 + k_{LS}), \\
  \textcolor{black}{\text{and}}~~~~  k_{SL} = - \frac{k_{LS}}{1 + k_{LS}}.~~~~~~~~~~~~~~~
    \end{align}
\label{eq:small_alpha_S_formulated}
\end{subequations}
}
\textcolor{black}{
Thus, for a multicomponent mixture, $\alpha_{ij}$ corresponding to the pair $(i,j)$ with size ratio $r_i/r_j < 1$ can be expressed as follows, 
\begin{align}
   \alpha_{ij} = -  \alpha_{ji}^{0} \frac{c_i}{c_j}  r_{ji}^2 ~(1 + k_{ji}) ~ \bigg(1 - \frac{k_{ji}}{1 + k_{ji}} (1- f_j) \bigg)
    \label{eq:small_alpha__fi}
\end{align}
Analogous to equation~\ref{eq:small_alpha_S_formulated} for a small particle of $i^{th}$ species relative to large particles of $j^{th}$ species, equation~\ref{eq:small_alpha__fi} can be expressed as
\begin{subequations}
\begin{align}
 \alpha_{ij} = \alpha_{ij}^{0}  \left[1 + k_{ij} (1- f_j)\right] \\
   \alpha_{ij}^{0} = -  \alpha_{ji}^{0} \frac{c_i}{c_j}  r_{ji}^2 ~(1 + k_{ji})\\
   k_{ij} = -\frac{k_{ji}}{1 + k_{ji}}~~~~~~~~~~~~
    \end{align}
     \label{eq:small_alpha__fi_simp}
\end{subequations}
}
The form of $\alpha_{SL}$ (given by equation~\ref{eq:small_alpha_S_formulated}) is similar to that of $\alpha_{LS}$ (given by equation~\ref{eq:alpha_LS_assymetry_dl}) and helps us write segregation flux expression for the small species (Equation~\eqref{eq:ch4_segflux_small}) in the case of a binary mixture similar to that of large species given by equation~\ref{eq:large_flux} as
\textcolor{black}{
\begin{equation}
    J^{seg}_{S} = \frac{ m_S g_y}{c_S \pi \eta d_S} \alpha_{SL}^{0} (1 + k_{SL}~f_S )  f_L f_S  .
\end{equation}
}
\textcolor{black}{The above form of $J^{seg}_{S}$ is consistent with the limiting cases of the volume fraction of small particles approaching either unity ($f_S \to 1$) or zero ($f_S \to 0$), both cases leading to $J_S^{seg} \to 0$. As $\alpha_{SL}^0$ is negative, the segregation flux $J^{seg}_{S}$ consequently becomes negative, implying the net downward motion of small particles.}
Since the segregation flux of small particles is derived using the species mass balance approach, it inherently ensures mass conservation of the species of the mixture. \textcolor{black}{Hence} solving the species transport equation for the small particles \textcolor{black}{in a binary mixture} should yield the same concentration profiles as those obtained from solving it for the large particles, not only at steady state but also for the transient evolution of the center of mass of species. 
\textcolor{black}{Such a} consistency is a fundamental requirement for any segregation model and serves as a good check for the correct implementation of the method. \textcolor{black}{We show in \ref{sec:comparison_small_large_v1} of supplementary material that the results from solving either of the species balance equation with the respective segregation flux are identical for five different mixture compositions confirming the consistency and accurate implementation of the model.}

\subsection{Size segregation model for multicomponent mixtures}
\label{subsec:size_multicomponent}
\begin{figure}
    \centering
\includegraphics[width=0.5\linewidth]{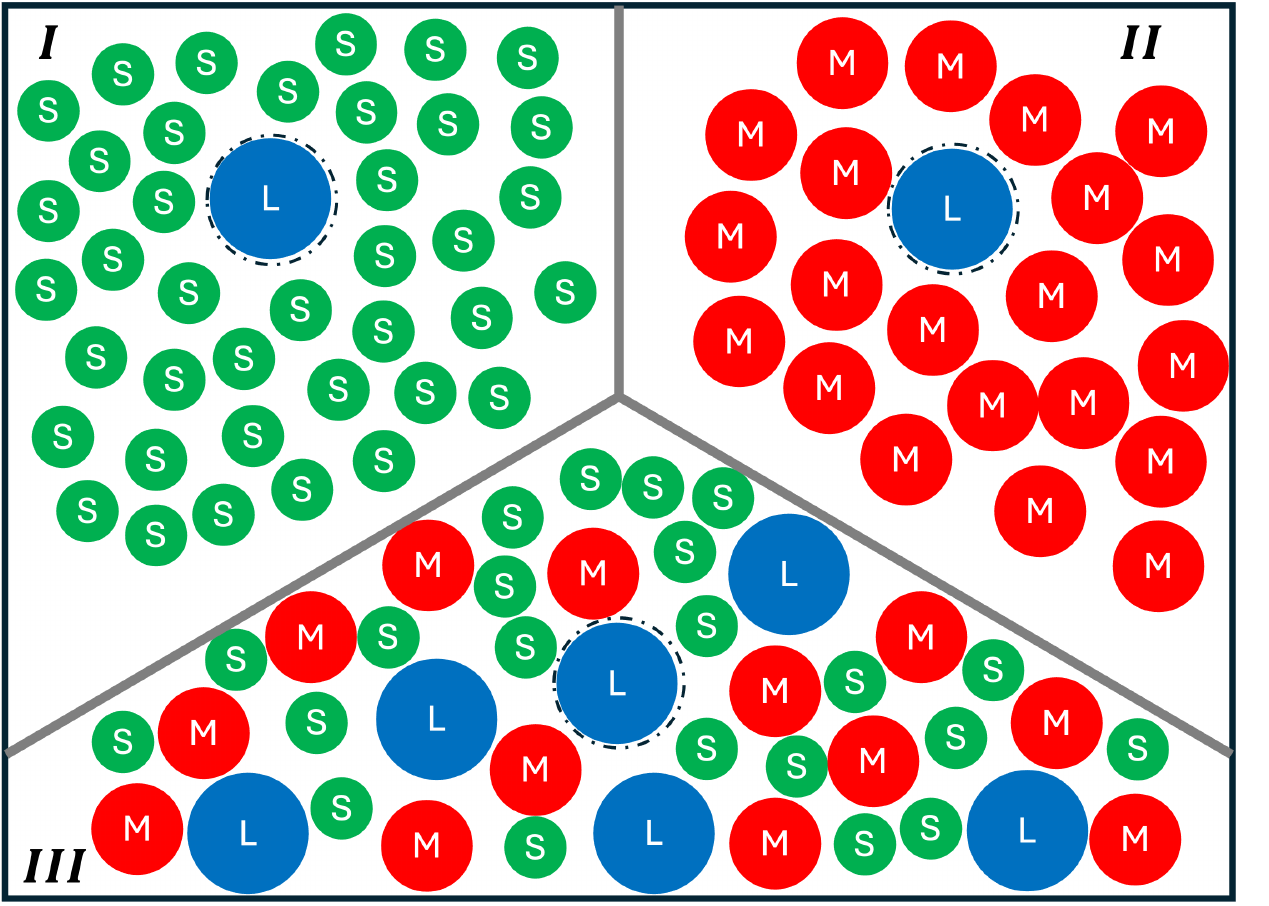}
    \caption{A schematic representation of large intruders in a ternary mixture consisting of small, medium, and large particles. Blue denotes the large intruder, red indicates medium-sized particles, and green represents the smaller particles in the mixture.
    }
    \label{fig:Single_Ternary}
\end{figure}
\textcolor{black}{
Consider a ternary mixture of large, medium, and small size grains. The total upward force on the large particle in the presence of only one of the two species (i.e., either medium size grains or only small size grain) can be written as
\begin{equation}
    F_{Li}=(1 + \alpha_{Li})m_L g_y,
\end{equation}
where $i$ is the index for the particular species present around the large particle and $\alpha_{Li}$ is obtained using equation~\ref{eq:alpha_Li_assymetry}. The total upward force on the large particle in limiting situations shown in region \Romannum{1} of Figure~\ref{fig:Single_Ternary} would be given as $ F_{LS}=(1 + \alpha_{LS}^{0})m_L g_y$
while that in region \Romannum{2} will be given as $F_{LM}=(1 + \alpha_{LM}^{0})m_L g_y$. \textcolor{black}{Recall that $\alpha_{LL}^{0}$ equals zero i.e.,} a large particle entirely surrounded by other large particles experiences an upward force exactly equal to its weight, leading to no net upward force. In the case of large particle surrounded by large, medium and small size particles (as in region \Romannum{3} of figure~\ref{fig:Single_Ternary}), the total upward force on the large particle would be between the limiting values for region \Romannum{1} and \Romannum{2}. We propose a local concentration averaged total upward force expression and write}
\begin{equation}
    F_{L} = f_S ~F_{LS} + f_M~ F_{LM}  + f_L ~F_{LL}.
    \label{eq:ch4_force_total_term}
\end{equation}
\textcolor{black}{Writing} the total upward force $F_L$ on a large intruder as $F_{L} = (1 + \alpha_{L}) m_L g_y$ \textcolor{black}{and substituting this} in equation~\eqref{eq:ch4_force_total_term}, we obtain  
\begin{equation}
    (1 + \alpha_{L}) m_L g_y = f_S (1 + \alpha_{LS}) m_L g_y  + f_M (1 + \alpha_{LM}) m_L g_y + f_L (1 + \alpha_{LL}) m_L g_y.
\end{equation}
The above expression simplifies to 
\begin{equation}
   1 +  \alpha_{L}  = f_S (1 + \alpha_{LS})   + f_M (1 + \alpha_{LM})  + f_L (1 + \alpha_{LL}) .
   \label{eq:ch4_simplified_force}
\end{equation}
As mentioned before, $\alpha_{LL} = 0$ and the total species fraction in the ternary mixture remains unity, i.e, $f_S + f_M + f_L = 1$. Substituting these conditions into equation~\eqref{eq:ch4_simplified_force}, we obtain
\begin{equation}
   \alpha_{L}  = f_S \alpha_{LS}   + f_M \alpha_{LM}. 
\end{equation}
The expression of dimensionless net upward force in vertical direction ($\alpha_L$) incorporates weighted contributions from both large–medium and large–small pairs. \textcolor{black}{In other words,} $\alpha_L$ in a ternary mixture is assumed to be the local volume concentration average of the two binary limits.
Substituting the expressions for $\alpha_{LS}$ and $\alpha_{LM}$ corresponding to a binary mixture with size ratios $r_{LS} = d_L/d_S$ and $r_{LM} = d_L/d_M$, and the associated interaction parameters \textcolor{black}{$k_{LS} (= r_{LS})$ and $k_{LM} (= r_{LM})$}, respectively, we obtain
\begin{equation}   
\alpha_L=\alpha_{LS}^{0} (1+k_{LS}~f_L) ~f_S+\alpha_{LM}^{0} (1+k_{LM}~f_L) ~f_M.
\label{eq:ch4_alpha_ternary}
\end{equation}
The parameters $\alpha_{LS}^{0}$ and $\alpha_{LM}^{0}$ correspond to the single large intruder limit in the small and medium size particles assembly, respectively. 
The above form of $\alpha_L$ (Equation~\eqref{eq:ch4_alpha_ternary}) ensures that in the limiting cases, such as when $f_M = 0$ (reducing the ternary system to a large–small binary mixture) or $f_S = 0$ (large-medium binary) - the expression correctly reduces to the respective binary forms. \textcolor{black}{The net upward force on large particles is balanced by drag force, which yields the segregation velocity $v_L = \frac{\alpha_L m_L g_y}{c \pi \eta d_L}$. Using definition of large particle segregation flux i.e., $J_L^{seg} = v_L f_L$, we derive the final expression $J^{seg}_{L}$ for large particles in a ternary mixture as}
\begin{equation}
    J^{seg}_{L}=\frac{m_Lg_y}{c_L\pi\eta d_L}[\alpha_{LS}^{0} (1+k_{LS}~f_L) ~f_S+\alpha_{LM}^{0} (1+k_{LM}~f_L) ~f_M] f_L.
    \label{eq:chp4_largefluxternary}
\end{equation}
\textcolor{black}{Since both $\alpha_{LS}^0$ and $\alpha_{LM}^0$ are positive along with $k_{LS}$ and $k_{LM}$, $J_L^{seg} >0$ for all values of $f_S$ and $f_M$, indicating that the large particles move towards the free surface irrespective of the relative concentration of species.} 

Following a similar approach, we extend the expression for $\alpha_S$ by accounting for the interactions of the small species with both the large and medium particles in the case of a ternary mixture. The net segregation tendency of small particles results from the combined effects of both size ratios: $r_{SL} = d_S/d_L$ and $r_{SM} = d_S/d_M$, and their corresponding interaction parameters $k_{SL}$ and $k_{SM}$.
As before, the expression incorporates weighted contributions from both large–small and medium–small pairs. This additive form captures the competing influences of large and medium particles on the small species's segregation tendency. Using equation~\ref{eq:small_alpha__fi_simp}a, mathematically this is reflected in the two terms of the extended expression:
\begin{align}
\alpha_S  = \alpha_{SL}^{0}  [1 + k_{SL}~(1 - f_L)] ~f_L + \alpha_{SM}^{0}  [1 + k_{SM} ~(1 - f_M)]~f_M. 
\label{eq:ch4_alpha_s}
\end{align}
\textcolor{black}{As discussed in section~\ref{sec:small_seg_flux},} the interaction parameters $k_{SL}$ and $k_{SM}$ are given in terms of $k_{LS}$ and $k_{MS}$ as $k_{SL} = -\frac{k_{LS}}{1 + k_{LS}}$ and $k_{SM} = -\frac{k_{MS}}{1 + k_{MS}}$. $\alpha_{SL}^{0}$ and $\alpha_{SM}^{0}$ are the particle size-dependent dimensionless measures of net upward force on small size particles with respect to large and medium sized particles, respectively. The expressions for these are given as \textcolor{black}{(following equation~\ref{eq:small_alpha__fi_simp}b)}
\begin{align}
\alpha_{SL}^{0} = - \alpha_{LS}^{0}~\frac{c_S}{c_L} r^2_{LS} (1+k_{LS}), ~~~~~~ \text{and} ~~~~~
\alpha_{SM}^{0} = - \alpha_{MS}^{0}~\frac{c_S}{c_M} r^2_{MS} (1+k_{MS}).
\label{eq:smallalpha_wrt_large_medium}
\end{align} 
This formulation of $\alpha_S$ in equation~\ref{eq:ch4_alpha_s} ensures that in the limiting cases, such as when $f_M = 0$ (reducing the ternary system to a large–small binary mixture) or $f_L = 0$ (medium–small binary) the expression correctly reduces to the respective binary forms. 
\textcolor{black}{Similar to force balance on large species, the force balance on small particles gives the segregation velocity of small species as $v_S = \frac{\alpha_S m_S g_y}{c \pi \eta d_S}$. Using the segregation flux definition for small particles $J_S^{seg} = v_S f_S$, we obtain the final expression $J^{seg}_{S}$ of small particles in a ternary mixture as}
\begin{align}
   J^{seg}_{S} = \frac{m_S g_y}{c_S\pi\eta d_S} \bigg(\alpha_{SL}^{0}  [1 + k_{SL}~(1-f_L)] ~f_L + \alpha_{SM}^{0}  [1 + k_{SM} ~(1 - f_M)]~f_M \bigg) f_S.
   \label{eq:chp4_smallfluxternary}
\end{align}
\textcolor{black}{Since $\alpha_{SL}^0$ and $\alpha_{SM}^0$ are negative (see equation~\ref{eq:smallalpha_wrt_large_medium}), equation~\ref{eq:chp4_smallfluxternary} establishes that $J_S^{seg} <0$ and the small particles percolate towards the base.}
The medium-size particles experience forces from both the small and large particles in the mixture. 
Accordingly, the net dimensionless force on the medium-size particle is given as 
\begin{equation}
    \alpha_M = \alpha_{ML}^{0} [1 + k_{ML}~(1 - f_L) ]~f_L
    +\alpha_{MS}^{0} (1 + k_{MS}~f_M )~f_S.
    \label{eq:alpha_M_mediumspecies_ternary}
\end{equation}
Using this expression for $\alpha_M$, the segregation flux of the medium species in a ternary mixture can then be expressed as follows:
\begin{align}
    J^{seg}_{M}  =\frac{m_M g_y}{c_M\pi\eta d_M} \bigg(\alpha_{ML}^{0} [1 + k_{ML}~(1 - f_L) ]~f_L
    +\alpha_{MS}^{0} (1 + k_{MS}~f_M )~f_S \bigg) f_M,
     \label{eq:chp4_mediumfluxternary}
\end{align}
\textcolor{black}{Since $r_L > r_M > r_S$,} the interaction parameters $k_{ML}$ and $k_{MS}$ are given as $k_{ML} = -\frac{k_{LM}}{1 + k_{LM}}$ \textcolor{black}{($r_M < r_L$)} and $k_{MS} = r_{MS}$ \textcolor{black}{($r_M > r_S$)}. The particle size-dependent dimensionless measure of net upward force on medium-sized particles due to large particles \textcolor{black}{$\alpha_{ML}^0$ is related to that of large particles in an assembly of small particles $\alpha_{LM}^0$} as $\alpha_{ML}^{0} =  - \alpha_{LM}^{0}~\frac{c_M}{c_L} r^2_{LM} (1+k_{LM})$. To summarize, the segregation flux expressions for species in a ternary mixture can be written as in Equations~\eqref{eq:chp4_largefluxternary}, \eqref{eq:chp4_smallfluxternary}, and \eqref{eq:chp4_mediumfluxternary}. To check the consistency of the ternary model, we reduce it to the binary-mixture limit in two ways. First, we set the concentration of one species to zero, which directly reduces the segregation-flux equations to their binary form as discussed above. Second, we consider the sizes of any two species to be identical, so that the ternary system effectively behaves as a binary mixture (see~\ref{sec:appendix_ternary_to_binary} of supplementary material). \textcolor{black}{Note that in contrast to the segregation flux of large and small particles, the segregation flux of the medium species depends on the local composition of the small and large species.} \textcolor{black}{This is because of the opposite signs of $\alpha_{ML}^0 (<0)$ and $\alpha_{MS}^0 (>0)$; the segregation flux of medium size species can be positive in regions rich in small particles (large values of $f_S$) but will be negative in regions rich in large particles (large values of $f_L$).} 

In order to extend this approach for multicomponent mixtures of $N$ species, we need to formulate the segregation flux expression for $N - 1$ species. The \textcolor{black}{segregation} flux for \textcolor{black}{each} species can be expressed as a sum of contributions arising from their interactions with \textcolor{black}{all other} species, following a similar pairwise flux formulation. The \textcolor{black}{generalized  expression of} segregation flux of the $i^{th}$ species in a polydisperse mixture of $N$ components is \textcolor{black}{given} as:
\begin{subequations}
\begin{align}
    J^{seg}_{i} = \frac{m_i g_y}{c_i \pi\eta d_i} f_i \sum_{ j=1, j \neq i}^{N} \alpha_{ij} f_j, ~~~~~~~~~~~~~~~~~~~~~~~~~~~~~~~~~~~~ \\
    r_{ij} = r_i / r_j > 1, \quad  k_{ij} = r_{ij},~~~ \text{and} ~~~ \alpha_{ij} = \alpha_{ij}^{0} (1 + k_{ij}f_i);~~~~~~~~~~~~\\
    r_{ij} = r_i / r_j < 1, \quad k_{ij} = -\frac{k_{ji}}{1 + k_{ji}} ~~~ \text{and} ~~~ \alpha_{ij} = \alpha_{ij}^{0}  \bigg(1 + k_{ij} (1 - f_j) \bigg).
    \end{align} 
    \label{eq:generalised_segflux_multi}
\end{subequations}
It is easy to verify that the system of equations represented by equation~\ref{eq:generalised_segflux_multi} reduces to equation~\ref{eq:chp4_largefluxternary}, \ref{eq:chp4_smallfluxternary}, and \ref{eq:chp4_mediumfluxternary} for a ternary mixture. In addition, we verify that the sum of the segregation fluxes is zero, that is $\sum_{i}^{N} J^{seg}_{i} = 0$ (see~\ref{sec:sum_segregation_zero} of the supplementary material). \textcolor{black}{One of the most important parameter in our particle force-based segregation model is the dimensionless upward force $\alpha_{ij}^0$ in the limit of dilute binary mixtures.} The expression of $\alpha_{ij}^{0}$ \textcolor{black}{can be} obtained by following~\cite{yennemadi2023drag} \textcolor{black}{who have proposed the following form for the dimensionless upward force} $\alpha_{ij}^{0} = [F(r_{ij}) - 1][1 + (A - Br_{ij}) \tan\theta]$. The term $(A - Br_{ij}) \tan\theta$ corresponds to the lift force \textcolor{black}{and typically accounts for $< 10\%$ to the total upward force. We find that ignoring this factor does not influence the predictions strongly (shown in figure~\ref{fig:lift_with_without}) and hence we chose to ignore} this lift force term in the continuum model. The function $F(r_{ij})$ is \textcolor{black}{found to be dependent on local solids fraction $\phi$ as well as size ratio $r_{ij}$,} given as 
\begin{equation}
 F(r_{ij}) =  \phi \left[ 
1 - \lambda_1 \exp \left( -\frac{r_{ij}}{R_1} \right) 
\right] 
\left[ 
1 + \lambda_2 \exp \left( -\frac{r_{ij}}{R_2} \right) 
\right] - 1.
\label{eq:alpha_function_DVK}
\end{equation}
We use the values of parameters as $\lambda_1 = 20.3$, $\lambda_2 = 2.11$, $R_1 = 0.25$, and $R_2 = 4.17$ reported in their study. 
\textcolor{black}{We note the} expression for $\alpha_{ij}^{0}$ \textcolor{black}{given by equation~\ref{eq:alpha_function_DVK}} is valid only for size ratios $r_i/r_j \geq 1$. Therefore, to evaluate the dimensionless net upward force on the small species relative to the large species, we employ \textcolor{black}{equation~\ref{eq:small_alpha_S_formulated}b to write the expression of} $\alpha_{ij}^{0}$ for $i^{th}$ small species with respect to $j^{th}$ larger species (i.e., $r_i < r_j$) as $\alpha_{ij}^{0} = - \alpha_{ji}^{0} \frac{c_i}{c_j}  r_{ji}^2 ~(1 + k_{ji})$. While \citet{yennemadi2023drag} showed that the Stokes drag coefficient is independent of the size ratio for $r_i/r_j \geq 1$, no data were reported for $r_i/r_j < 1$. Note that the formulation given by equation~\ref{eq:generalised_segflux_multi} does not require the value of the Stokes drag coefficient for particles with $r_i/r_j < 1$; \textcolor{black}{instead the values of $c_i$ for size ratio $>1$ suffice}. \textcolor{black}{Similar to binary granular mixtures, the consistency of the model for ternary mixtures is examined by solving the species equations for two different combinations, namely large and small, and large and medium. The model predictions obtained from both approaches are identical, as shown in figure~\ref{fig:LM_LS} of the supplementary material. Furthermore, we verify that when the governing equations for a quaternary mixture are reduced to the ternary case, the resulting predictions are consistent with those obtained by directly solving the ternary mixture equations (predictions for a few cases are shown in figure~\ref{fig:Quat_ternary}.)}

\subsection{Coupled system of equations for segregation and rheology}
We employ the generalized transport equation that incorporates advection, segregation, and diffusion fluxes. Using this equation, the evolution of the concentration $f_i$ of the $i^{th}$ species is given as
\begin{equation}
   \frac{\partial f_i}{\partial t} +   \nabla \cdot (\textbf{v} f_i) +  \nabla \cdot (\textbf{J}_{i}^{seg}) + \nabla \cdot (\textbf{J}_{i}^{diff}) = 0,
   \label{eq:ch4_adv_diff_seg_vectoreqn}
\end{equation}
where, $\textbf{v}$ is the mixture velocity. $\textbf{J}_{i}^{seg}$ and $\textbf {J}_{i}^{diff}$ are segregation and diffusion fluxes of $i^{th}$ species, respectively. To investigate the applicability of the particle force-based approach for multicomponent size mixtures, we impose periodic boundary conditions in our DEM simulation in both the flow and the vorticity directions. This setup corresponds to the classical assumption of fully developed flow in the continuum framework, where velocity and concentration profiles vary only in the direction normal to the flow (i.e., in the $y$ direction). Under these simplifying assumptions, the general three-dimensional transport equation \textcolor{black}{given by} Equation~\eqref{eq:ch4_adv_diff_seg_vectoreqn} reduces to a one-dimensional, time-dependent segregation-diffusion equation in the vertical $y-$ direction as:
\begin{equation}
\frac{\partial f_i}{\partial t} = -\frac{\partial}{\partial y} \left( J_{i}^{seg} + J_{i}^{diff} \right).
\label{eq:ch4_timedependesegdiff}
\end{equation}
\textcolor{black}{Following~\cite{barker_2021_gray}, diffusion flux in multicomponent mixtures can be written as $J_{i}^{diff} = \sum_{j,j \neq i}^{N} D_{ij}(f_j \nabla f_i - f_i \nabla f_j)$. Assuming symmetric diffusion coefficient $D_{ij} = D_{ji}$, and further assuming that the diffusion coefficients for different pair are equal to each other i.e., $D_{ij} = D_{ik} = D$, this expression becomes analogous to the classical form of Fickian diffusion as $J_{i}^{diff} = -D\frac{\partial f_i}{\partial y}$}, where $D$ denotes the diffusion coefficient. In the context of granular shear flows, it has been demonstrated~\citep{tripathi2013density,fry2019,Duan2021,sahu_kumawat_agrawal_tripathi_2023} that $D$ depends on the local shear rate ($\dot \gamma$) and local volume average diameter ($d_{mix}$). Following these studies, we use the spatially varying diffusivity as $D = b \dot \gamma d_{mix}^2$, where $b =  0.041$ is an empirical parameter. For mixtures with particles of varying sizes, $d_{mix}$ is calculated using a volume fraction weighted average \textcolor{black}{diameter of all the species given} as $d_{mix} = \sum_i^N f_i d_i $. Substituting the expression for diffusion flux into Equation~\eqref{eq:ch4_timedependesegdiff}, we obtain
\begin{equation}
\frac{\partial f_i}{\partial t} = -\frac{\partial}{\partial y} \left[ \frac{m_i g_y}{c_i \pi\eta d_i} f_i \sum_{ j=1, j \neq i}^{N} \alpha_{ij} f_j - b \dot{\gamma}d_{mix}^2 \frac{\partial f_i}{\partial y} \right].
\label{eq:ch4_finalsegdiff}
\end{equation}
The initial and boundary conditions needed to solve Equation~\eqref{eq:ch4_finalsegdiff} are given as follows:
\begin{subequations}
\begin{align}
     &f_{i}(y,0) = f_{i,ini}(y); ~~~ i\in [1:N -1],\\
     &J_{i}^{S}(0,t) + J_{i}^{D}(0,t) = 0; ~~~ i\in [1:N - 1],\\
     &J_{i}^{S}(h,t) + J_{i}^{D}(h,t) = 0; ~~~ i\in [1:N - 1].
    \end{align}
\label{eq:ch4_IC_BC_concentration_multi}
\end{subequations}
Here, $f_{i,ini}(y)$ is the initial concentration profile of the $i^{th}$ species \textcolor{black}{and the two boundary conditions correspond to zero flux condition at the free surface and base}. Equation~\eqref{eq:ch4_finalsegdiff} is a non-linear partial differential equation for $i^{th}$ species concentration ($f_i$). 
This equation accounts for the effects of local shear rate and viscosity on the concentration evolution. Since the viscosity of dense granular flow depends on both the local shear rate and pressure, the prediction of species concentration requires knowledge of the flow kinematics as well as pressure. 
For this, we solve the momentum balance equations for unidirectional, fully developed flow over a surface inclined at an angle $\theta$, which simplify to
\begin{align}
   &\rho_b \frac{\partial v_x }{\partial  t} =   \rho_b g \sin\theta - \frac{\partial \tau_{yx}}{\partial y},
   \label{eq:ch4_mombal_x}\\
  & P(y,t) =  g \cos \theta (1 - a \tan \theta )\int_{h}^y  \rho_b(y,t) dy.
    \label{eq:ch4_mombal_y}
\end{align}
Here, $v_x$ represents the mixture velocity in the $x-$ direction, $\rho_b$ is the bulk density, $\tau_{yx}$ denotes the shear stress, and $y = 0$ and $y = h$ represent the chute base and free surface, respectively. For mixtures of different-sized particles with the same density, the bulk density is given by $\rho_b = \phi \rho_p$, where $\phi$ is the local solids fraction and $\rho_p$ is the particle density.
The local packing fraction $\phi$ is obtained using the dilatancy law with the generalized inertial number for the granular mixtures (\cite{tripathi2011rheology}) as follows: 
\begin{equation}
    \phi(y,t)=\phi_{max}- \beta I_{mix}. 
    \label{eq:ch4_packing_fraction}
\end{equation}
$\phi_{max}$ and $\beta$ are the dilatancy law parameters obtained from DEM simulation data. The expression of inertial number for granular mixture having different size and identical density particles is given by $ I_{mix}=|\dot\gamma| d_{mix} /\sqrt {P/\rho_{p}}$, $\dot \gamma =dv_x/dy$ being the local shear rate. The shear stress ($\tau_{yx}$) and pressure ($P$) relate to each other by means of the inertial number dependent effective friction coefficient $\mu(I_{mix}) = |\tau_{yx}|/ P$. 
The JFP model by \cite{jop2006constitutivenature} empirically relates the variation of $\mu$ with $I$ as 
\begin{equation}
\mu(I_{mix}) = \mu_{s}+\frac{\mu_{m}-\mu_{s}}{1+I_{0}/I_{mix}}, 
\label{eq_5:mu_I_JFP}
\end{equation}
where, $\mu_s$, $\mu_m$, and $I_0$ are the rheological model parameters. The values of these parameters are taken from \cite{tripathi2013density}. In our simulations, the flow is started from an initial condition of zero velocity and we incorporate a rough, bumpy base to ensure the no-slip boundary condition at the base.  
Hence, we use the following initial and boundary conditions to solve the partial differential Equation~\eqref{eq:ch4_mombal_x}:
\begin{subequations}
   \begin{align}
        &IC: v_{x}(y,0) = 0,\\
        &BC1: v_{x}(0,t) = 0,\\
        &BC2: \tau_{yx}(h,t) =0.
   \end{align} 
\label{eq:ch4_IC_BC_momentum}
\end{subequations}
Further, the species concentration and velocity fields are obtained by solving the segregation-diffusion equation (Equation~\eqref{eq:ch4_finalsegdiff}) and momentum balance equation (Equation~\eqref{eq:ch4_mombal_x}) simultaneously using the PDEPE solver along with the initial and boundary conditions given in Equation~\eqref{eq:ch4_IC_BC_concentration_multi} and Equation~\eqref{eq:ch4_IC_BC_momentum}, respectively. We account for the dilatancy effects and update the layer height at any time using the relation $h(t) = h_{\text{min}} \phi_{\text{max}} / \phi_{\text{avg}}(t)$, as discussed in \cite{kumawat2025sizetransient}. To facilitate comparison with DEM results, we also perform simulations of spherical particles of identical density and different sizes flowing down an inclined plane. The key details of DEM simulations are reported in~\ref{sec:simulationMethod} of supplementary materials.

\section{Result and discussion}
\label{sec:results_discussion}
We now present a comparison between the predictions of the continuum model and discrete element method (DEM) simulations for multicomponent mixtures comprising more than two species. We begin by discussing the results for a \textcolor{black}{well-mixed configuration}, followed by those for \textcolor{black}{small-near-base and large-near-base configurations}.
\subsection{Well-mixed configuration}
\begin{figure}
    \centering
    \includegraphics[scale=0.4]{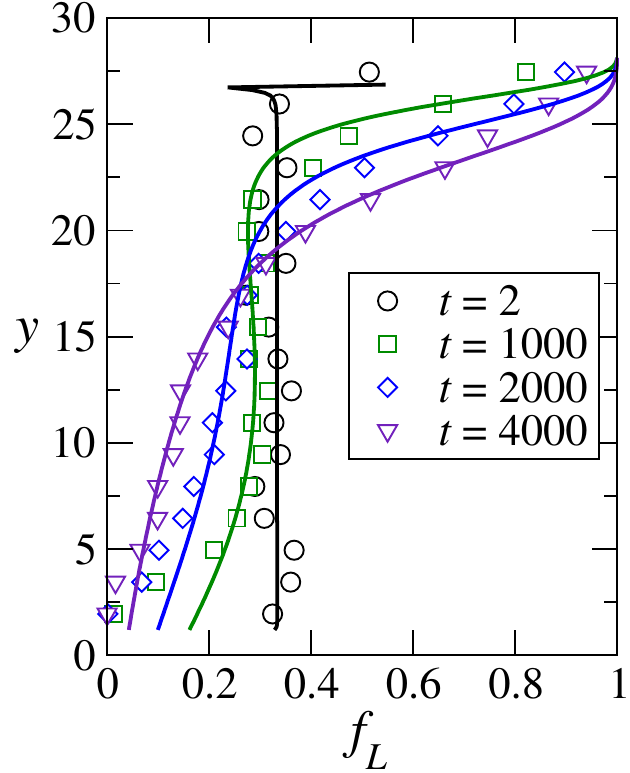}\put(-120,150){(a)} \quad 
    \includegraphics[scale=0.4]{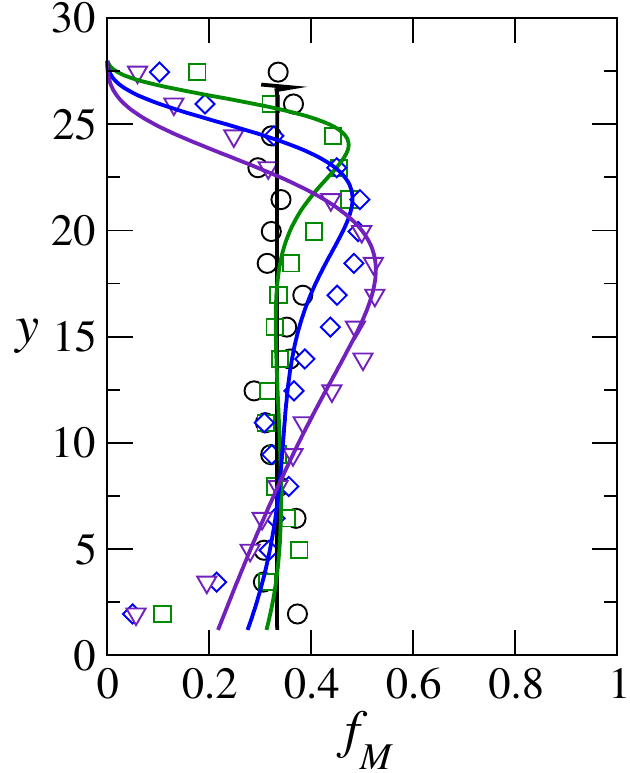}\put(-120,150){(b)} \quad
    \includegraphics[scale=0.4]{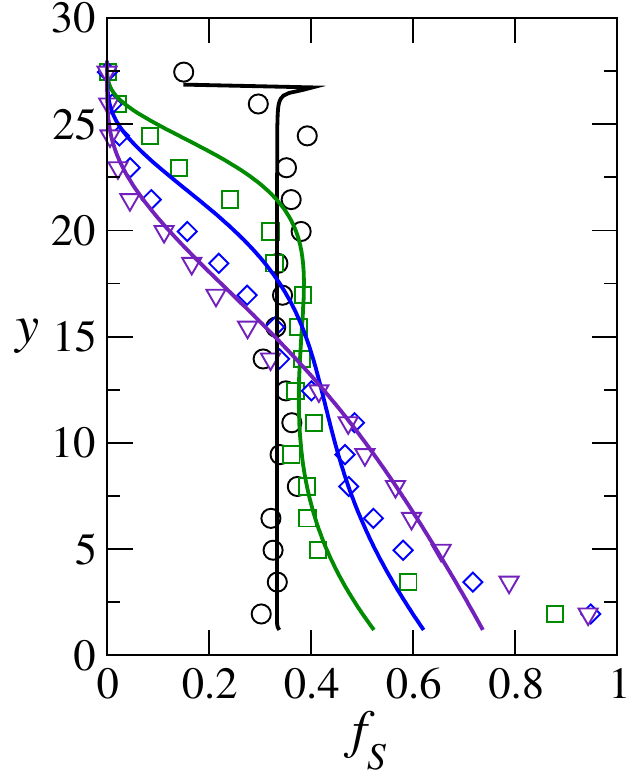}\put(-120,150){(c)}
    \caption{Instantaneous concentration profiles of (a) large, (b) medium, and (c) small particles in an equal composition ($f^T_L = f^T_M = f^T_S = 1/3$) ternary mixture with size ratio $ 1.5:1.25:1.0$ flowing down an inclined plane at an inclination angle $\theta = 25^\circ$. Symbols represent the DEM data, while solid lines correspond to continuum model predictions.}
\label{fig:instantaneous_r_1.5_1.25_1_theta_25}
\end{figure}
\begin{figure}
    \centering
    \includegraphics[width=0.45\linewidth]{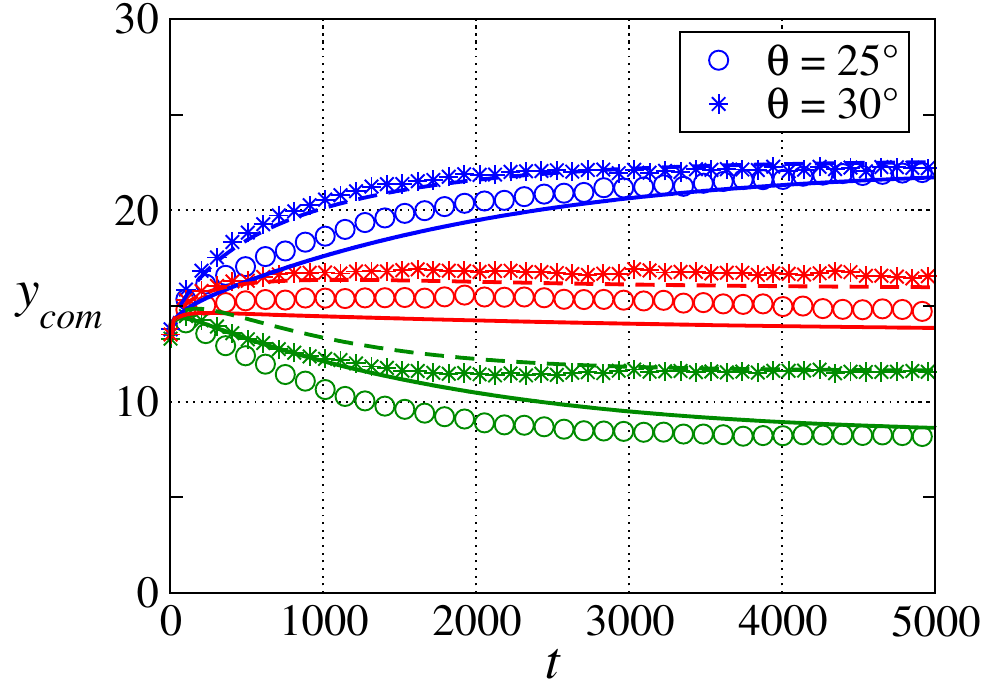}\put(-163,120){(a)} \quad  
    \includegraphics[width=0.45\linewidth]{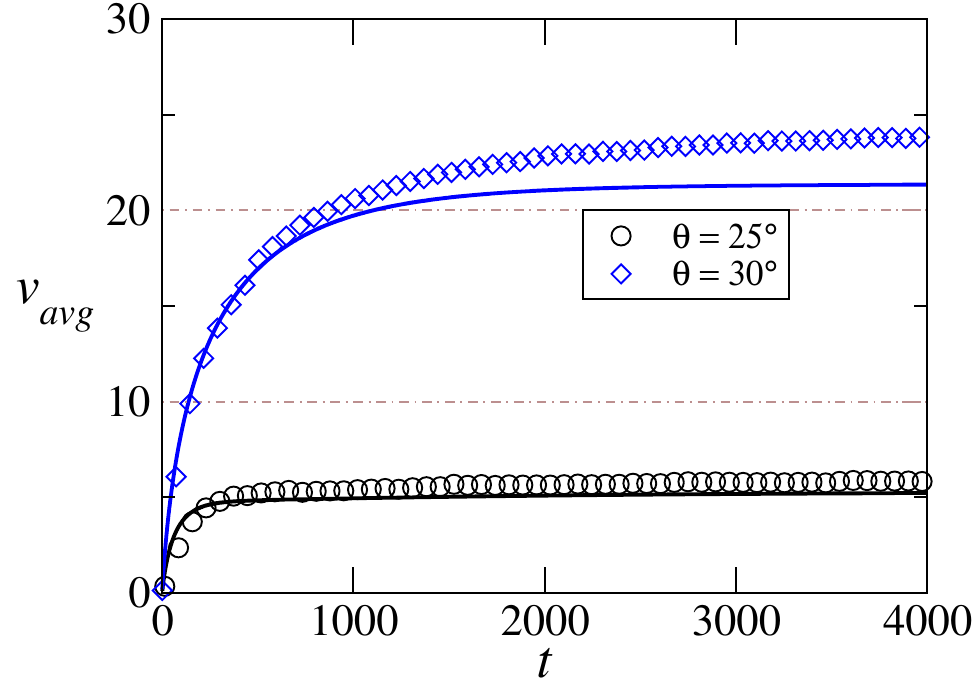}\put(-165,120){(b)}
    \caption{Effect of inclination angle: (a) Comparison of time evolution of $y_{com}$ for equal composition ternary mixture of size ratio $ 1.5:1.25:1.0$ at two different inclination angles $\theta = 25^\circ$ and $30^\circ$. (b) Time evolution of average mixture velocity. Symbols represent the DEM data, while solid lines correspond to continuum model predictions.}
\label{fig:y_com_1.5_1.25_1_theta_effect}
\end{figure}

Figure~\ref{fig:instantaneous_r_1.5_1.25_1_theta_25} presents the instantaneous concentration profiles for a ternary granular mixture with particle size ratios $1.5:1.25:1.0$, flowing down an inclined plane at an angle $\theta = 25^\circ$. The flow starts from a fully mixed initial condition, with equal volume fractions of large, medium, and small particles.
Figures~\ref{fig:instantaneous_r_1.5_1.25_1_theta_25}a -~\ref{fig:instantaneous_r_1.5_1.25_1_theta_25}c show the concentration \textcolor{black}{profiles} for large, medium, and small particles, respectively. The DEM simulation results are denoted by symbols, while the corresponding predictions from the continuum model are represented by solid lines \textcolor{black}{of same color}. 
As the flow evolves, size-driven segregation becomes evident. Large particles tend to move toward the free surface due to \textcolor{black}{the} net upward force, resulting in \textcolor{black}{their increased concentration} near the \textcolor{black}{free surface}, as shown in Figure~\ref{fig:instantaneous_r_1.5_1.25_1_theta_25}a. In contrast, small particles percolate downward, leading to their enrichment near the bottom of the flowing layer (Figure~\ref{fig:instantaneous_r_1.5_1.25_1_theta_25}c). The medium-sized particles tend to rise upwards near the basal region but move downwards in the region near the free surface. Due to this reason, the medium species exhibits non-monotonic concentration with maximum concentration at \textcolor{black}{some} distance from the free surface (Figure~\ref{fig:instantaneous_r_1.5_1.25_1_theta_25}b). The agreement between the \textcolor{black}{same color lines and symbols} indicates that the continuum model successfully captures the spatial and temporal evolution of species segregation in a moderately polydisperse mixture. Slightly larger deviations, however, are observed for the smallest size species in the regions of low concentration. We note that this slight difference in the concentration profile for low concentration of small species was observed by~\cite{tripathi2021size} in case of binary mixtures as well. \textcolor{black}{The good match of DEM results with the model predictions can be more conveniently observed in figure~\ref{fig:y_com_1.5_1.25_1_theta_effect}a which shows the time evolution of the center of mass height $y_{com}$ for each species in the ternary mixture using circles for this case.}
\textcolor{black}{In addition, we report results for $\theta = 30^\circ$ to explore the effect of inclination angle on segregation behavior for the same mixture. The time evolution of the species center of mass height ($y_{com}$) obtained from DEM data for $\theta = 30^\circ$ (shown using $\star$ symbols) is again well predicted by the continuum model for all the three species.} 
Differences in the evolution of the center of mass $y_{com}$ are observed for the small (green), medium (red) species, and large (blue) species, clearly at the two inclination angles. \textcolor{black}{Since our approach solves for the momentum balance equations by intercoupling the rheology and segregation, we are also able to predict the evolution of the average mixture velocity $v_{avg}$ with time (Figure~\ref{fig:y_com_1.5_1.25_1_theta_effect}b).} A significant increase in $v_{avg}$ is observed at $\theta = 30^\circ$, where the average velocity is more than four times higher than that at $\theta = 25^\circ$. The flow becomes substantially faster as the inclination angle increases, and the model predictions for the average velocity are found to be somewhat smaller than those observed in DEM data. This difference can be mitigated by accounting for the variation of the normal stress difference parameter $a$ with inertial number as done in the work of \citet{sahu_kumawat_agrawal_tripathi_2023}. However, since its effect on segregation is small, we continue with the simpler choice of constant $a$ value in this work.

\begin{figure}
    \centering
     \includegraphics[scale=0.4]{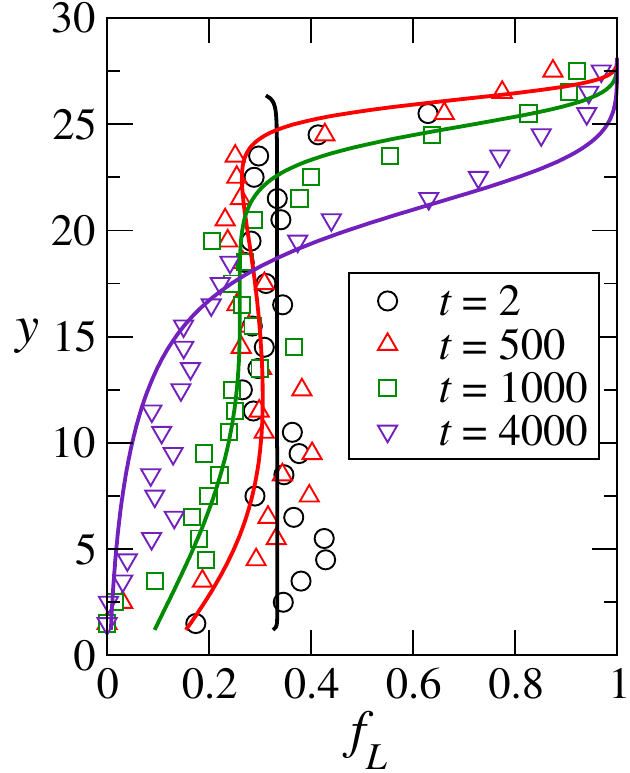}\put(-120,150){(a)} \quad 
       \includegraphics[scale=0.4]{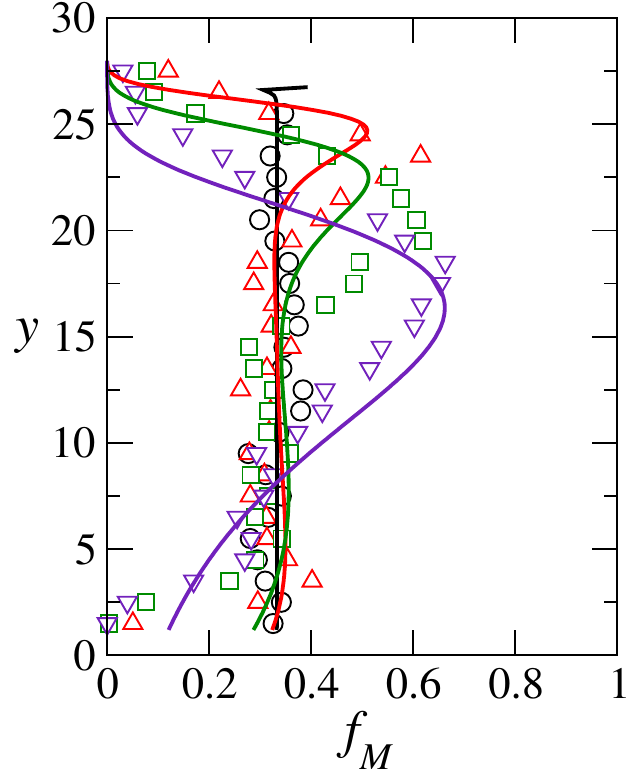}\put(-120,150){(b)} \quad 
         \includegraphics[scale=0.4]{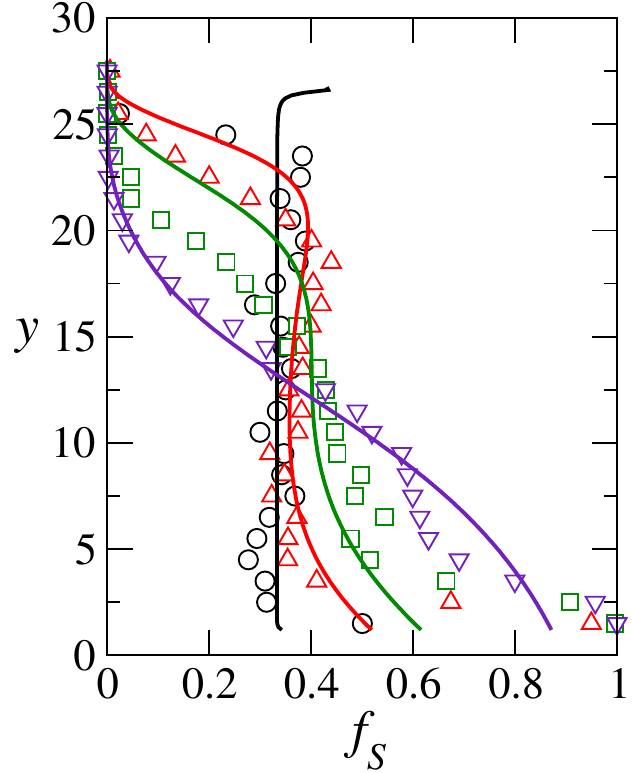}\put(-120,150){(c)}
    \caption{Instantaneous concentration profiles of (a) large, (b) medium, and (c) small particles in an equal composition ($f^T_L = f^T_M = f^T_S = 1/3$) ternary mixture with size ratio $2.0:1.5:1.0$ flowing down an inclined plane at an inclination angle $\theta = 25^\circ$. Symbols represent the DEM data while solid lines correspond to continuum model predictions.}
\label{fig:instantaneous_r_2.0_1.5_1_theta_25}
\end{figure}
\begin{figure}
    \centering
    \includegraphics[scale=0.12, trim=550 60 590 60, clip]{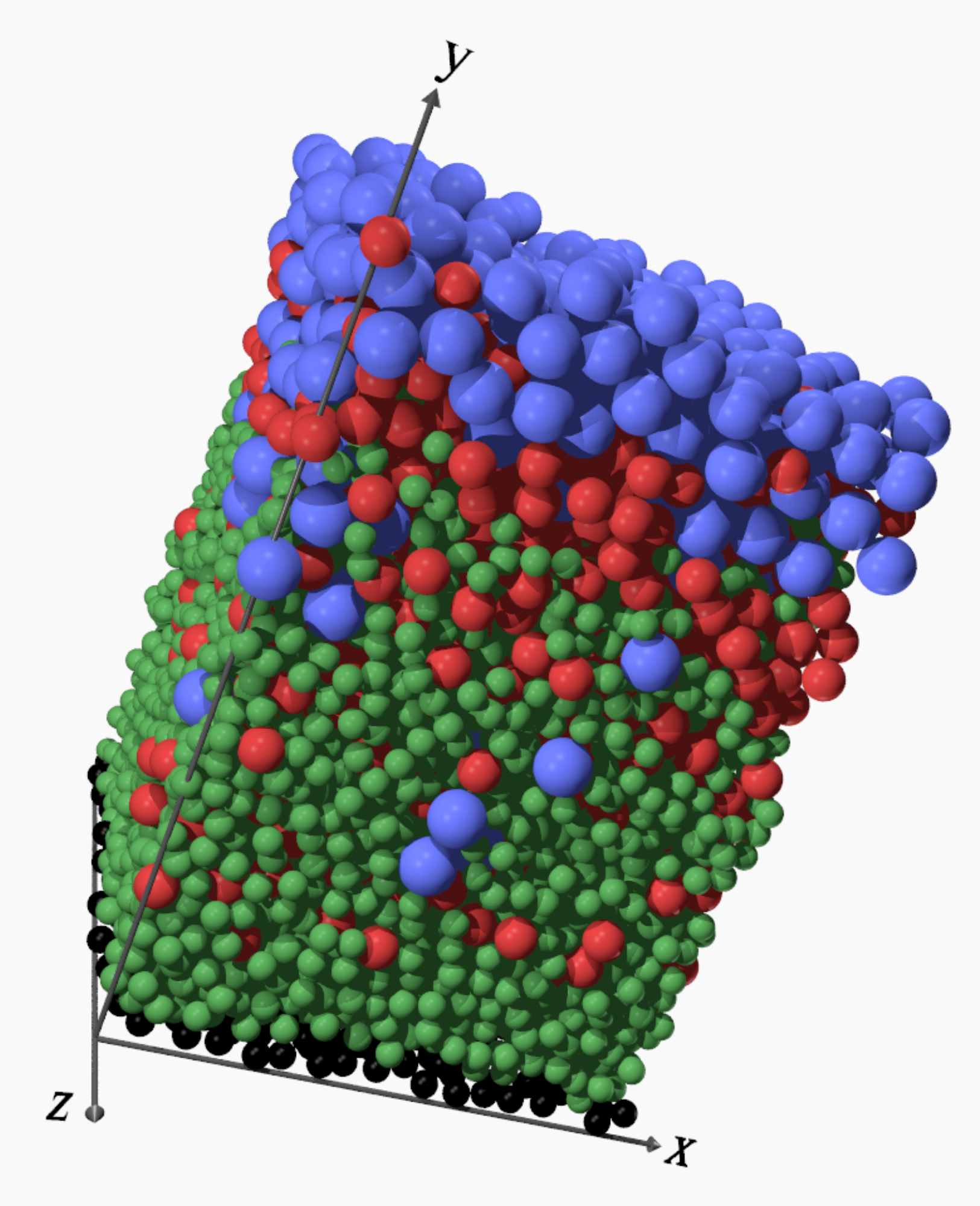}\put(-90,110){(a)} \quad \quad \quad 
     \includegraphics[scale=0.12, trim=550 60 590 60, clip]{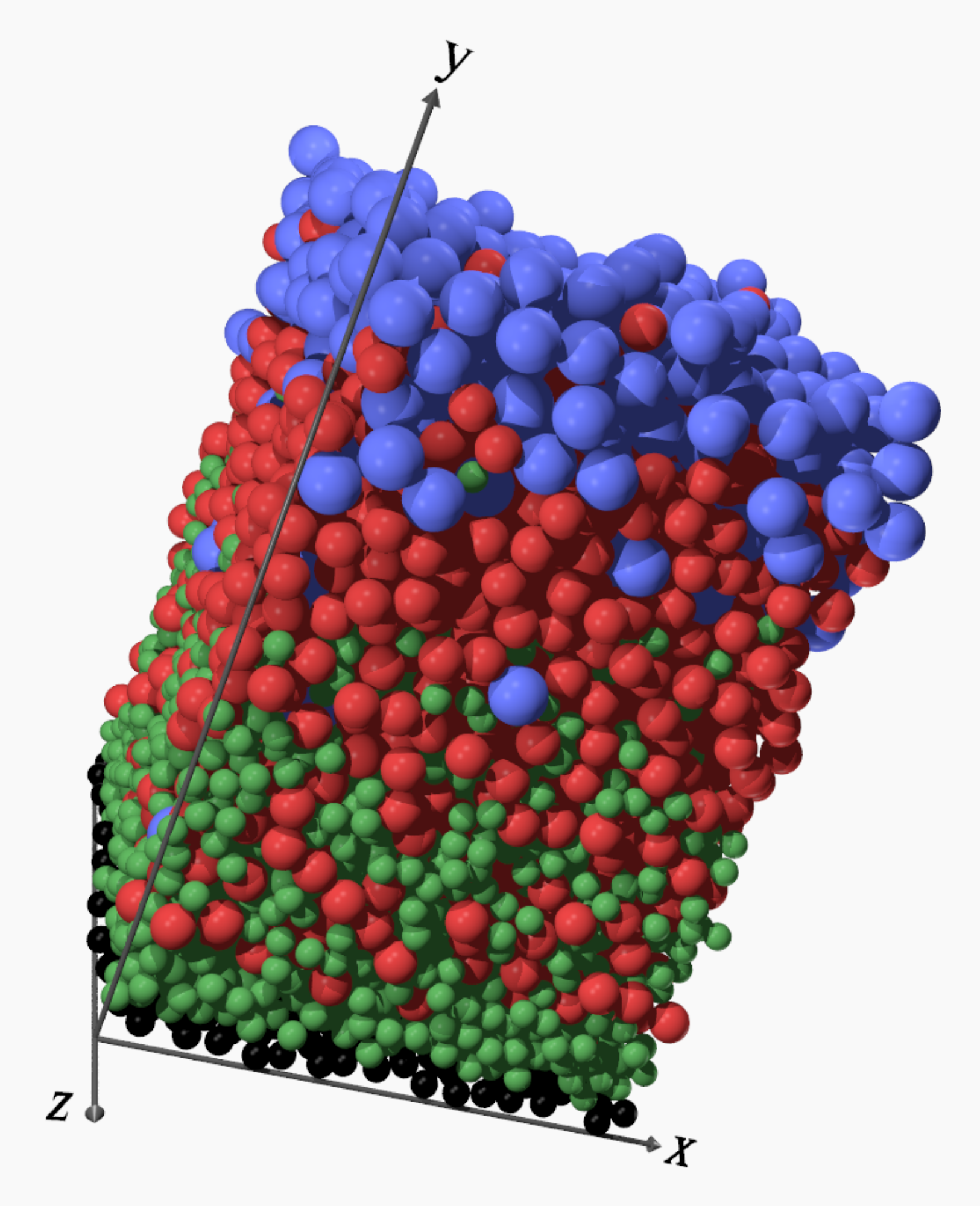}\put(-90,110){(b)}\quad \quad  \quad 
      \includegraphics[scale=0.12, trim=550 60 590 60, clip]{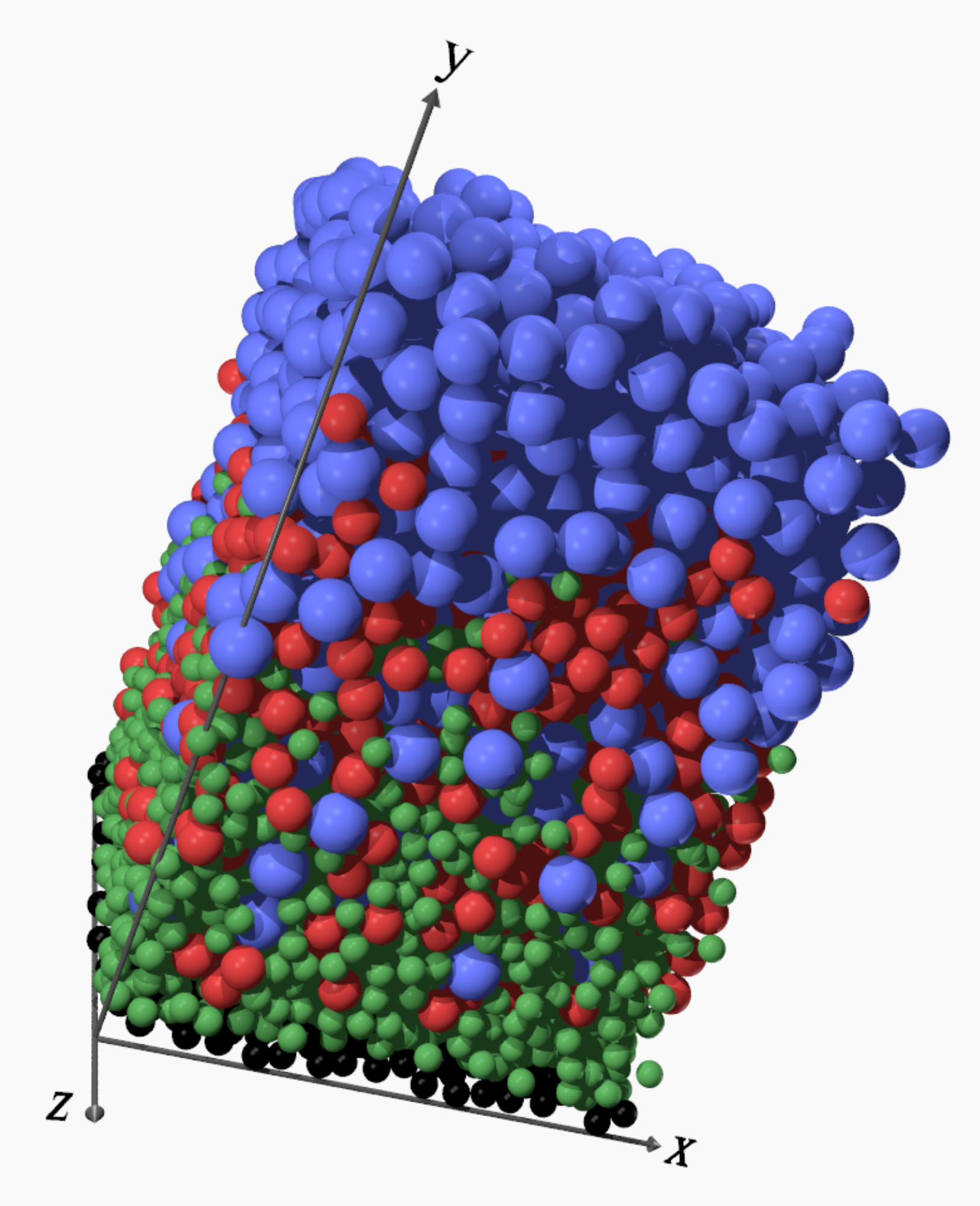}\put(-90,110){(c)} \quad
       \vspace{0.2cm}
     \includegraphics[scale=0.27, trim=0 0 0 20, clip]{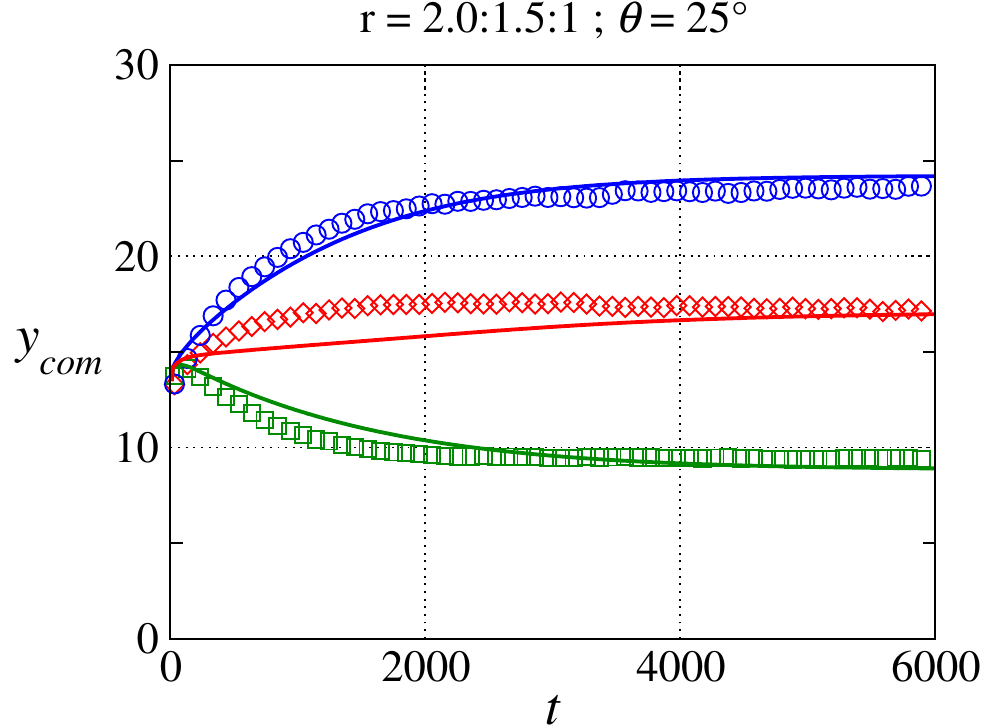}\put(-120,93){(d)}\put(-100,90){\tiny $f_L = 0.25; f_M = 0.25; f_S = 0.50$} \hfill 
    \includegraphics[scale=0.27, trim=0 0 0 20, clip]{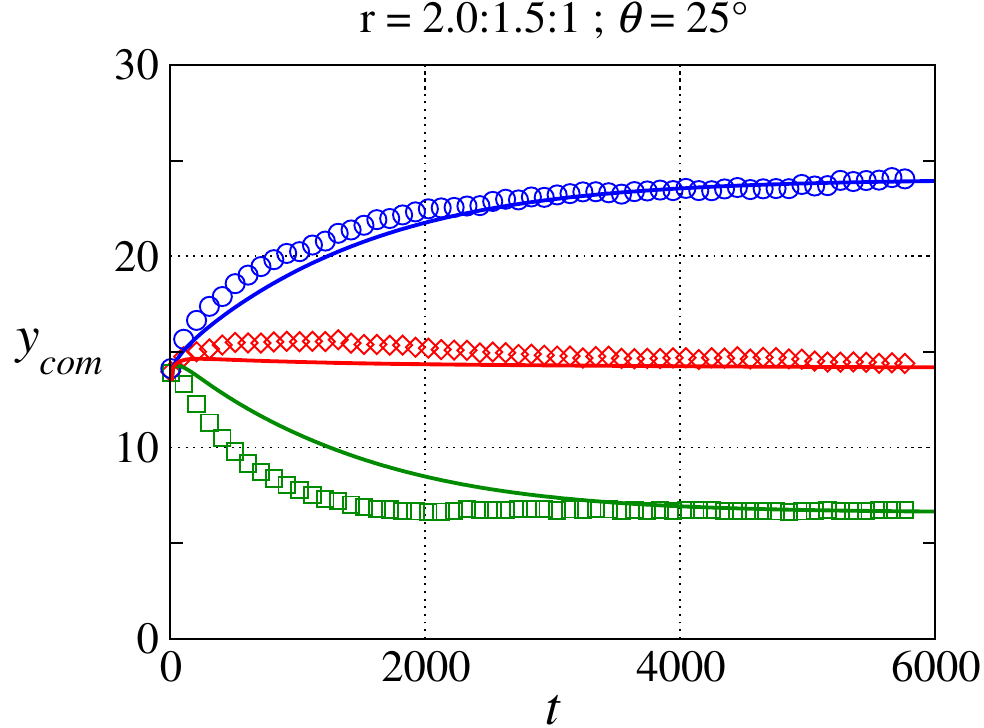}\put(-120,93){(e)}\put(-100,90){\tiny $f_L = 0.25; f_M = 0.50; f_S = 0.25$} \hfill
    \includegraphics[scale=0.27, trim=0 0 0 20, clip]{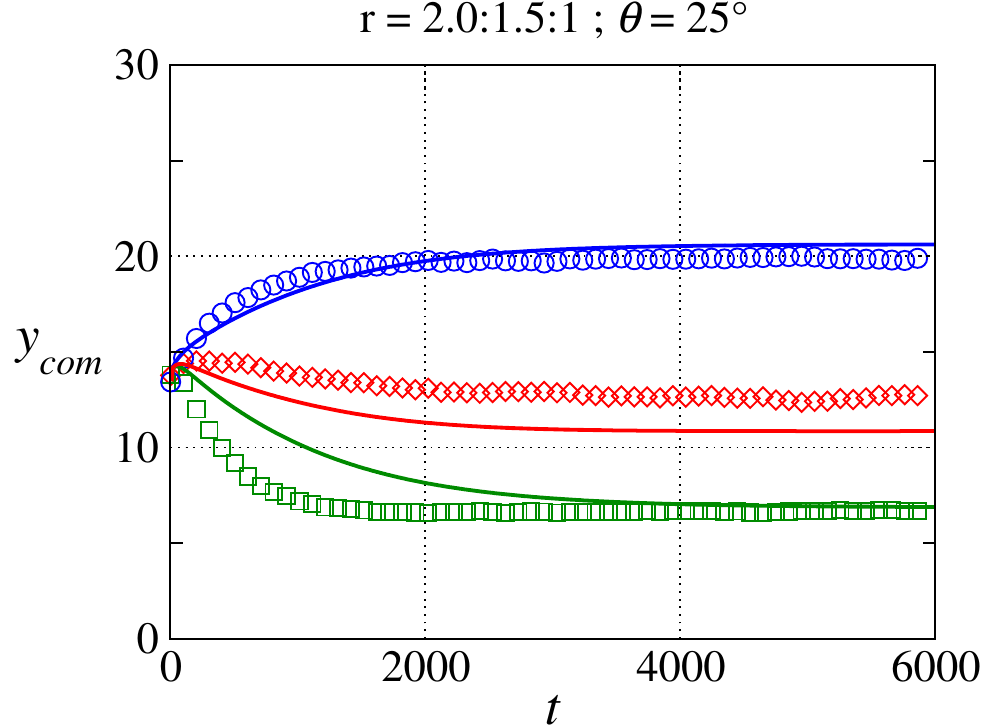}\put(-120,93){(f)} \put(-100,90){\tiny $f_L = 0.50; f_M = 0.25; f_S = 0.25$}\hfill
    \vspace{0.2cm}
     \includegraphics[scale=0.27, trim=0 0 0 20, clip]{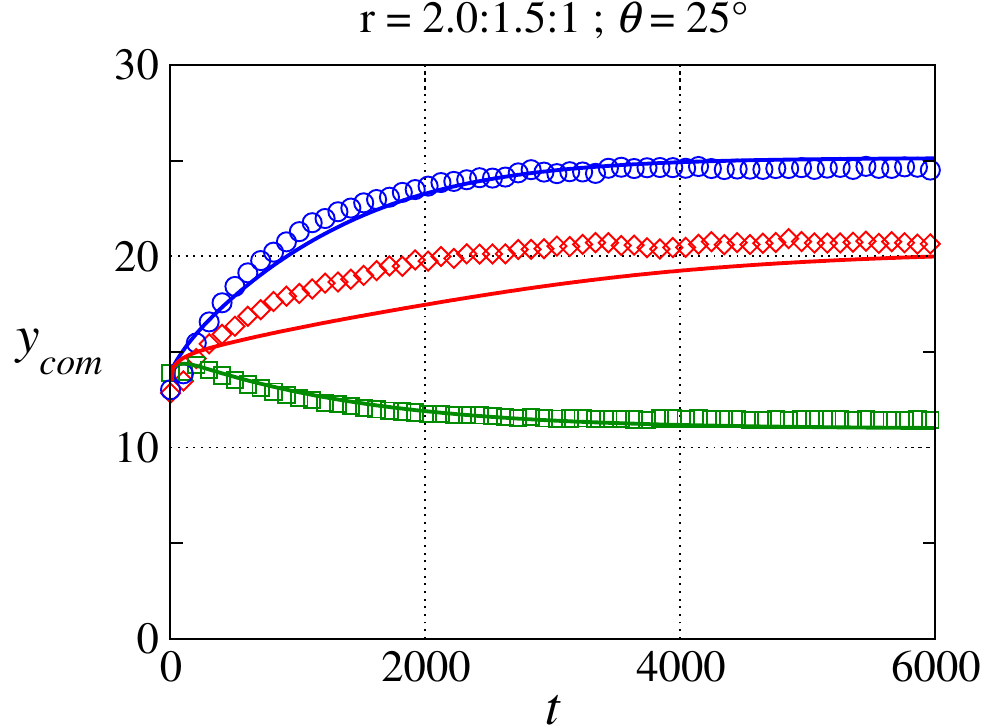}\put(-120,93){(g)}\put(-90,90){\tiny $f_L = 0.2; f_M = 0.1; f_S = 0.7$} \hfill
    \includegraphics[scale=0.27, trim=0 0 0 20, clip]{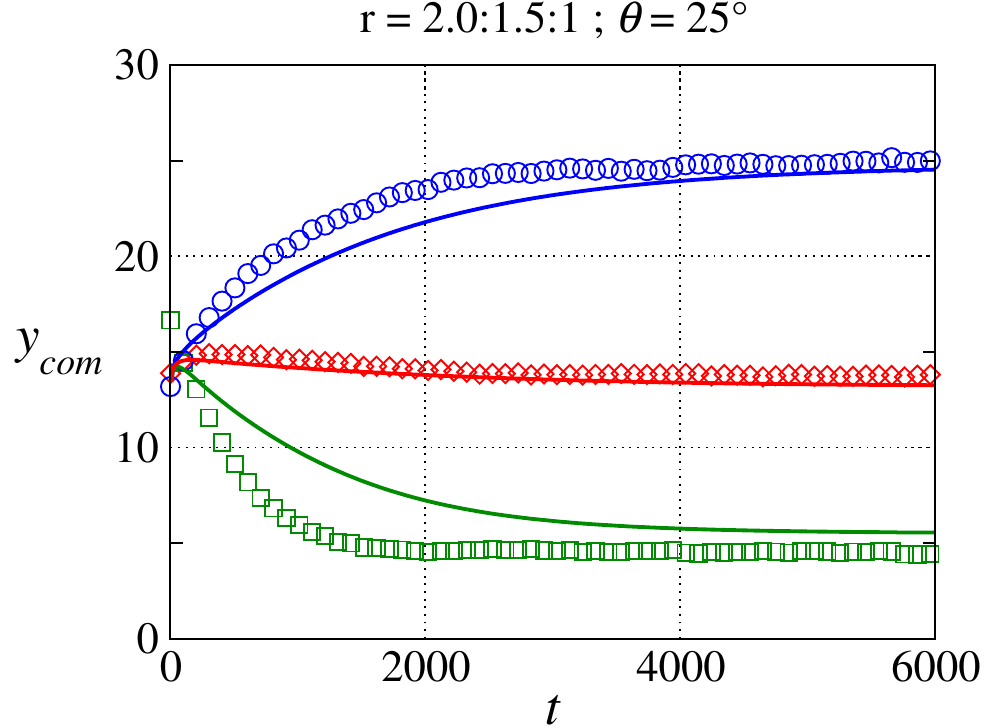}\put(-120,93){(h)}\put(-90,90){\tiny $f_L = 0.2; f_M = 0.7; f_S = 0.1$} \hfill 
    \includegraphics[scale=0.27, trim=0 0 0 20, clip]{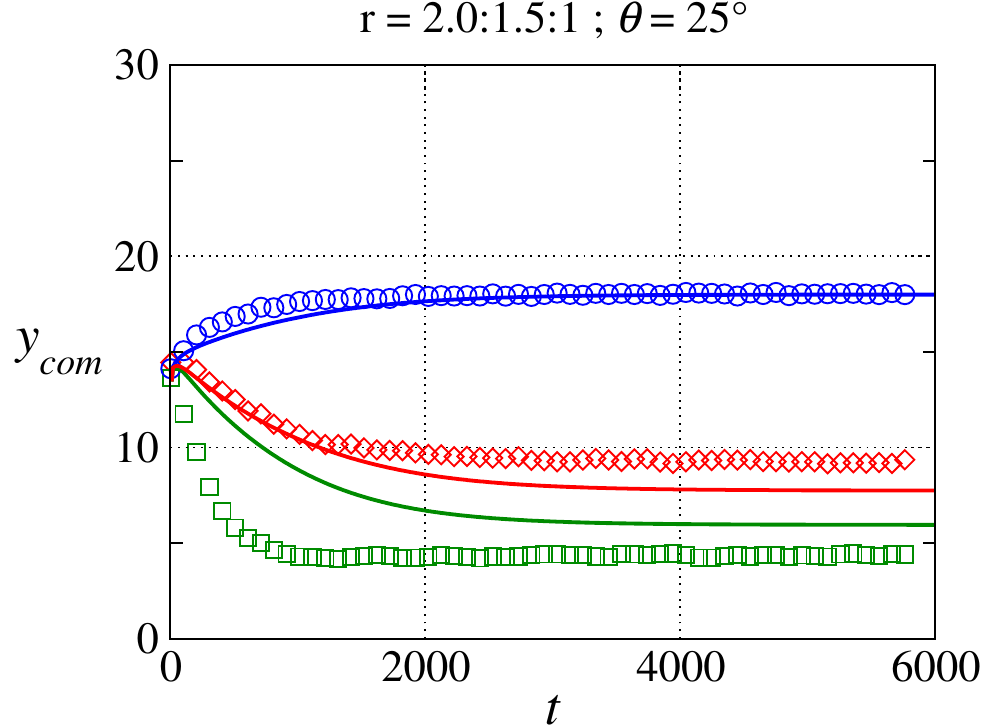}\put(-120,93){(i)} \put(-90,90){\tiny $f_L = 0.7; f_M = 0.2; f_S = 0.1$}
    \vspace{0.3cm}
     \includegraphics[scale=0.12, trim=550 60 590 60, clip]{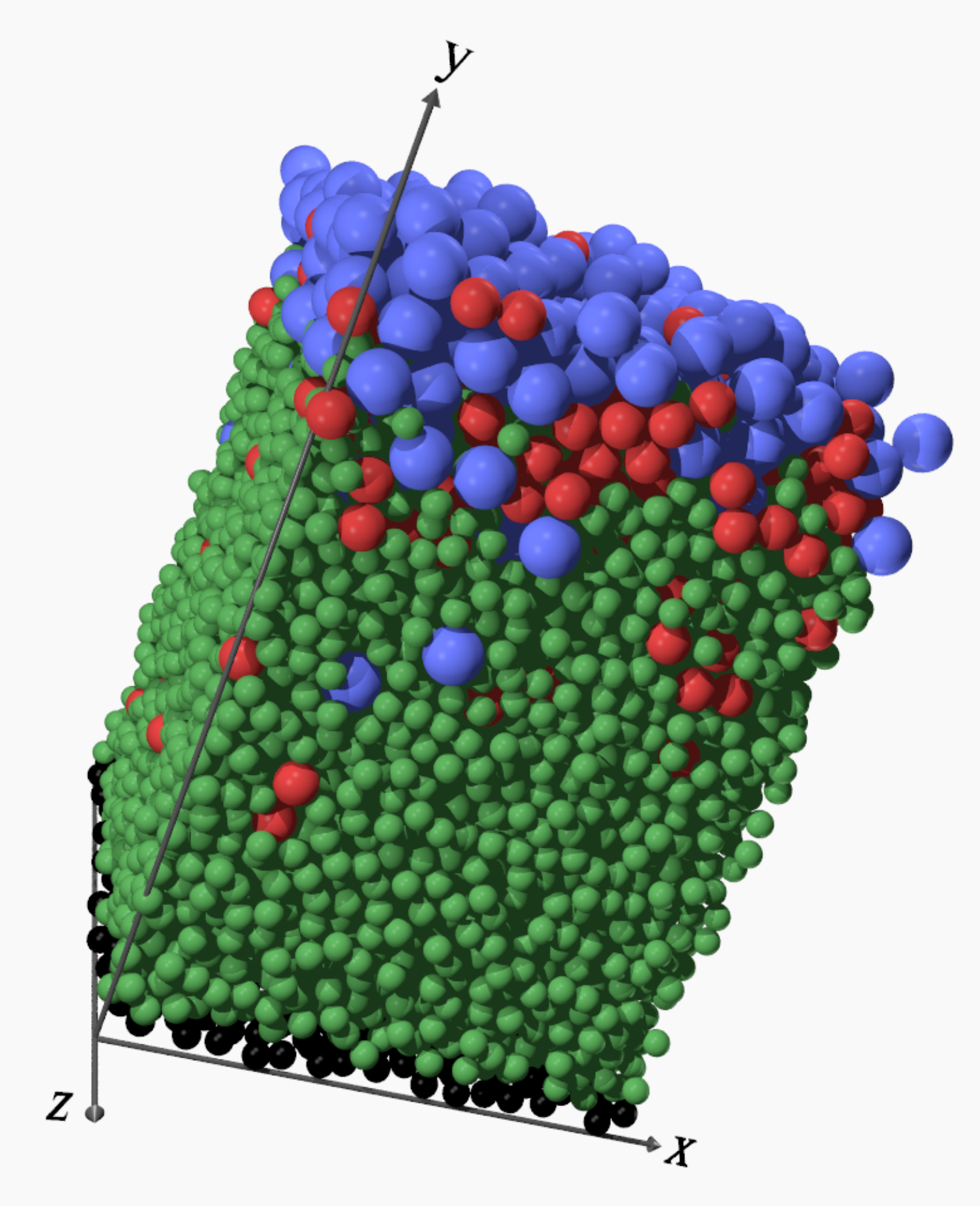}\put(-90,110){(j)} \quad \quad \quad 
     \includegraphics[scale=0.12, trim=550 60 590 60, clip]{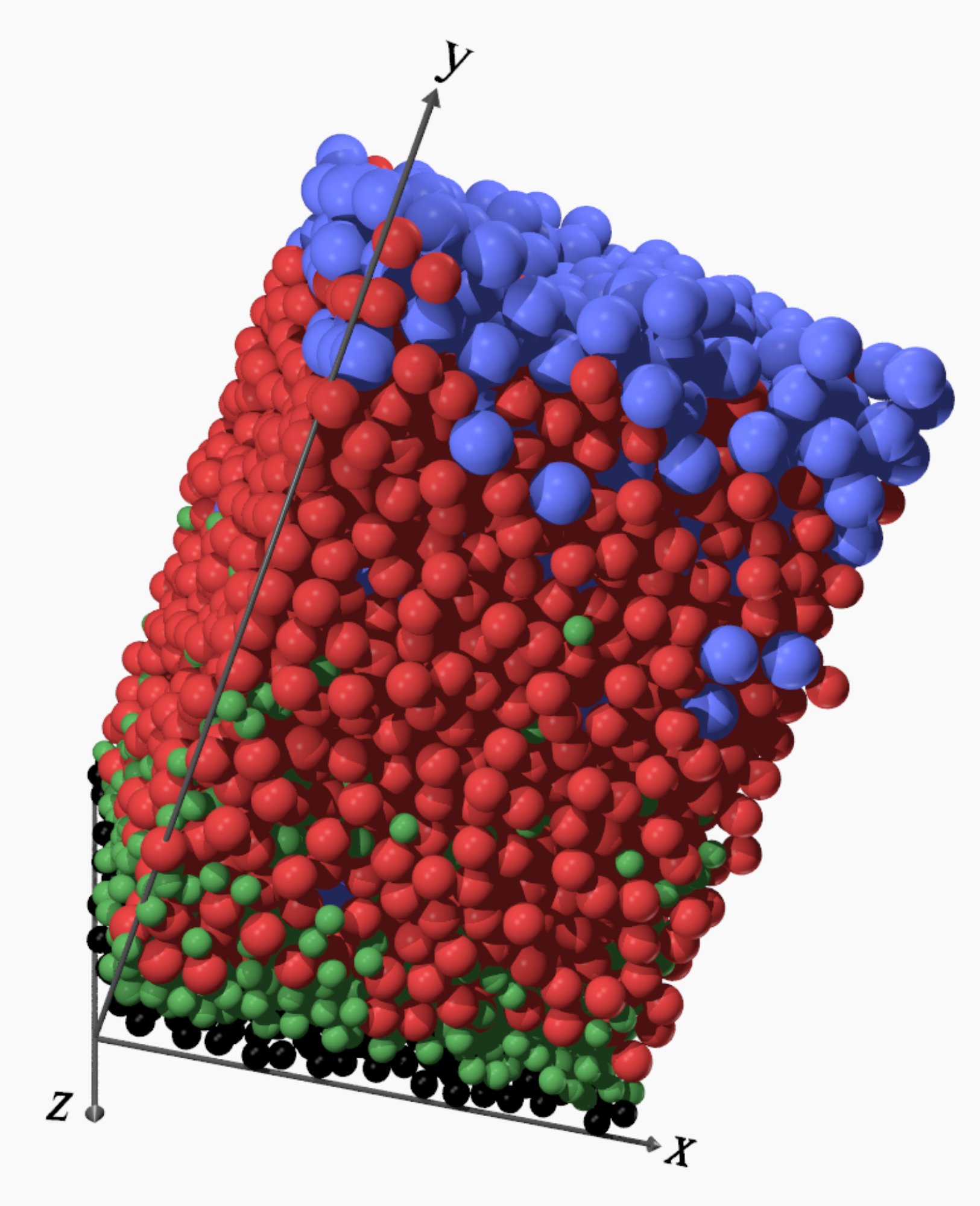} \put(-90,110){(k)}\quad \quad  \quad 
      \includegraphics[scale=0.12, trim=550 60 590 60, clip]{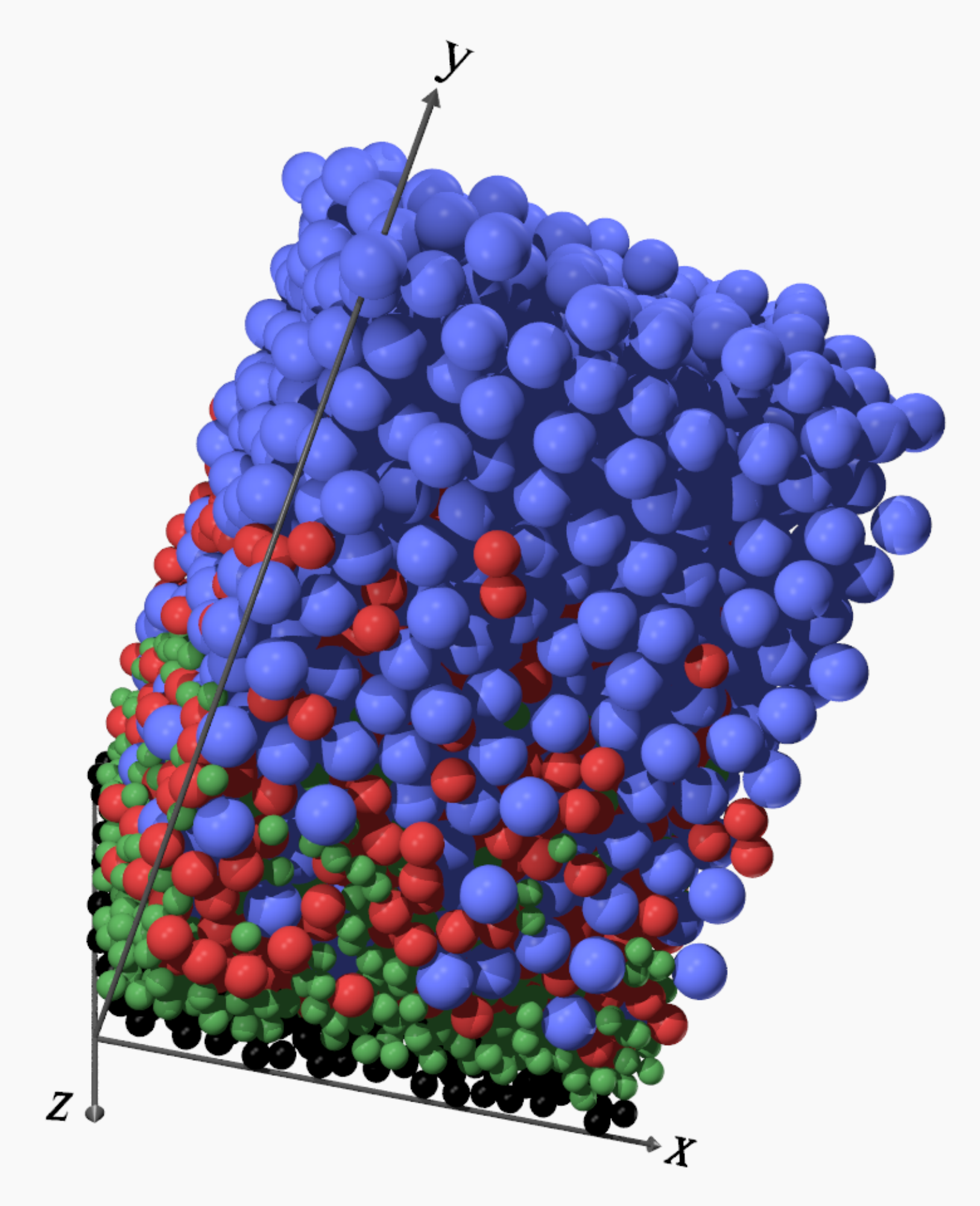} \put(-90,110){(l)}\quad
    \caption{ Time evolution of the center of mass height ($y_{com}$) for ternary granular mixtures with particle size ratios $2.0:1.5:1.0$, flowing down an inclined plane at $\theta = 25^\circ$. Panels (d)–(i) correspond to different volume fraction combinations of large (L), medium (M), and small (S) particles: (d) $f_L = 0.25; f_M = 0.25; f_S = 0.50$, (e) $f_L = 0.25; f_M = 0.50; f_S = 0.25$, (f) $f_L = 0.50; f_M = 0.25; f_S = 0.25$, (g) $f_L = 0.2; f_M = 0.1; f_S = 0.7$, (h) $f_L = 0.2; f_M = 0.7; f_S = 0.1$, and (i) $f_L = 0.7; f_M = 0.2; f_S = 0.1$ flowing at inclination angle $\theta = 25^\circ$. Panels (a) - (c) present the steady-state DEM snapshots corresponding to (d) - (f), while panels (j) - (l) correspond to (g) - (i).}
    \label{fig:y_com_2_1.5_1_diff_comp}
\end{figure}

Figure~\ref{fig:instantaneous_r_2.0_1.5_1_theta_25} shows the instantaneous concentration profiles for a ternary granular mixture with particle size ratios $2.0:1.5:1.0$, flowing down an inclined plane at an angle $\theta = 25^\circ$. \textcolor{black}{As before,} the flow is started from a fully mixed initial condition, with equal volume fractions of large, medium, and small particles.
Figures~\ref{fig:instantaneous_r_2.0_1.5_1_theta_25}a -~\ref{fig:instantaneous_r_2.0_1.5_1_theta_25}c show the concentration distributions for large, medium, and small particles, respectively. The DEM simulation results are denoted by symbols, while the corresponding predictions from the continuum model are represented by solid lines. 
\textcolor{black}{As} observed in figure~\ref{fig:instantaneous_r_1.5_1.25_1_theta_25}, the concentration profiles in figure~\ref{fig:instantaneous_r_2.0_1.5_1_theta_25} \textcolor{black}{also} show that large particles move toward the free surface and small particles settle near the base. However, the degree of segregation is noticeably enhanced in this case due to higher size ratios. Again, the continuum model appears to do a \textcolor{black}{reasonable} predictions in this case as well, \textcolor{black}{though slightly larger deviations are observed for the smallest species.}

We next consider the effect of initial mixture composition on the segregation dynamics of a ternary granular mixture with particle size ratio $ 2.0:1.5:1.0$. As before, the results are reported for a mixture flowing down an inclined plane at $\theta = 25^\circ$. 
Figures~\ref{fig:y_com_2_1.5_1_diff_comp}a -\ref{fig:y_com_2_1.5_1_diff_comp}c show the DEM snapshots for the final segregated state for mixtures in which two species \textcolor{black}{have} volume fractions equal to $0.25$, while the third is present in a large proportion with $50\%$ of overall concentration. Specifically, figure~\ref{fig:y_com_2_1.5_1_diff_comp}a shows the steady state DEM snapshots for compositions of the ternary mixture as $f_L = 0.25$, $f_M = 0.25$, $f_S = 0.50$. Similarly, Figure~\ref{fig:y_com_2_1.5_1_diff_comp}b corresponds to $f_L = 0.25$, $f_M = 0.50$, $f_S = 0.25$, while figure~\ref{fig:y_com_2_1.5_1_diff_comp}c corresponds to $f_L = 0.50$, $f_M = 0.25$, $f_S = 0.25$. The corresponding time evolution of the center of mass, $y_{\text{com}}$, for each species is shown in figures~\ref{fig:y_com_2_1.5_1_diff_comp}d,~\ref{fig:y_com_2_1.5_1_diff_comp}e and \ref{fig:y_com_2_1.5_1_diff_comp}f, \textcolor{black}{respectively,} showing the upward motion of large particles and the downward movement of small particles. The behavior of intermediate-sized particles center of mass depends on the relative proportions of the small and large species in the mixture.
These figures provide insight into how increasing or decreasing the proportion of species affects the overall segregation behavior of the mixture. Similarly, figures~\ref{fig:y_com_2_1.5_1_diff_comp}g - \ref{fig:y_com_2_1.5_1_diff_comp}i show $y_{com}$ variation corresponding to mixture compositions in which all three species have different volume fractions, representing mixtures with very little composition of two species along with a significantly dominant species. Figures~\ref{fig:y_com_2_1.5_1_diff_comp}g, \ref{fig:y_com_2_1.5_1_diff_comp}h, and \ref{fig:y_com_2_1.5_1_diff_comp}i show time evolution of $y_{com}$ for different mixtures with $f_L = 0.2, f_M = 0.1, f_S = 0.7$; $f_L = 0.2, f_M = 0.7, f_S = 0.1$; and $f_L = 0.7, f_M = 0.2, f_S = 0.1$ respectively. In all cases, the trajectories of $y_{com}$ show that the larger particles migrate toward the free surface and the smaller ones settle near the base. As expected, the $y_{com}$ values at steady state are different for different cases. \textcolor{black}{This is also evident from} the DEM snapshots at steady state corresponding to these mixtures shown in Figures~\ref{fig:y_com_2_1.5_1_diff_comp}j - \ref{fig:y_com_2_1.5_1_diff_comp}l.
The continuum model predictions (solid lines) again are found to match reasonably well with the DEM data (symbols) for nearly all the cases, except for \textcolor{black}{ small/medium species in} mixtures having low concentration of the small/medium species. 
\begin{figure}
    \centering
    \includegraphics[scale=0.45, trim=70 0 40 0, clip]{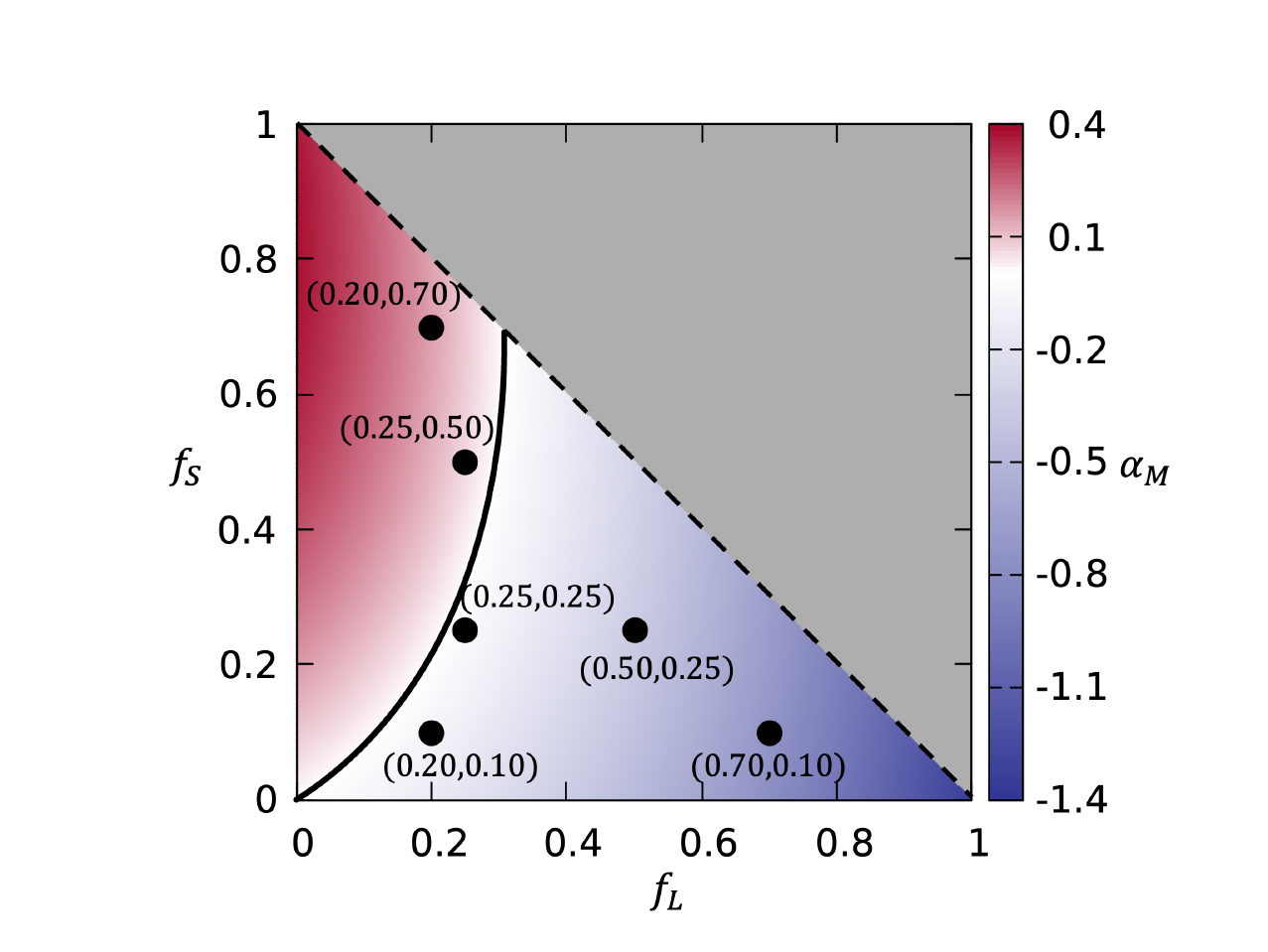} \put(-125,155){$f_L + f_S > 1$}
\caption{Variation of $\alpha_M (f_L,f_S)$ for medium-sized species in a ternary mixture having size ratio $ 2.0:1.5:1$. Solid line represents the contour for $\alpha_M (f_L,f_S) = 0$ and the dashed line corresponds to $f_L + f_S = 1$ with $f_M = 0$.}
\label{fig:mediumspecies_alpha_variation}
\end{figure}

The centre of mass evolution of the medium species in figures~\ref{fig:y_com_2_1.5_1_diff_comp} can be better understood by looking at figure~\ref{fig:mediumspecies_alpha_variation} which shows the variation of dimensionless net upward force on the medium size species $\alpha_M$ with the local composition of the ternary mixture at inclination angle $\theta = 25^\circ$. In this figure, a ternary mixture of composition $f_L,f_S$, \& $f_M$ is represented by a point ($f_L,f_S$) on the $f_L - f_S$ plane and the value of $f_M$ can be calculated as $f_M = 1 - (f_L + f_S)$. Only region below the line $f_L + f_S = 1$ can represent realistic mixture compositions since the grey region corresponds to $f_L + f_S >1$. The lower left vertex corresponds to $f_L = f_S = 0$ with $f_M = 1$ i.e., pure medium size species. Similarly, lower right vertex corresponds to pure large size species $f_L = 1$ and the top left vertex with $f_S = 1$ corresponds to pure small size species. While the vertices of the triangle color map correspond to a pure species state, binary mixtures are represented by the edge of the triangular color map that joins the pure species vertices. Thus, horizontal line at $f_S = 0$ represents all possible binary mixtures of medium and large species with the concentration of medium species given as $f_M = 1 - f_L$. Similarly, vertical edge with $f_L = 0$ represents all possible binary mixtures of small and medium species with the medium species concentration $f_M = 1 - f_S$. The dashed line (given by $f_S = 1 - f_L$) joining the vertex ($0,1$) and vertex ($1,0$) represents binary mixture of large and small species and $f_M = 0$ for all points on this line. The solid black line represents the no segregation contour for the medium species, obtained from $\alpha_M (f_L,f_S) = 0$ using equation~\ref{eq:alpha_M_mediumspecies_ternary}. Note that $\alpha_M$ remains positive (red region) in the dilute limit of large particles and becomes negative (blue region) in the dilute limit of small particles. The white band near the zero-contour represents compositions for which $\alpha_M \approx 0$, implying weak segregation of the medium species.
In figures~\ref{fig:y_com_2_1.5_1_diff_comp}d and~\ref{fig:y_com_2_1.5_1_diff_comp}g, the medium species move upward, indicating $\alpha_M>0$. This agrees with the corresponding composition points $(0.25,0.50)$ and $(0.20,0.70)$, which lie in the positive-$\alpha_M$ region of figure~\ref{fig:mediumspecies_alpha_variation}. In contrast, figures~\ref{fig:y_com_2_1.5_1_diff_comp}e and~\ref{fig:y_com_2_1.5_1_diff_comp}h show that the centre of mass of the medium species remains nearly unchanged with time. This suggests that $\alpha_M$ is close to zero, consistent with the points $(0.25,0.25)$ and $(0.20,0.10)$ falling near the zero-contour region. Finally, in figures~\ref{fig:y_com_2_1.5_1_diff_comp}f and~\ref{fig:y_com_2_1.5_1_diff_comp}i, the downward motion of the medium species reflects $\alpha_M<0$, as also observed at the corresponding points $(0.50,0.25)$ and $(0.70,0.20)$ in figure~\ref{fig:mediumspecies_alpha_variation}.

\begin{figure}
    \centering
      \includegraphics[scale=0.125, trim=560 0 590 0, clip]{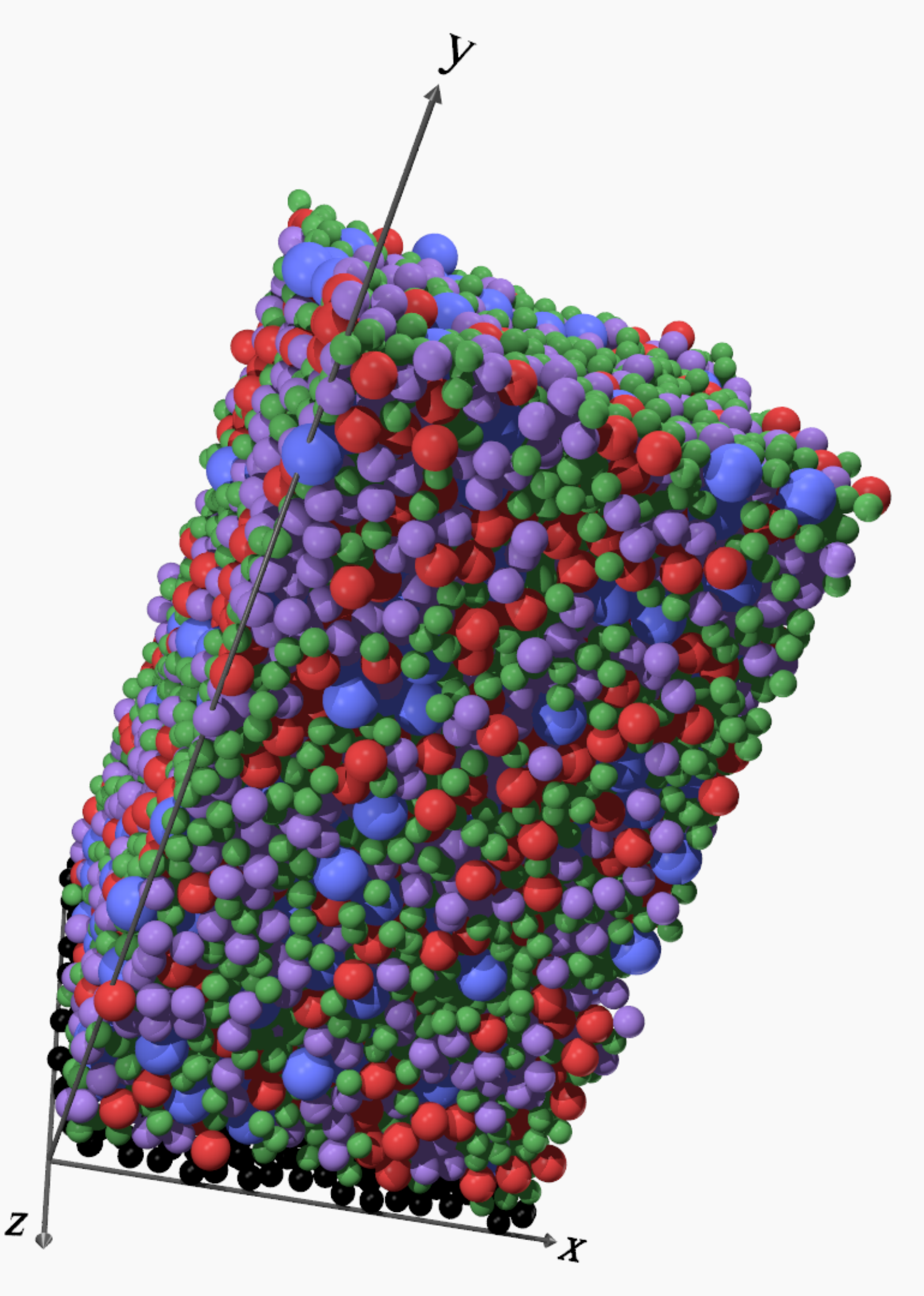} \put(-75,140){Initial state}\put(-95,125){(a)}
\includegraphics[scale=0.37, trim=0 0 0 20, clip]{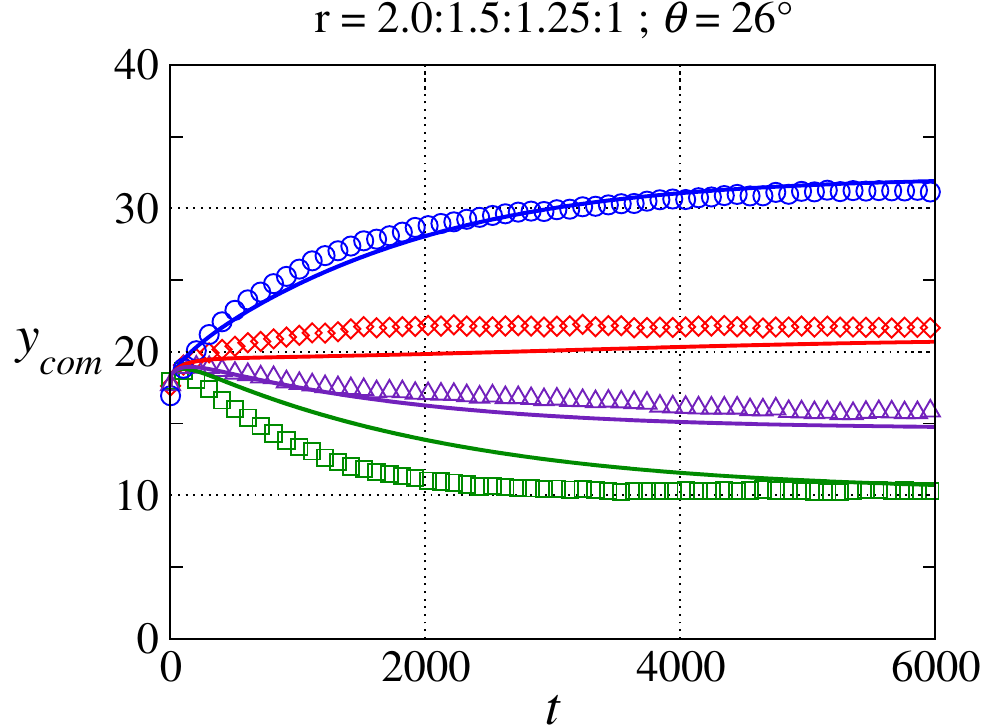}\put(-160,125){(b)} \quad 
    \includegraphics[scale=0.125, trim=550 0 570 0, clip]{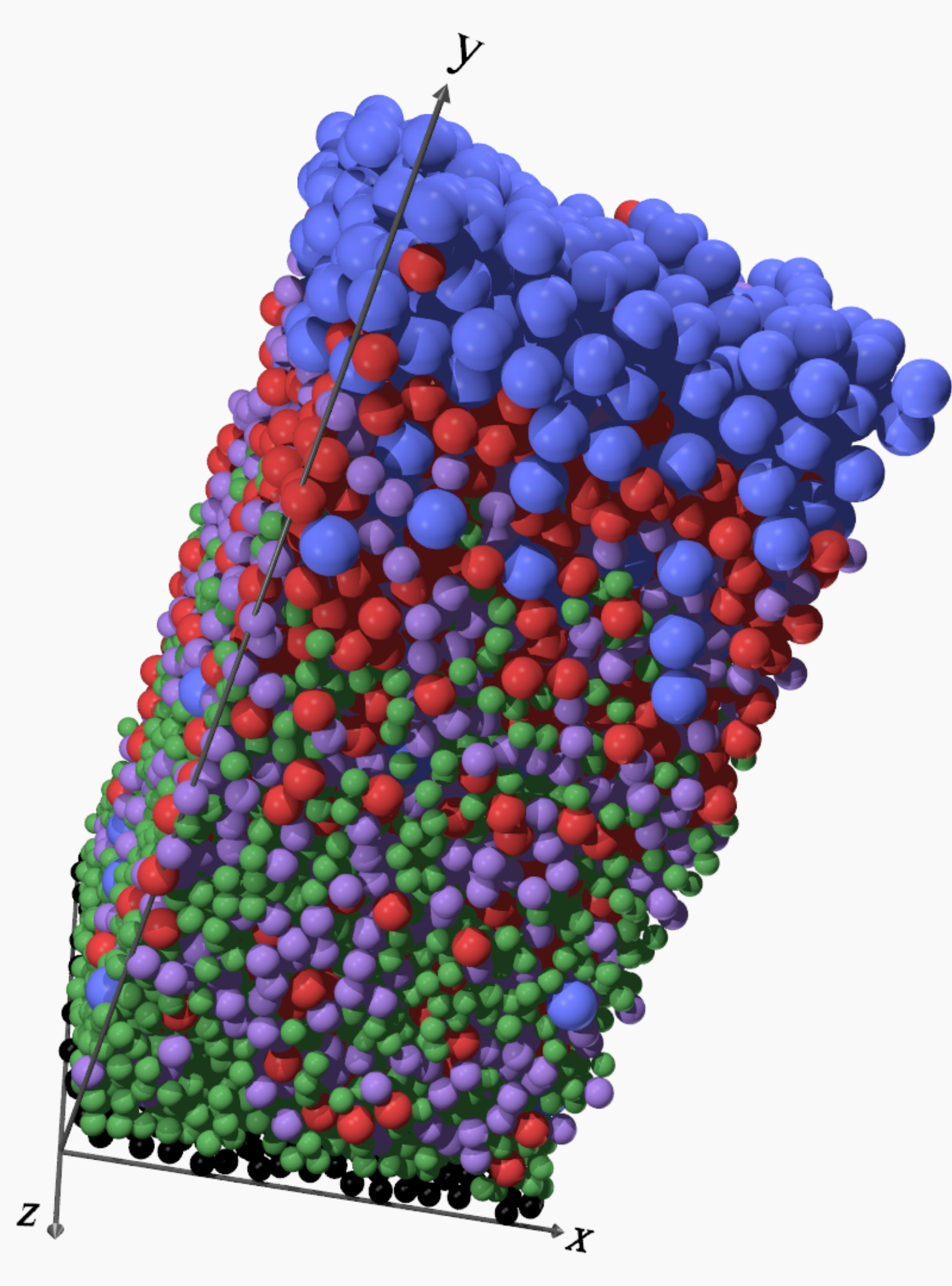}\put(-75,140){Steady state}\put(-95,125){(c)}
    \\
    \vspace{0.4cm}
   \includegraphics[scale=0.315]{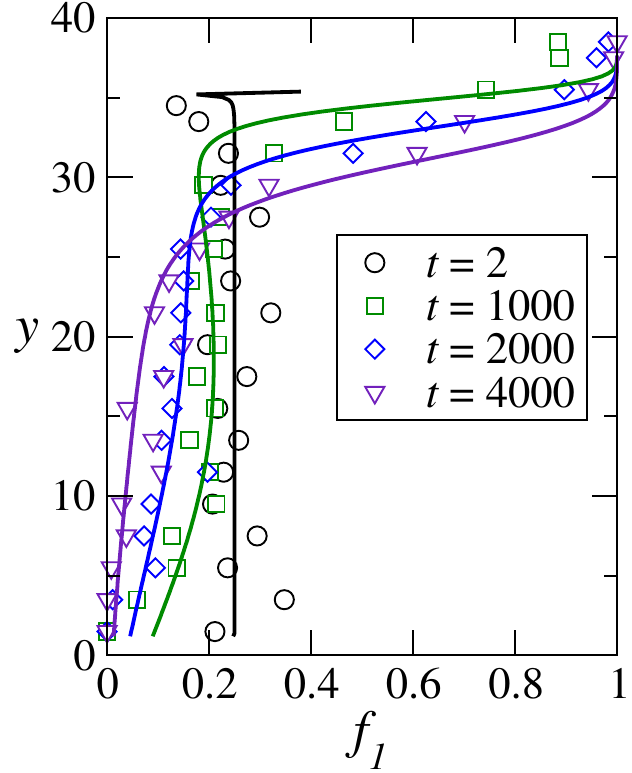}\put(-95,120){(d)}
   \includegraphics[scale=0.315]{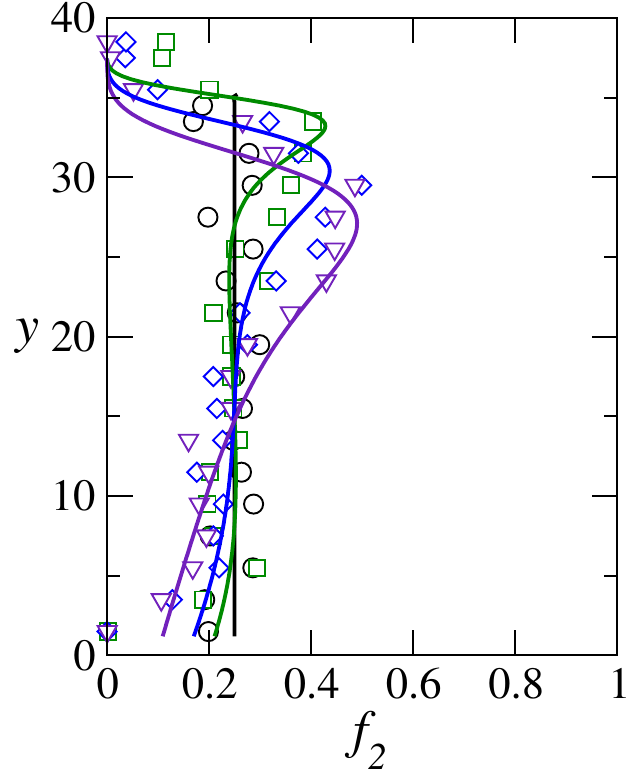}\put(-95,120){(e)}
   \includegraphics[scale=0.315]{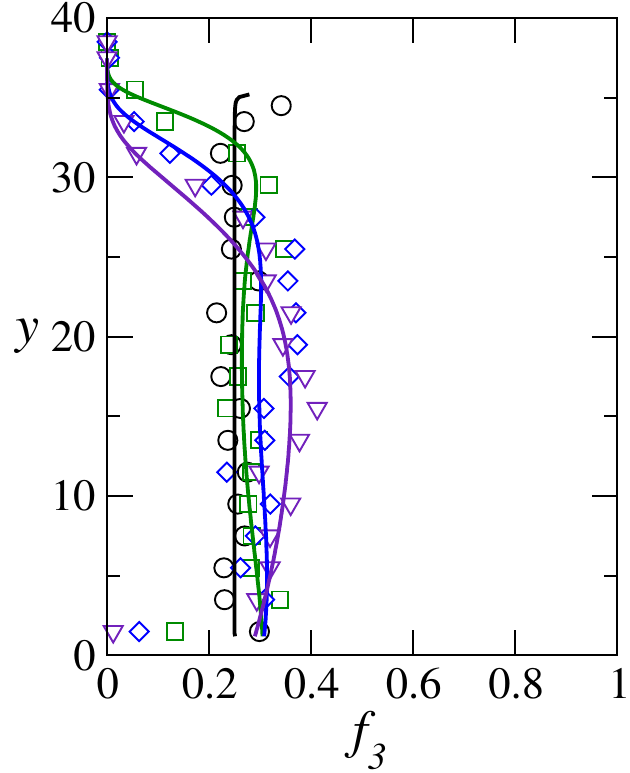}\put(-95,120){(f)}
   \includegraphics[scale=0.315]{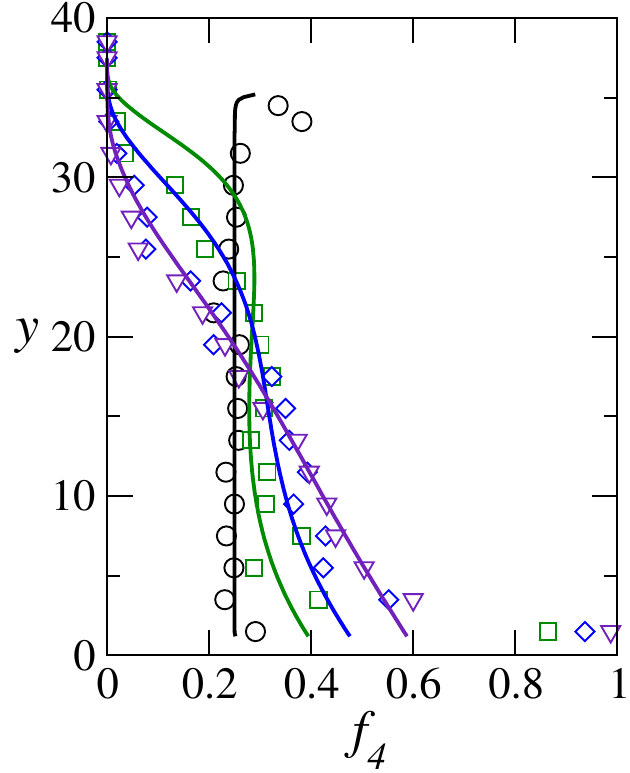}\put(-95,120){(g)}
    \caption{DEM snapshots of an equal composition quaternary mixture consisting of particles with size ratios $d_1 : d_2 : d_3 : d_4 = 2.0 : 1.5 : 1.25 : 1.0$ flowing down an inclined plane at $\theta = 26^\circ$, (a) Initial state, (b) Steady state. (c) Time evolution of the center of mass $y_{\text{com}}$ for each species. Instantaneous concentration profiles at different times for (d) species $1$ ($d_1 = 2d_s$), (e) species $2$ ($d_2 = 1.5d_s$), (f) species $3$ ($d_3 = 1.25d_s$), and (g) species $4$ of diameter $d_4 = d_s$.}
    \label{fig:y_com_2_1.5__1.25_1}
\end{figure}

We now report the results for quaternary mixtures of four different sizes particles. Figure~\ref{fig:y_com_2_1.5__1.25_1}a -~\ref{fig:y_com_2_1.5__1.25_1}g  show the results for a quaternary granular mixture with particle size ratios $2.0:1.5:1.25:1.0$, flowing down an inclined plane at an angle of $\theta = 26^\circ$. We consider equal volume fractions ($25\%$) of each species in the mixture (see figure~\ref{fig:y_com_2_1.5__1.25_1}a), \textcolor{black}{leading to} identical $y_{com}$ positions of all the species at $t=0$, in figure~\ref{fig:y_com_2_1.5__1.25_1}b. Once the flow begins, segregation occurs as the largest size particles rise to the top, the smallest \textcolor{black}{size particles} settle near the base. This is evident by the time evolution of center of mass ($y_{com}$) shown in figure~\ref{fig:y_com_2_1.5__1.25_1}b. The intermediate-sized particles however accumulate in between. The DEM snapshot for segregated state is shown in the figure~\ref{fig:y_com_2_1.5__1.25_1}c. A good agreement of continuum model predictions with the DEM is again observed for this mixture with four different sizes. Interestingly, the flowing layer exhibits noticeable dilation (evident from the comparison of figure~\ref{fig:y_com_2_1.5__1.25_1}a and figure~\ref{fig:y_com_2_1.5__1.25_1}c) and the height of the layer increases from $\simeq 35d_1$ at $t = 0$ to $\simeq 39d_1$ at $t = 6000$. This behavior is also well captured by our continuum model due to the coupling of the particle force-based segregation model with the $\mu$ – $I$ and $\phi - I$ mixture rheological model along with the simplified layer dilation model.

Figures~\ref{fig:y_com_2_1.5__1.25_1}d – \ref{fig:y_com_2_1.5__1.25_1}g show the instantaneous concentration profiles of all species in the mixture. 
Species $1$, being the largest, segregates toward the top of the flowing layer (Figure~\ref{fig:y_com_2_1.5__1.25_1}d), whereas species $4$, the smallest, concentrates near the base (Figure~\ref{fig:y_com_2_1.5__1.25_1}g). Figures~\ref{fig:y_com_2_1.5__1.25_1}e and \ref{fig:y_com_2_1.5__1.25_1}f show intermediate size species (species $2$ and $3$) settle between the largest and smallest particles depending on the particle size.
The continuum model predictions (solid lines) match \textcolor{black}{reasonably} with DEM simulations (symbols), demonstrating the model’s ability to capture the key features of segregation dynamics in a polydisperse system. As in figure~\ref{fig:instantaneous_r_1.5_1.25_1_theta_25}, differences in the concentration profile of small species in the regions of low concentration is observable for this case as well.

\subsection{Different initial configurations}
Next, we explore the applicability of generalized particle force-based segregation model to predict the segregation for different initial configurations. 
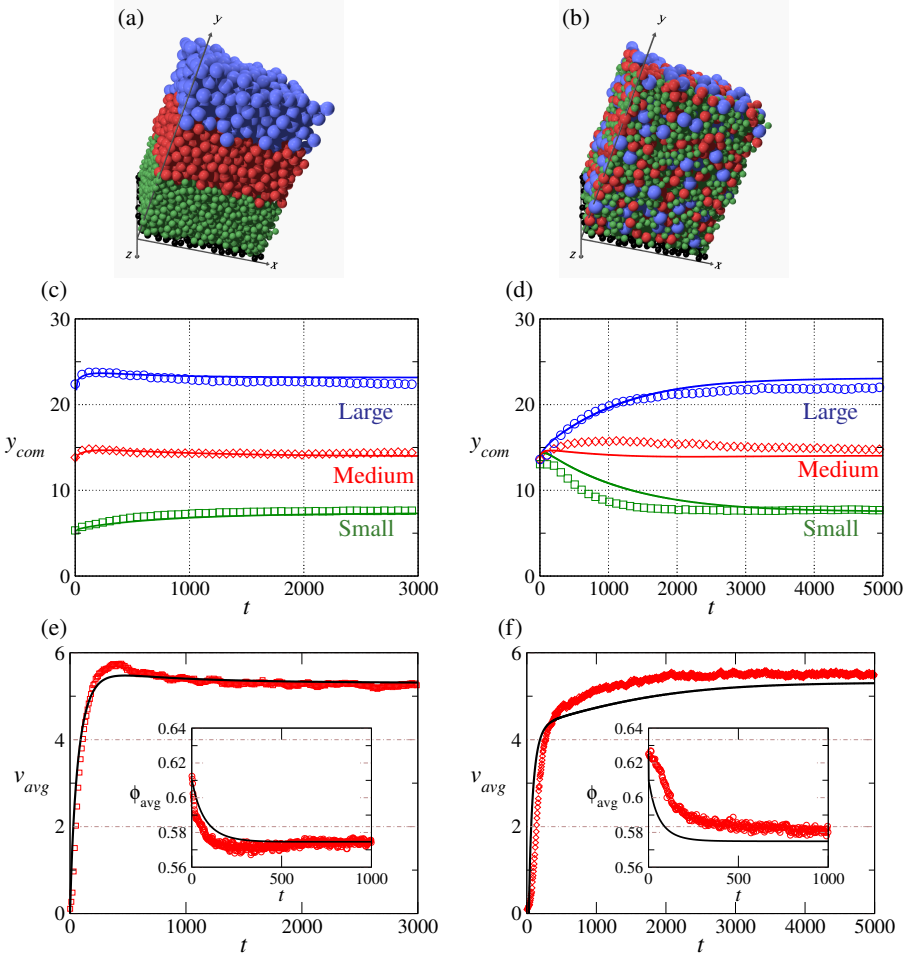
\begin{figure}
    \centering
    \includegraphics[scale=0.11, trim=550 60 590 60, clip]{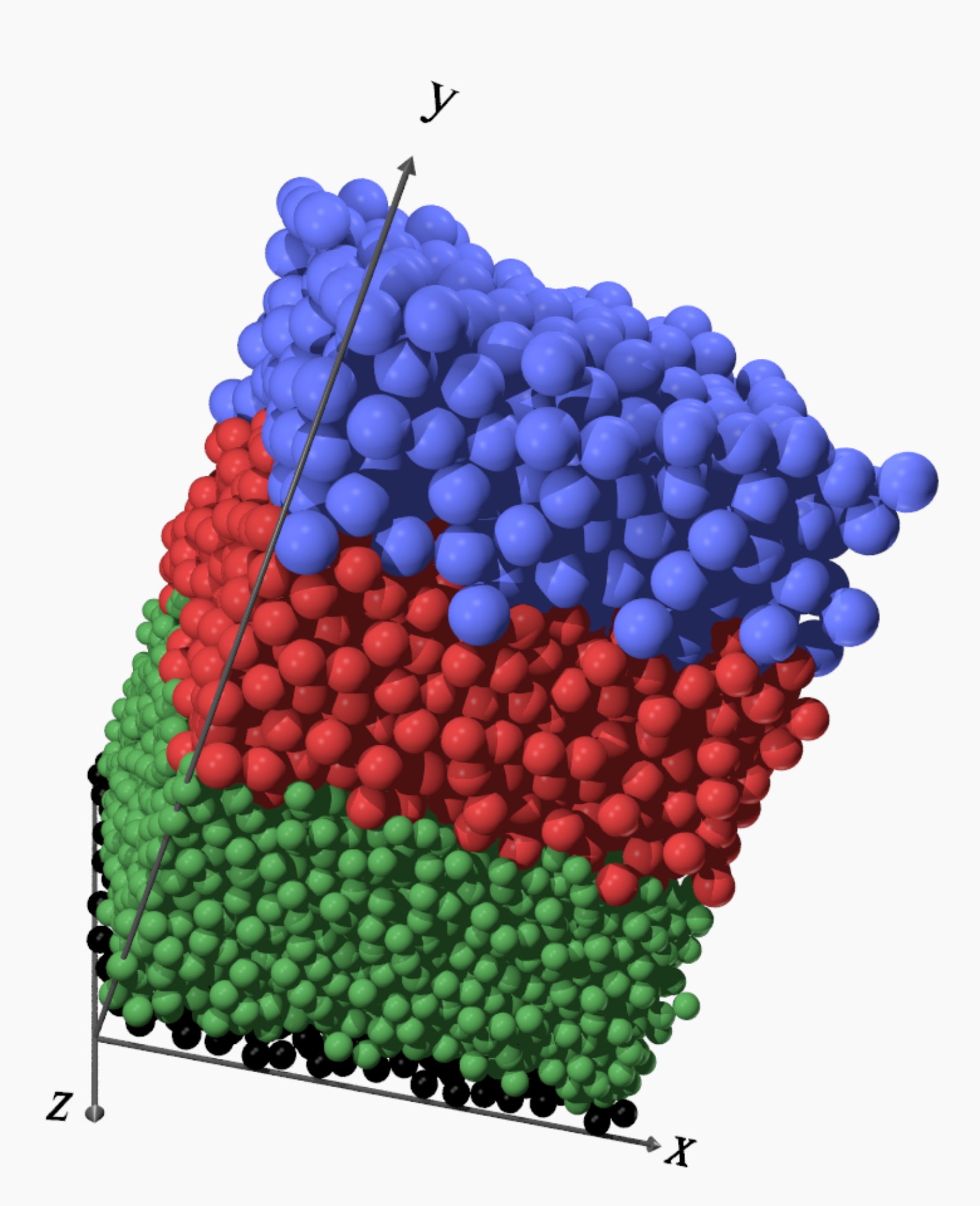} \put(-85,95){(a)} \quad \quad \quad \quad \quad \quad \quad \quad \quad 
    \includegraphics[scale=0.11, trim=550 60 590 60, clip]{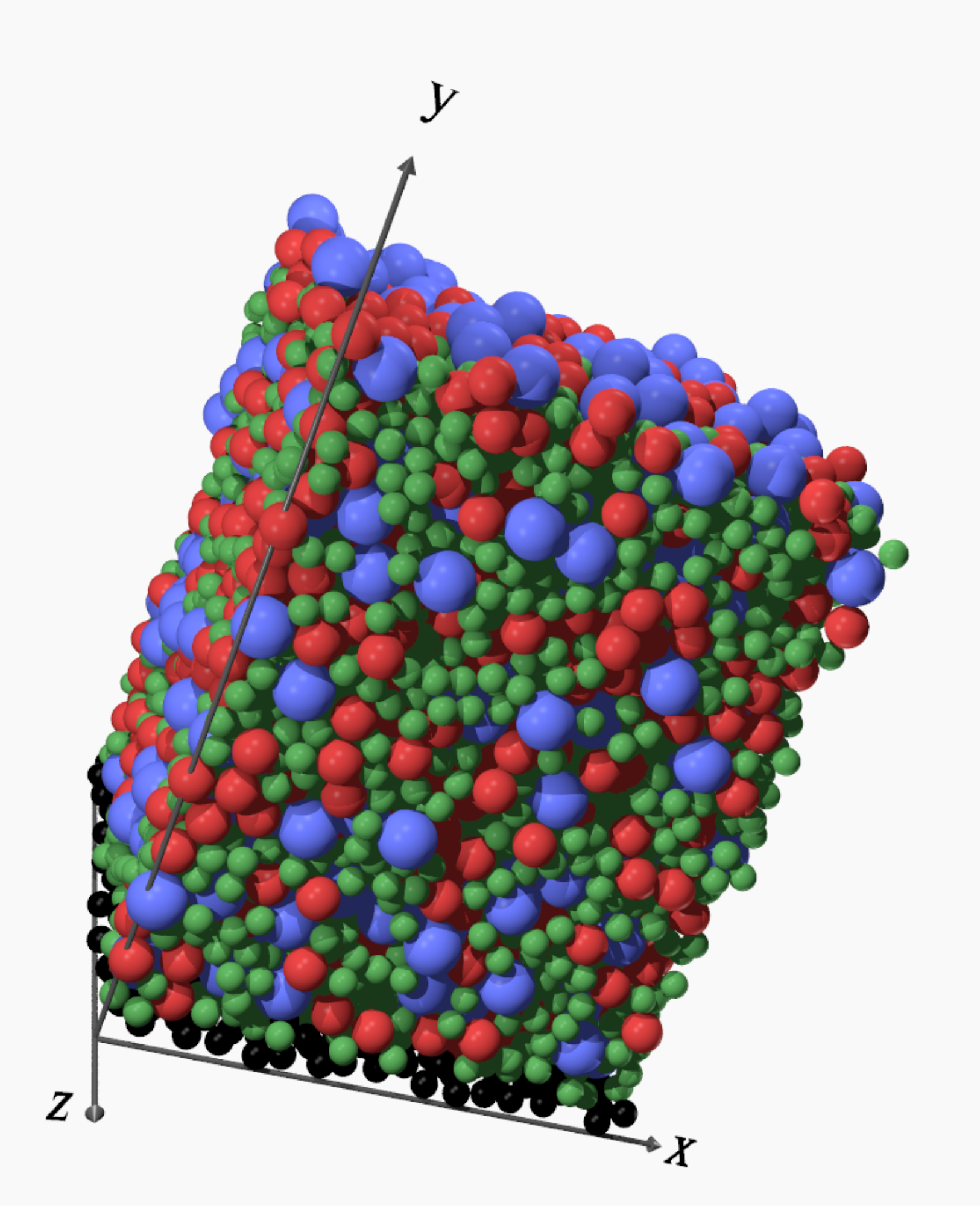} \put(-85,95){(b)}
     \\
    \includegraphics[scale=0.35]{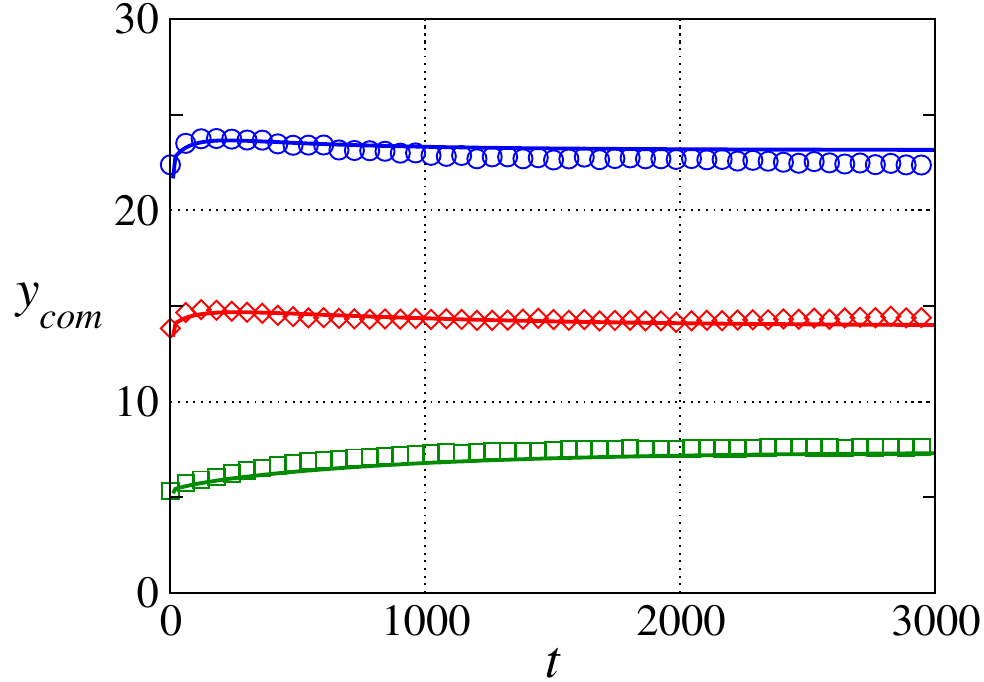}\put(-150,120){(c)}\put(-38,30){\textcolor{OliveGreen}{Small}}\put(-38,74){\textcolor{Blue}{Large}}\put(-40,50){\textcolor{red}{Medium}}\quad 
     \includegraphics[scale=0.35]{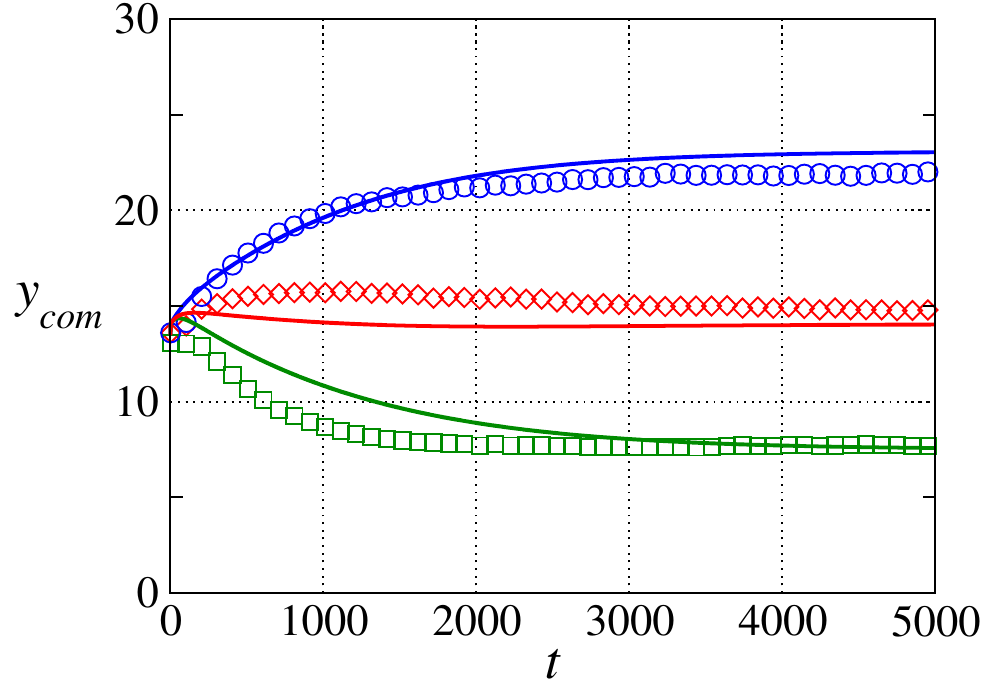}\put(-150,120){(d)}\put(-38,30){\textcolor{OliveGreen}{Small}}\put(-38,74){\textcolor{Blue}{Large}}\put(-40,52){\textcolor{red}{Medium}}
     \\
    \includegraphics[scale=0.355]{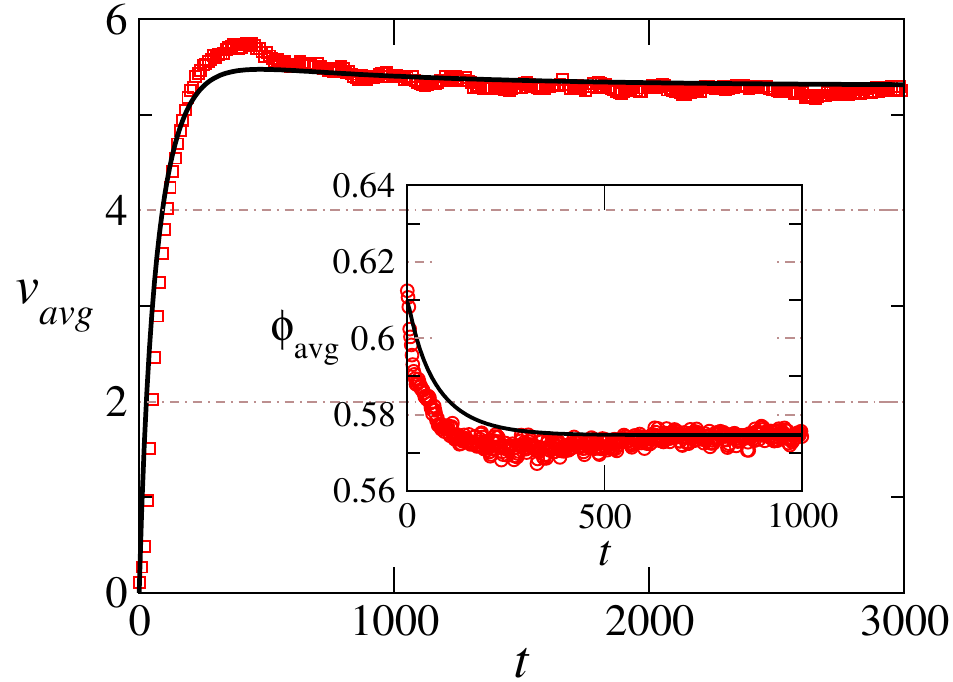}\put(-150,120){(e)} \quad 
    \includegraphics[scale=0.355]{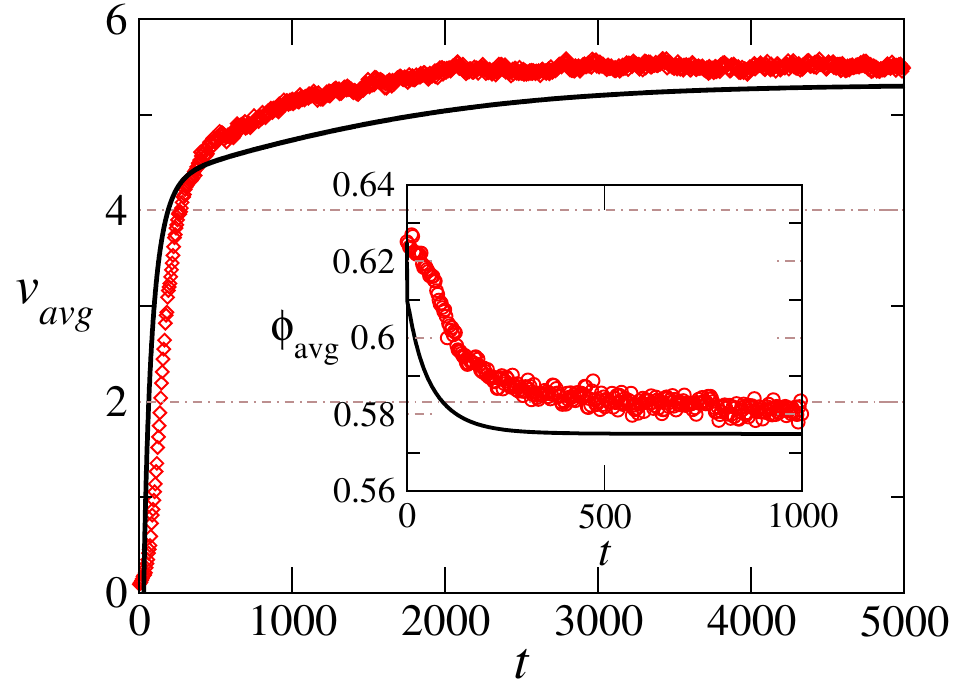}\put(-150,120){(f)} 
\caption{DEM snapshots of the initial states of an equal composition ternary mixture with size ratio $2.0:1.5:1.0$ flowing at an inclination angle $\theta=25^\circ$ for (a) small-near-base (SNB) and (b) well-mixed configurations. Panels (c) and (d) show the corresponding temporal evolution of the centre of mass positions $y_{com}$ for the large, medium, and small species. Panels (e) and (f) show the evolution of the average mixture velocity $v_{avg}$ for the SNB and well-mixed configurations, respectively. The insets show the corresponding evolution of the average solid fraction $\phi_{avg}$. Symbols represent the DEM data, while solid lines represent the model predictions.}
    \label{fig:modified_th_LNB_SNB}
\end{figure}
Figure~\ref{fig:modified_th_LNB_SNB} shows the results for equal composition ternary mixture with size ratio $2:1.5:1.0$ flowing at inclination angle $\theta = 25^\circ$. Figures~\ref{fig:modified_th_LNB_SNB}a and~\ref{fig:modified_th_LNB_SNB}b show DEM snapshots of the initial states corresponding to small-near-base (SNB) and well-mixed configurations, respectively. The corresponding evolution of $y_{com}$ position of species is shown in figures~\ref{fig:modified_th_LNB_SNB}c and~\ref{fig:modified_th_LNB_SNB}d, respectively. For the SNB configuration, the particles are initially in an already segregated state (figure~\ref{fig:modified_th_LNB_SNB}a). Therefore, the evolution is mainly governed by diffusion at the interfaces, which is well captured by the particle force-based segregation model, as shown in figure~\ref{fig:modified_th_LNB_SNB}c. The average mixture velocity for this case increases with time up to $t \approx 500$ and then decreases slightly before reaching a nearly constant value, as shown in figure~\ref{fig:modified_th_LNB_SNB}e. This trend is captured by the model as well, though the peak in velocity is not as pronounced as \textcolor{black}{in the case of} DEM data. \textcolor{black}{The inset in figure~\ref{fig:modified_th_LNB_SNB}e shows that the evolution of average solids fraction $\phi_{avg}$ in the bulk region (excluding data near base and free surface) of the layer with time is also captured reasonably well with the model.} 

Figure~\ref{fig:modified_th_LNB_SNB}d shows the temporal evolution of the species centre of mass positions $y_{com}$ for the well-mixed initial configuration, where the $y_{com}$ values of all species are nearly identical at $t=0$. \textcolor{black}{Despite small deviations between the model predictions (solid lines) and DEM data (symbols), the model predictions are in reasonable agreement with DEM.} Figure~\ref{fig:modified_th_LNB_SNB}f shows the evolution of the average mixture velocity, $v_{avg}$ for this mixed configuration, while the inset shows the temporal evolution of the average solid fraction $\phi_{avg}$. \textcolor{black}{The value of $\phi_{avg}$ at $t = 0$ is found to be higher in the well-mixed case compared to the segregated case, indicating the importance of composition dependence of the solids fraction for different size mixtures.} The model predictions for $\phi_{avg}$ deviate from the DEM data at early times, which contributes to the corresponding deviations observed in the predicted $y_{com}$ and $v_{avg}$ values.

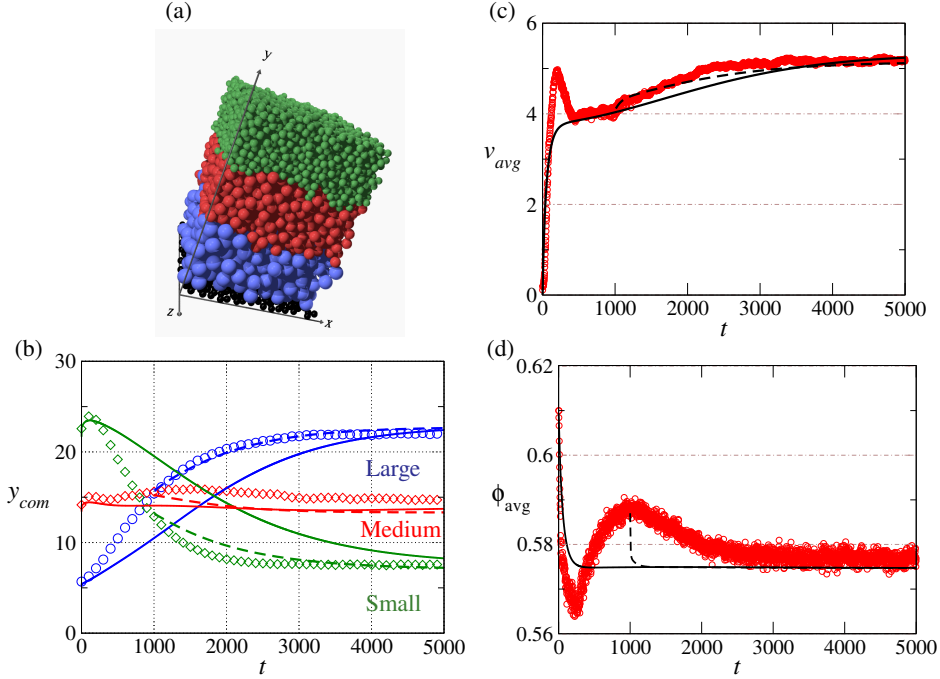
\begin{figure}
    \centering
    \quad \quad \quad 
     \includegraphics[scale=0.12, trim=550 60 590 60, clip]{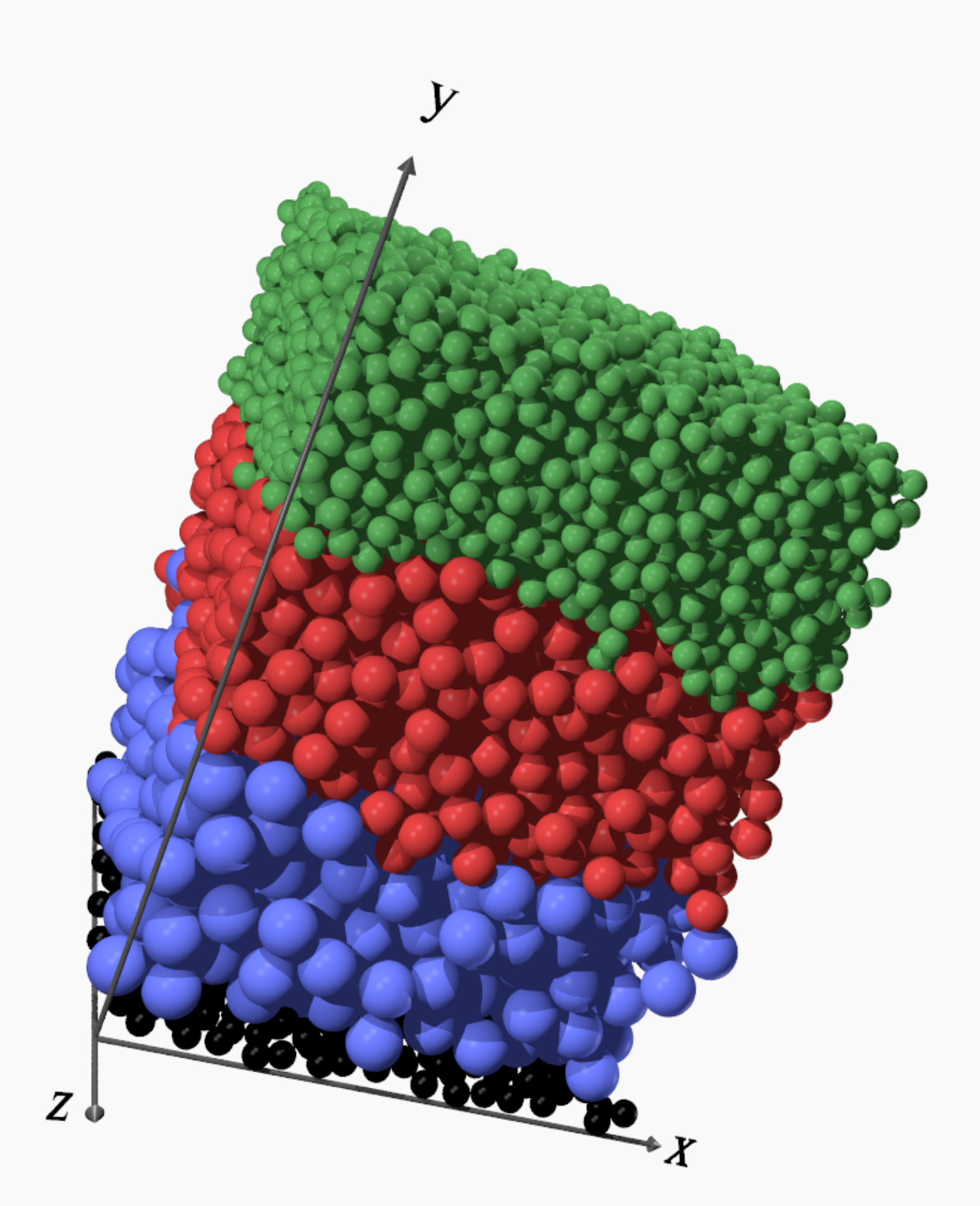} \put(-90,120){(a)} \quad \quad \quad   
     \includegraphics[scale=0.37]{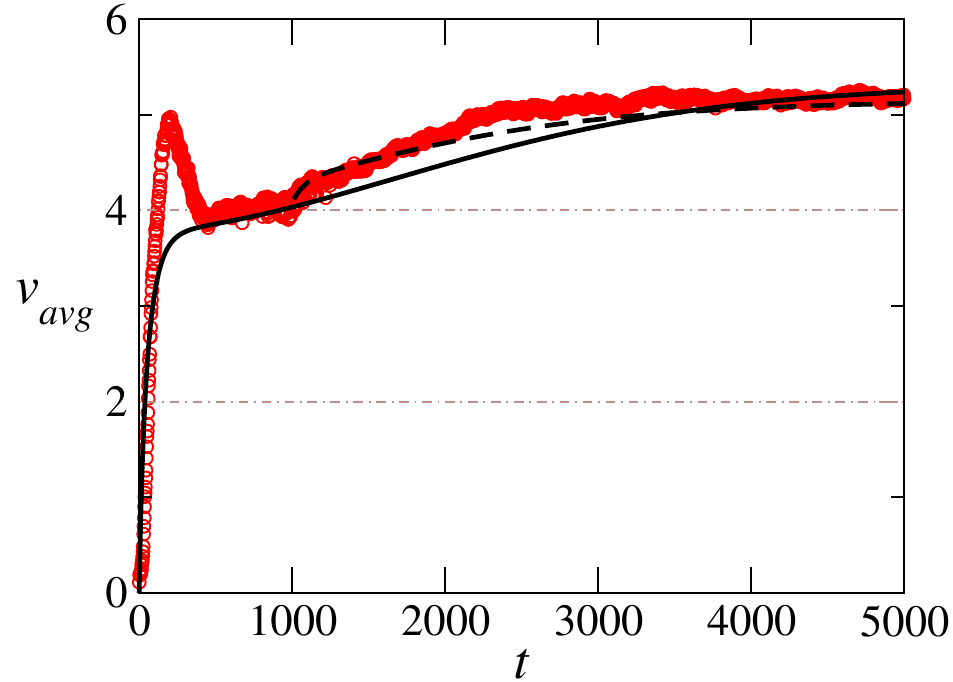}\put(-165,120){(c)}\\
     \includegraphics[scale=0.37]{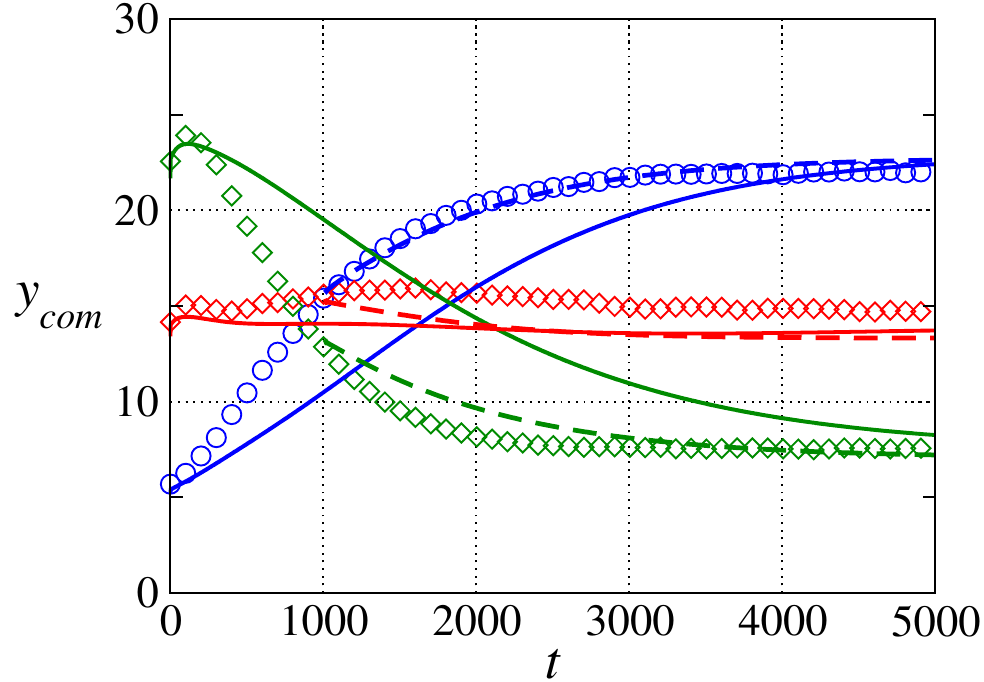}\put(-170,120){(b)}\put(-38,24){\textcolor{OliveGreen}{Small}}\put(-38,75){\textcolor{Blue}{Large}}\put(-40,52){\textcolor{red}{Medium}} \quad 
    \includegraphics[scale=0.365]{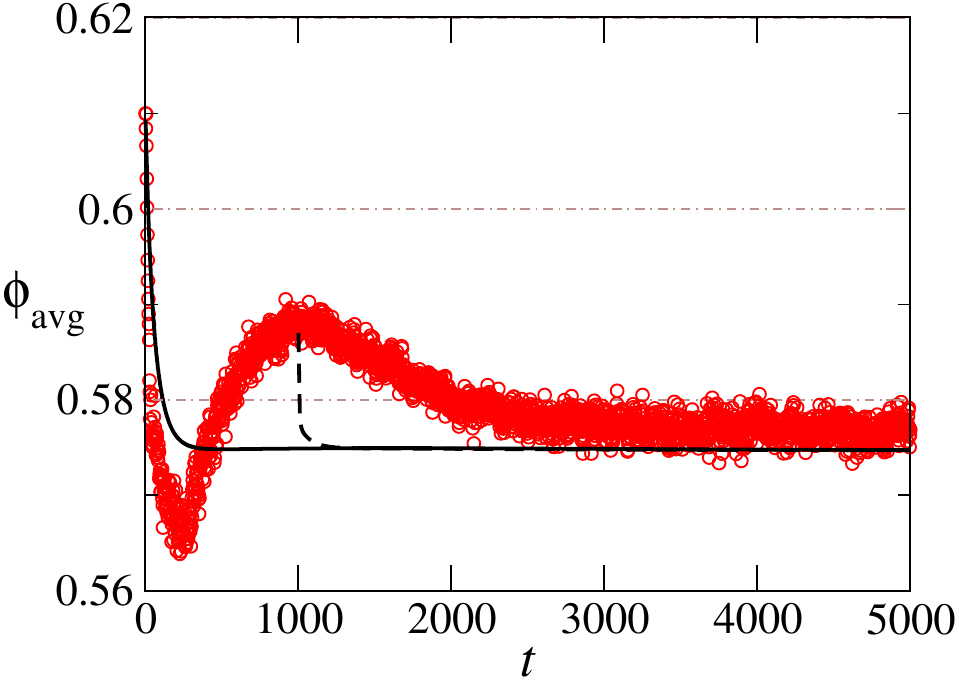}\put(-170,120){(d)} \quad \quad \quad \quad 
\caption{(a) DEM snapshot of the initial state for large-near-base (LNB) configurations of an equal composition ternary mixture with size ratio $r=2.0:1.5:1.0$ flowing at an inclination angle $\theta=25^\circ$. (b) The corresponding temporal evolution of the centre of mass positions $y_{com}$ for the large, medium, and small species. Panels (c) and (d) show the evolution of the average mixture velocity $v_{avg}$ and average solid fraction $\phi_{avg}$, respectively. Symbols represent the DEM data while solid lines represent the model predictions. The dashed lines correspond to predictions using properties at $t =1000$ as the initial condition in continuum model.}
    \label{fig:modified_th_LNB}
\end{figure}

Figure~\ref{fig:modified_th_LNB} presents the results for the large-near-base configuration. The corresponding DEM snapshot of the initial state is shown in figure~\ref{fig:modified_th_LNB}a \textcolor{black}{ and the evolution of centre of mass positions $y_{com}$ for this case in figure~\ref{fig:modified_th_LNB}b.} The continuum model predictions (solid lines) show significant deviations from the DEM data (symbols) \textcolor{black}{for this configuration}. \textcolor{black}{This difference} can be attributed to the non-monotonic evolution of the average mixture velocity $v_{avg}$ shown in figure~\ref{fig:modified_th_LNB}c \textcolor{black}{and that of $\phi_{avg}$ shown in figure~\ref{fig:modified_th_LNB}.} \textcolor{black}{We note that} $v_{avg}$ initially increases with time \textcolor{black}{and reaches a local maximum around} $t \approx 250$, \textcolor{black}{after which it starts to decrease till $t \approx 500$ and remains} approximately constant up to $t \approx 1000$. \textcolor{black}{After time $t \approx 1000$}, $v_{avg}$ increases again \textcolor{black}{slowly} and approaches the steady state value. This non-monotonic variation in velocity is accompanied by a corresponding variation in the average solid fraction $\phi_{avg}$ as shown in figure~\ref{fig:modified_th_LNB}d. As the velocity increases initially, $\phi_{avg}$ decreases \textcolor{black}{sharply and shows a local minima at $t \approx 250$,} indicating dilation of the flowing layer. Subsequently, as the velocity decreases, $\phi_{avg}$ increases \textcolor{black}{up to $t=500$. Interestingly, while the velocity remains nearly constant from $t = 500$ to $1000$, \textcolor{black}{the bulk layer solids fraction} $\phi_{avg}$ keeps increasing and achieves a local maxima at $t \approx 1000$}.
If this initial non-monotonic \textcolor{black}{velocity and solids fraction variation} up to $t=1000$ is disregarded and the segregation evolution is predicted \textcolor{black}{after $t = 1000$}, the continuum model predictions (dashed lines) for $y_{com}$ and $y_{avg}$ agree well with the DEM data. \textcolor{black}{The average solids fraction $\phi_{avg}$, however, still shows significant deviation up to $t \approx2000$, beyond which $\phi_{avg}$ reaches a constant value.} \textcolor{black}{The results shown in figure~\ref{fig:modified_th_LNB}b confirm that our $1D$ continuum model is unable to account for the segregation at early times ($t <1000$) in this configuration. During this time, the segregated layers of large, medium, and small species interdiffuse with each other and reach a relatively well-mixed configuration (evident by nearly the same value of $y_{com}$ for each species) around $t \approx 1000$. Due to this intermixing, the solids fraction $\phi_{avg}$ in the bulk layer increases (after the quick initial drop due to layer dilation) during this period and achieves a local maximum.} \textcolor{black}{The solid fraction in polydispersed mixture has a complex, non-monotonic dependency on the concentration of species. As the segregation in the layer evolves, the average solids fraction in the layer reduces for $t \geq 1000$ and slowly reaches the steady value. Since our $1D$ model does not account for the composition dependence of $\phi_{avg}$, the solids fraction DEM data is not very well captured despite using the concentration profiles and velocity profile at $t = 1000$ as the initial conditions in our $1D$ continuum model.}
\textcolor{black}{We also note that} the early-time deviations like this were observed in the case of multicomponent density segregation study of~\cite{kumawat2025transient} for light species near base configuration. In that case, it was observed that a Rayleigh-Taylor like instability leads to two dimensional concentration evolution which could not be captured using $1D$ continuum model. \textcolor{black}{While we do not clearly observe such an instability for the results shown in figure~\ref{fig:modified_th_LNB}}, in the next section, we show that a similar instability is observed in case of different size mixtures for larger size ratios in this large-near-base configuration.

\section{Results for larger size ratios}
\label{sec:solidsfraction_fl}
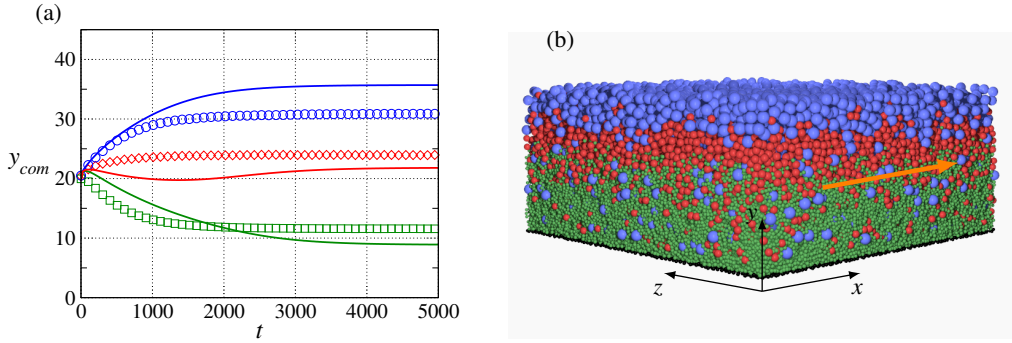
\begin{figure}
    \centering
\includegraphics[width=0.45\linewidth, trim=0 0 0 20, clip]{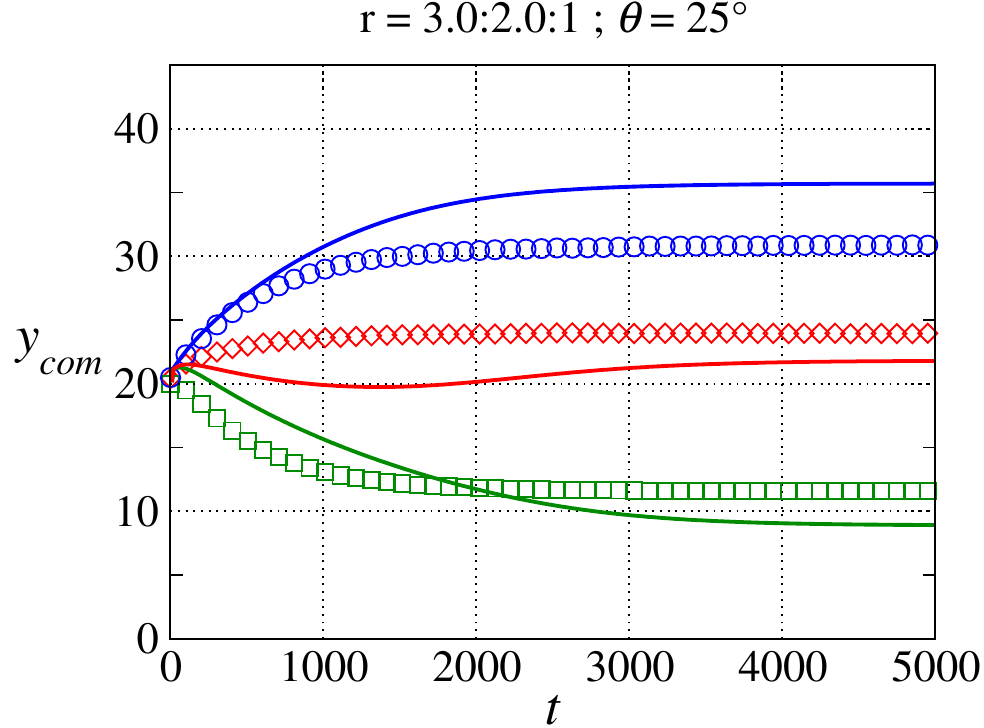}\put(-160,120){(a)}
    \quad \quad
    \includegraphics[scale=0.23, trim=550 280 550 300, clip]{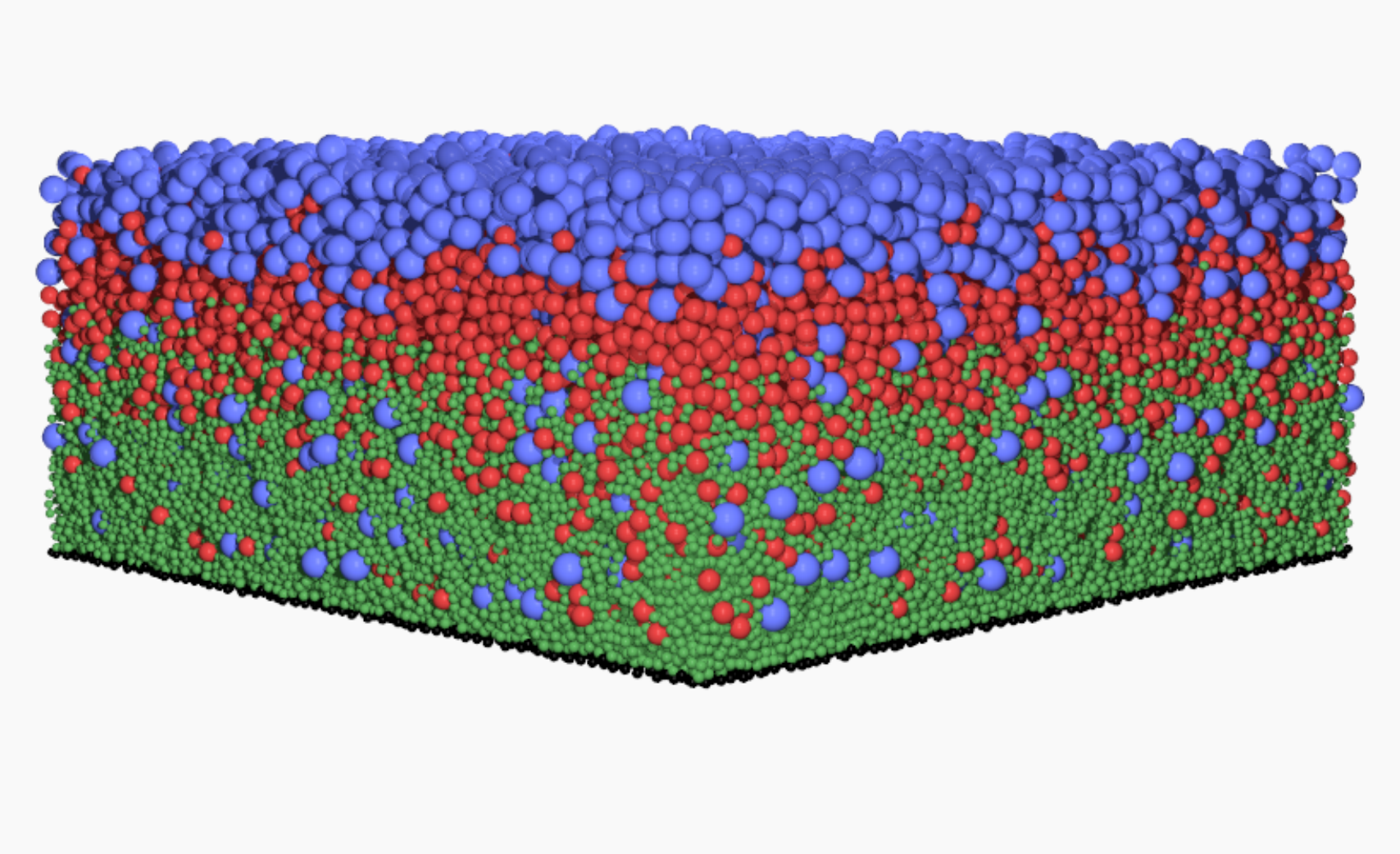} \put(-175,110){(b)} \put(-100,45){$y$}\put(-60,17){$x$}\put(-135,17){$z$}
      \begin{tikzpicture}[overlay, remember picture]
        \draw[->, orange, very thick, line width=1.7pt, >=latex] (-2.5,2.0) -- (-0.7,2.33); 
         \draw[->, black, very thick, line width=0.5pt, >=latex] (-3.3,0.63) -- (-3.3,1.6); 
        \draw[->, black, very thick, line width=0.5pt, >=latex] (-3.3,0.63) -- (-2.0,0.87); 
        \draw[->, black, very thick, line width=0.5pt, >=latex] (-3.3,0.63) -- (-4.6,0.87); 
    \end{tikzpicture}
    \caption{(a) Time evolution of $y_{com}$ for equal composition ternary mixture $ 3:2:1$ at inclination angle $\theta = 25^\circ$ \textcolor{black}{starting from well-mixed initial configuration}. Symbols represent the DEM data, while solid lines correspond to continuum model predictions. (b) DEM snapshot showing the particle arrangement at steady state. \textcolor{black}{The orange arrow denotes the direction of the flow.}}
\label{fig:y_com_3_2_1}
\end{figure}
Next, we check the validity of \textcolor{black}{our continuum} model for higher size ratios \textcolor{black}{of the species}. Figure~\ref{fig:y_com_3_2_1}a shows the variation of $y_{com}$ for an equal composition ternary mixture having size ratio $3:2:1$ flowing at inclination angle $\theta = 25^\circ$. \textcolor{black}{Due to the large size of grains, we utilize a much bigger simulation box of $100d_s \times 100d_s$ in the $x$ and $z$ directions for this case. The flow starts from an initially uniform mixed state and figure~\ref{fig:y_com_3_2_1}b shows the DEM snapshots for the final segregated state.} The continuum model predictions (solid lines) show significant deviations from the DEM data (symbols) \textcolor{black}{for these larger size ratios. A comparison of figure~\ref{fig:y_com_3_2_1}a with figure~\ref{fig:modified_th_LNB_SNB}d (both start from well-mixed state) shows that the evolution of the species centre of mass for the three species is quite similar, not only for the DEM data (symbols), but also for the continuum model predictions (lines). The continuum model is able to predict the segregation evolution qualitatively for the large size ratio as well. The discrepancy between the DEM results and model predictions shown in figure~\ref{fig:y_com_3_2_1}a, however, are found to be enhanced for the size ratio of $3:2:1$. To explain this enhanced discrepancy between DEM and model predictions, we refer to the recent study by~\cite{kumawat2025sizetransient}.} The authors have shown that for binary mixtures with $r > 1.5$ the local solid packing fraction $\phi$ is strongly influenced by the local species concentration. \textcolor{black}{Specifically, Figure~4 of~\cite{kumawat2025sizetransient} shows that neglecting the concentration dependence of the solid fraction leads to substantial deviations between the model predictions and DEM data for binary mixtures.}

\begin{figure}
    \centering
     \includegraphics[scale=0.35]{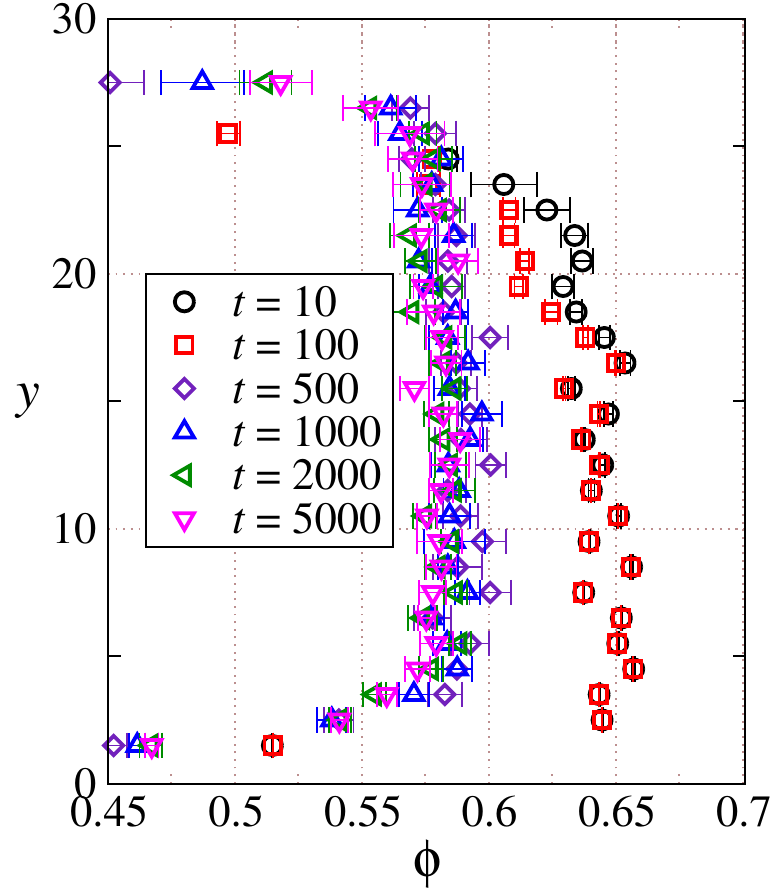}\put(-135,150){(a)} \quad \quad 
    \includegraphics[scale=0.35]{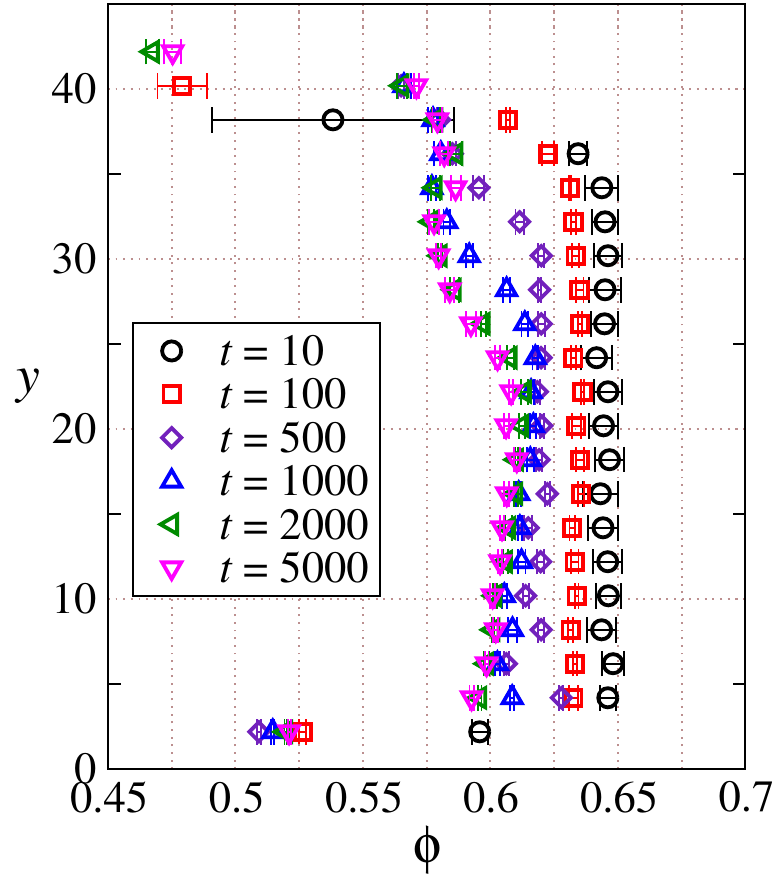}\put(-135,150){(b)}
    \caption{Instantaneous profiles of solid fraction along the flowing layer for equal \textcolor{black}{composition} ternary mixtures \textcolor{black}{starting from a well-mixed initial configuration with} size ratios (a) $ 2:1.5:1$ and (b) $3:2:1$ at inclination angle $\theta = 25^\circ$.}
    \label{fig:solidfraction_instant}
\end{figure}

In order to check whether the variation of solid fraction across the layer at different times gets affected for large size ratios, we report the DEM simulation data for the instantaneous solid fraction profiles of two different ternary mixtures having equal species composition in figure~\ref{fig:solidfraction_instant}. The first case \textcolor{black}{shown in figure~\ref{fig:solidfraction_instant}a} corresponds to a moderate size ratio $2.0:1.5:1$, while \textcolor{black}{that in figure~\ref{fig:solidfraction_instant}b corresponds to the mixture} with larger size ratio $ 3:2:1$. \textcolor{black}{For size ratio $2:1.5:1$, the solid fraction $\phi$ becomes nearly uniform throughout the bulk region of the flowing layer and shows little variation with layer height for $t \geq 500$ (figure~\ref{fig:solidfraction_instant}a)}. In contrast, \textcolor{black}{figure~\ref{fig:solidfraction_instant}b shows that} for the larger size ratio of $ 3:2:1$, the variation in $\phi$ across the flowing layer \textcolor{black}{becomes significant for $t>500$ and} persists even at steady state \textcolor{black}{at $t = 5000$}. The solids fraction variation across the layer, shown in figure~\ref{fig:solidfraction_instant}b, confirm that the lower part of the layer is more densely packed compared to the upper part of the layer. As segregation progresses, the lower portion of the layer \textcolor{black}{becomes rich in small and medium species. Hence, it} \textcolor{black}{resembles} more like a binary mixture of small and medium species (relative size ratio $r_{MS} = 2.0$). The upper part of the layer becomes rich in large and medium species with very few small particles. Thus, the upper portion of the layer \textcolor{black}{resembles} more like a binary mixture with size ratio $r_{LM} = 1.5$. \cite{tripathi2011rheology} and \cite{kumawat2025sizetransient} have shown that the solid fraction for binary mixture of $r = 2.0$ is larger than that for $r = 1.5$. This suggests that in ternary mixtures with large size differences, the influence of concentration on the solids fraction becomes important. \textcolor{black}{As mentioned earlier, the current $1D$ model does not account for this dependency.} In contrast to a binary mixture, where the variation of solids fraction depends on the concentration of only one species $\phi = \phi(f_L)$; for these $N$ component mixtures we need to accurately account for $\phi = \phi(f_1,f_2,...,f_{N-1})$ in terms of concentration of $N-1$ species. Thus, for a ternary mixture, $\phi(f_L,f_M)$ needs to be determined by performing DEM simulations across a wide range of mixture compositions. 
\textcolor{black}{In our particle force-based segregation model}, the dimensionless upward force \textcolor{black}{$\alpha$} depends on the local solids fraction $\phi$. \textcolor{black}{Hence, it is possible that accounting for the solids fraction dependency on the local species concentration may lead to relatively more accurate predictions of the segregation of polydisperse mixtures at larger size ratios due to the modifications of the net force on each species.}

\begin{figure}
    \centering
     \includegraphics[scale=0.14, trim=520 180 520 320, clip]{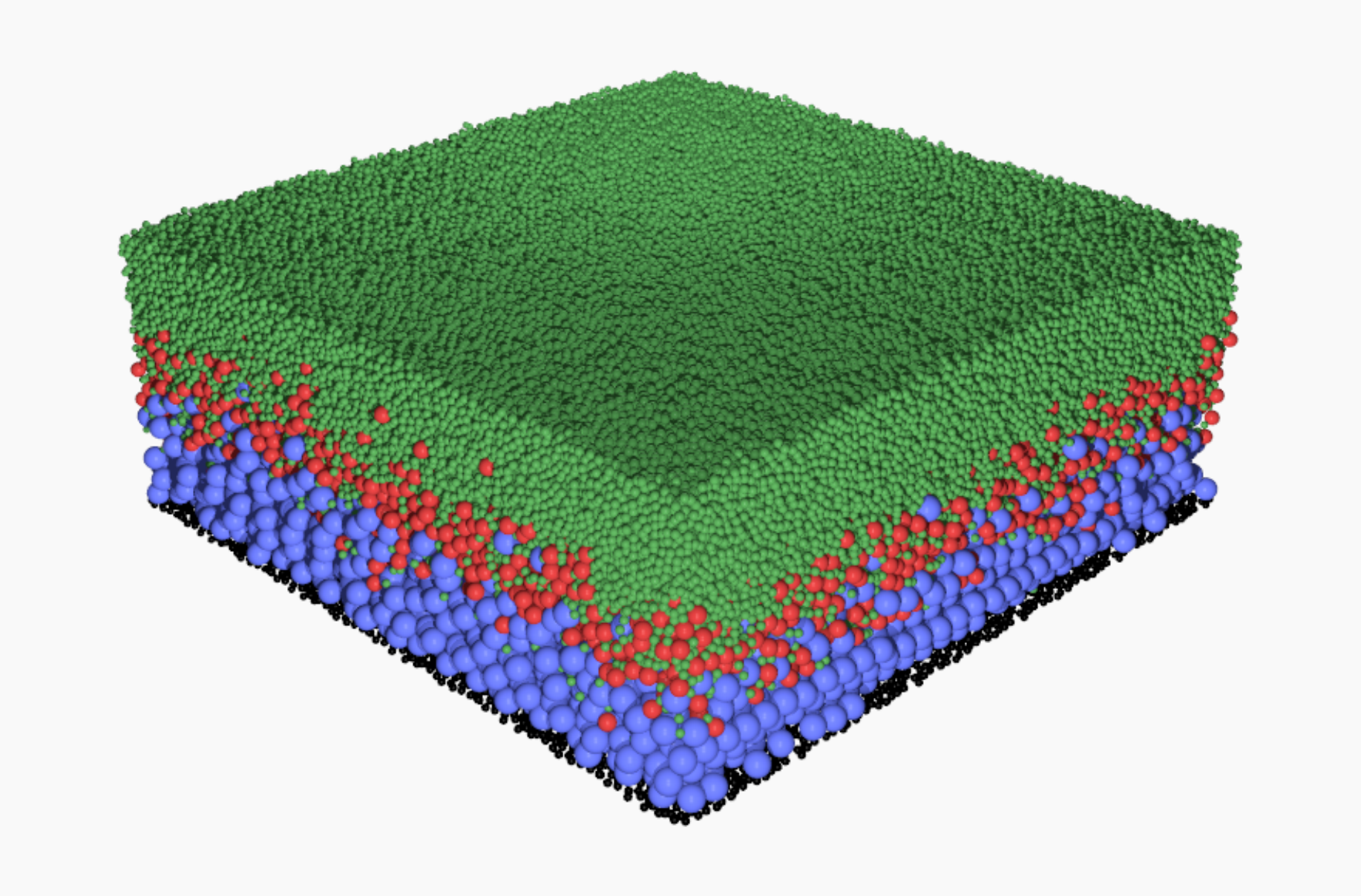}\put(-120,70){(a)}\put(-70,80){$t = 200$}\put(-65,30){$y$}\put(-36,13){$x$}\put(-90,12){$z$}\hfill
       \begin{tikzpicture}[overlay, remember picture]
         \draw[->, orange, very thick, line width=1.7pt, >=latex] (-2.0,1.05) -- (-0.8,1.7); 
         \draw[->, black, very thick, line width=0.5pt, >=latex] (-2.25,0.05) -- (-2.25,1.0); 
        \draw[->, black, very thick, line width=0.5pt, >=latex] (-2.25,0.05) -- (-1.4,0.55); 
        \draw[->, black, very thick, line width=0.5pt, >=latex] (-2.25,0.05) -- (-3.1,0.55); 
    \end{tikzpicture}
    \includegraphics[scale=0.14, trim=520 180 520 320, clip]{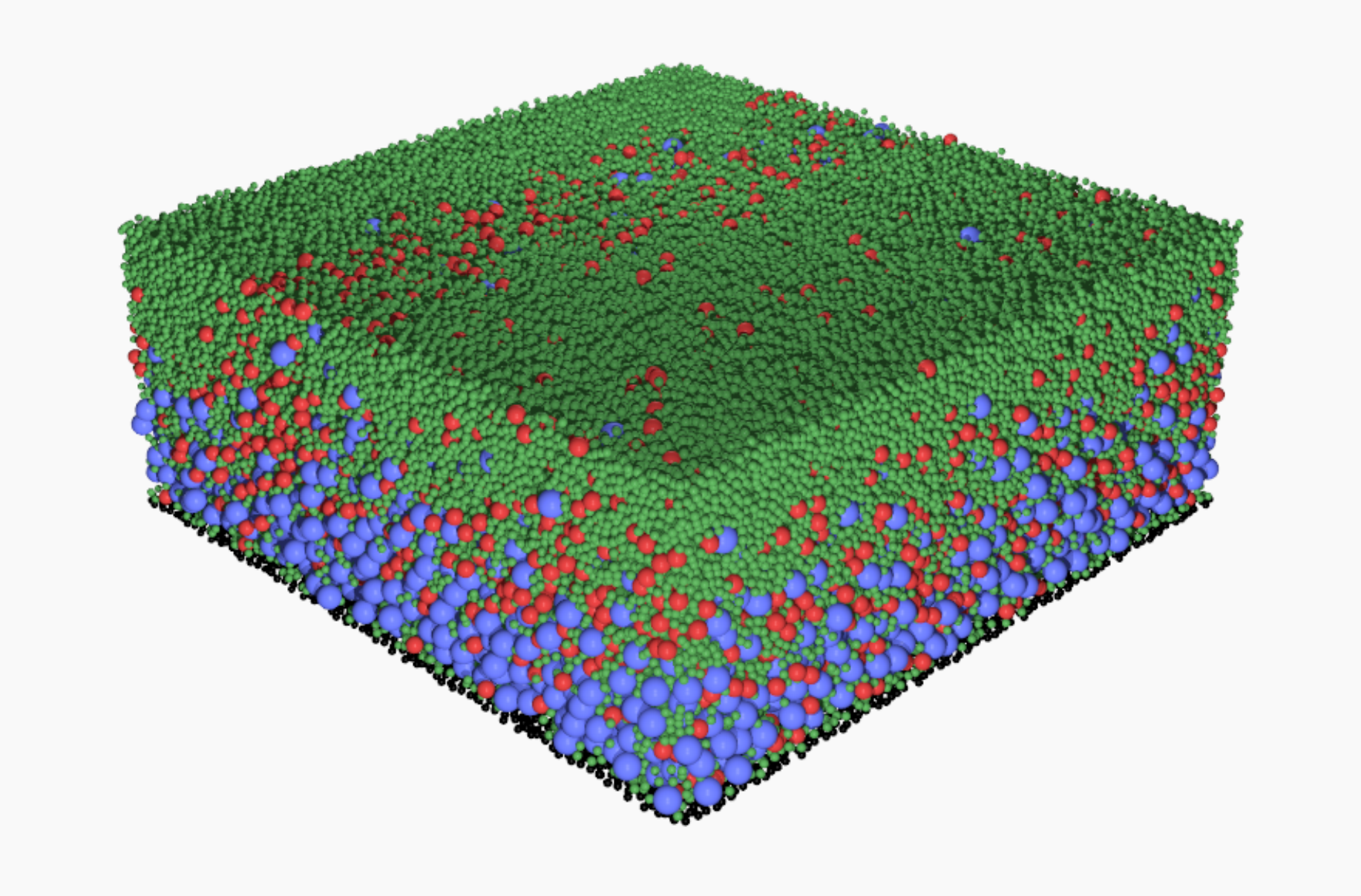}\put(-120,70){(b)}\put(-70,80){$t = 400$}\put(-65,30){$y$}\put(-36,13){$x$}\put(-90,12){$z$}\hfill
    \begin{tikzpicture}[overlay, remember picture]
         \draw[->, orange, very thick, line width=1.7pt, >=latex] (-2.0,1.05) -- (-0.8,1.7); 
         \draw[->, black, very thick, line width=0.5pt, >=latex] (-2.25,0.05) -- (-2.25,1.0); 
        \draw[->, black, very thick, line width=0.5pt, >=latex] (-2.25,0.05) -- (-1.4,0.55); 
        \draw[->, black, very thick, line width=0.5pt, >=latex] (-2.25,0.05) -- (-3.1,0.55); 
    \end{tikzpicture}
    \includegraphics[scale=0.14, trim=520 180 520 320, clip]{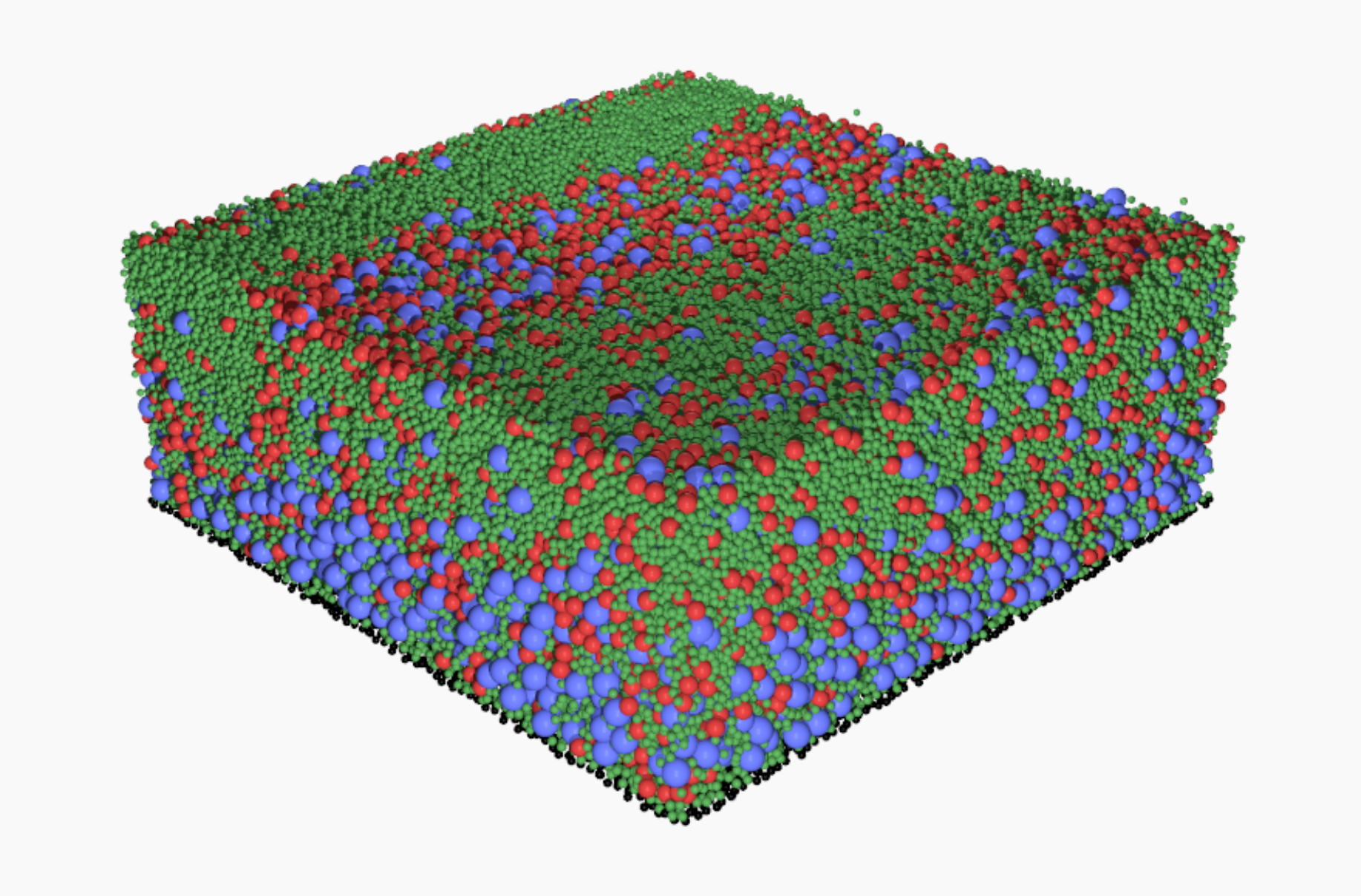}\put(-120,70){(c)}\put(-70,80){$t = 500$}\put(-65,30){$y$}\put(-36,13){$x$}\put(-90,12){$z$}\hfill
    \begin{tikzpicture}[overlay, remember picture]
         \draw[->, orange, very thick, line width=1.7pt, >=latex] (-2.0,1.05) -- (-0.8,1.7); 
         \draw[->, black, very thick, line width=0.5pt, >=latex] (-2.25,0.05) -- (-2.25,1.0); 
        \draw[->, black, very thick, line width=0.5pt, >=latex] (-2.25,0.05) -- (-1.4,0.55); 
        \draw[->, black, very thick, line width=0.5pt, >=latex] (-2.25,0.05) -- (-3.1,0.55); 
    \end{tikzpicture}
    \includegraphics[scale=0.14, trim=520 180 520 320, clip]{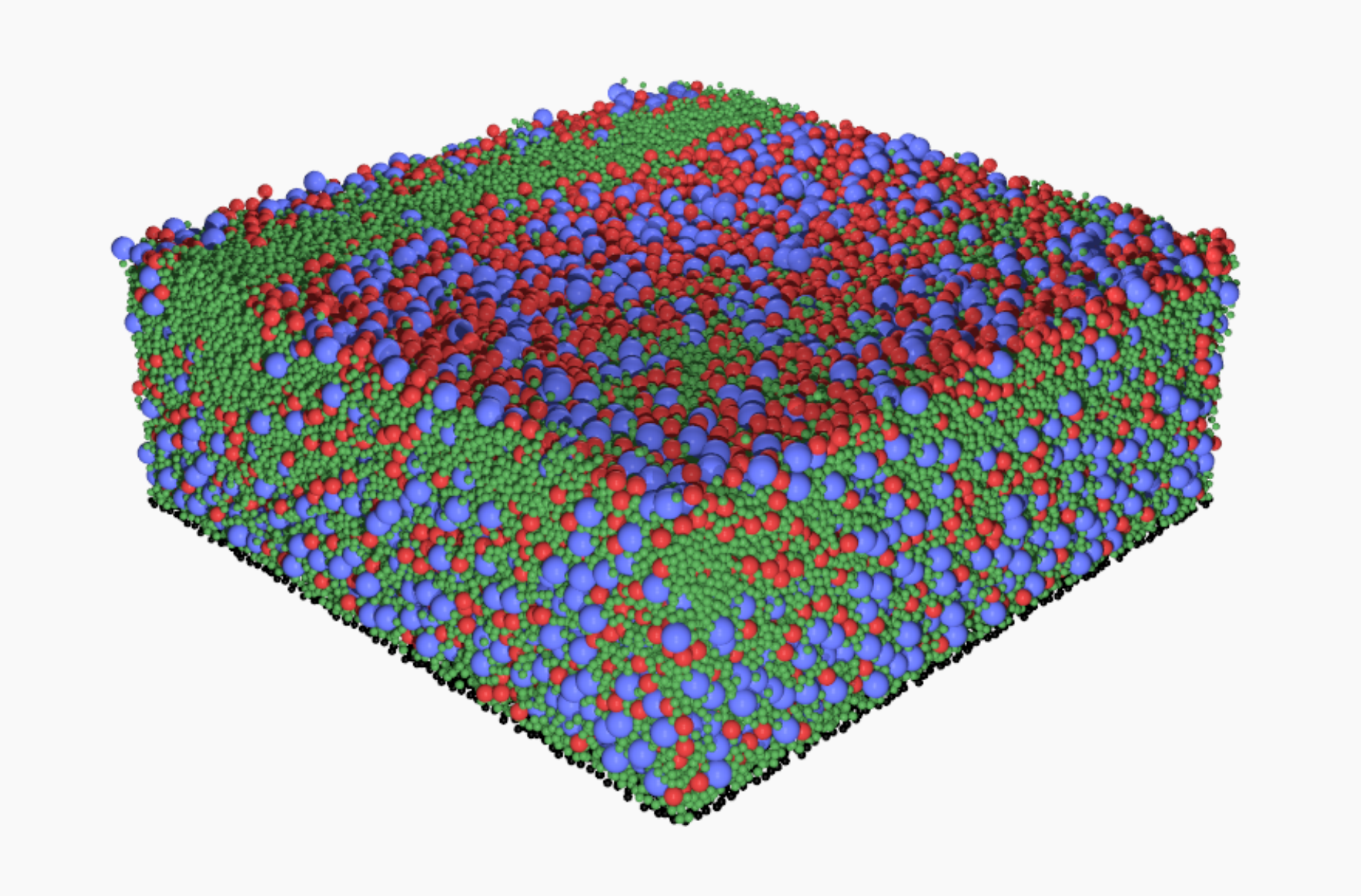}\put(-120,70){(d)}\put(-70,80){$t = 600$}\put(-65,30){$y$}\put(-36,13){$x$}\put(-90,12){$z$}\hfill
    \begin{tikzpicture}[overlay, remember picture]
         \draw[->, orange, very thick, line width=1.7pt, >=latex] (-2.0,1.05) -- (-0.8,1.7); 
         \draw[->, black, very thick, line width=0.5pt, >=latex] (-2.25,0.05) -- (-2.25,1.0); 
        \draw[->, black, very thick, line width=0.5pt, >=latex] (-2.25,0.05) -- (-1.4,0.55); 
        \draw[->, black, very thick, line width=0.5pt, >=latex] (-2.25,0.05) -- (-3.1,0.55); 
    \end{tikzpicture}
    \includegraphics[scale=0.14, trim=520 180 520 320, clip]{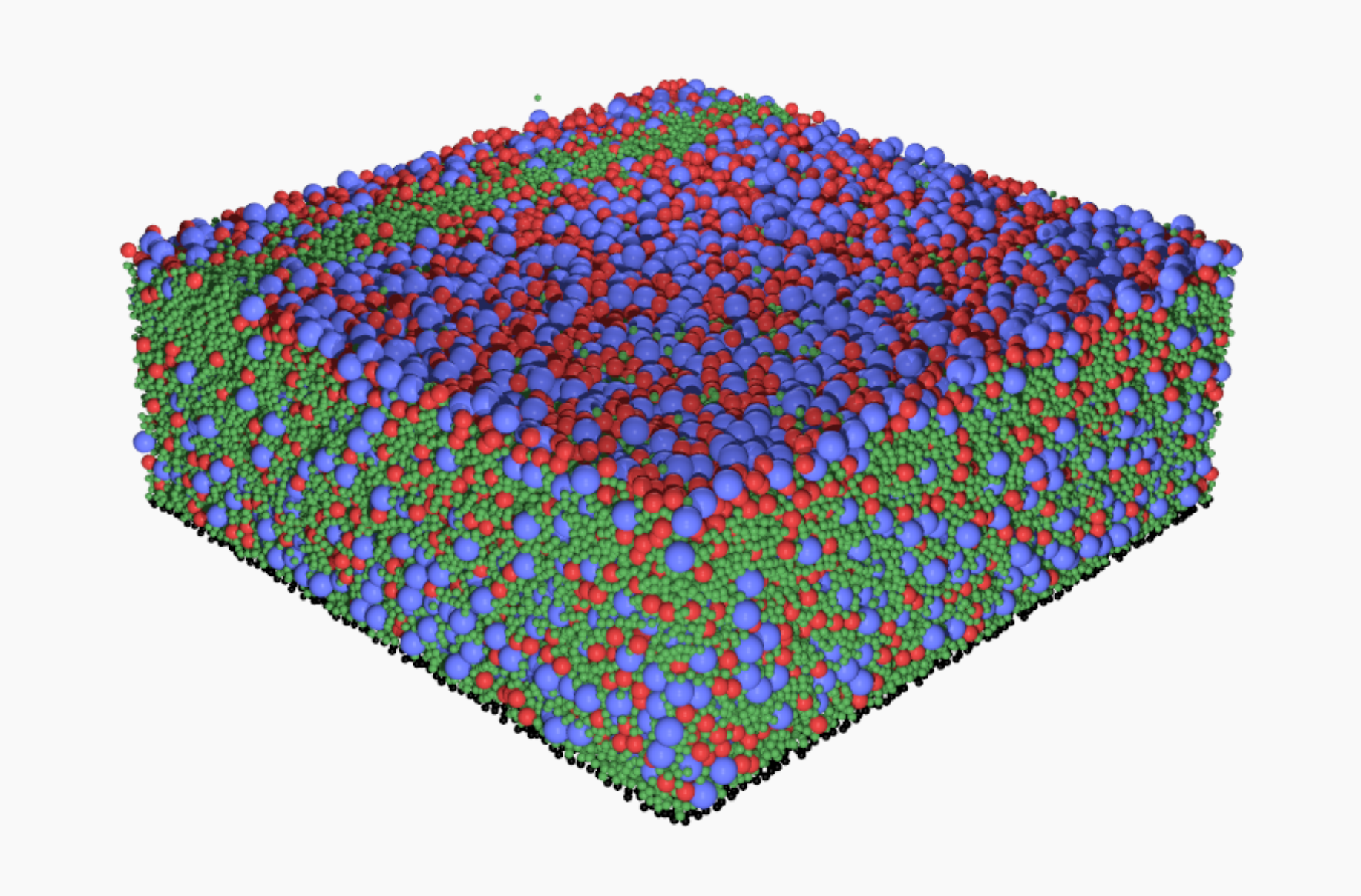}\put(-120,70){(e)}\put(-70,80){$t = 700$}\put(-65,30){$y$}\put(-36,13){$x$}\put(-90,12){$z$}\hfill
    \begin{tikzpicture}[overlay, remember picture]
         \draw[->, orange, very thick, line width=1.7pt, >=latex] (-2.0,1.05) -- (-0.8,1.7); 
         \draw[->, black, very thick, line width=0.5pt, >=latex] (-2.25,0.05) -- (-2.25,1.0); 
        \draw[->, black, very thick, line width=0.5pt, >=latex] (-2.25,0.05) -- (-1.4,0.55); 
        \draw[->, black, very thick, line width=0.5pt, >=latex] (-2.25,0.05) -- (-3.1,0.55); 
    \end{tikzpicture}
    \includegraphics[scale=0.14, trim=520 180 520 320, clip]{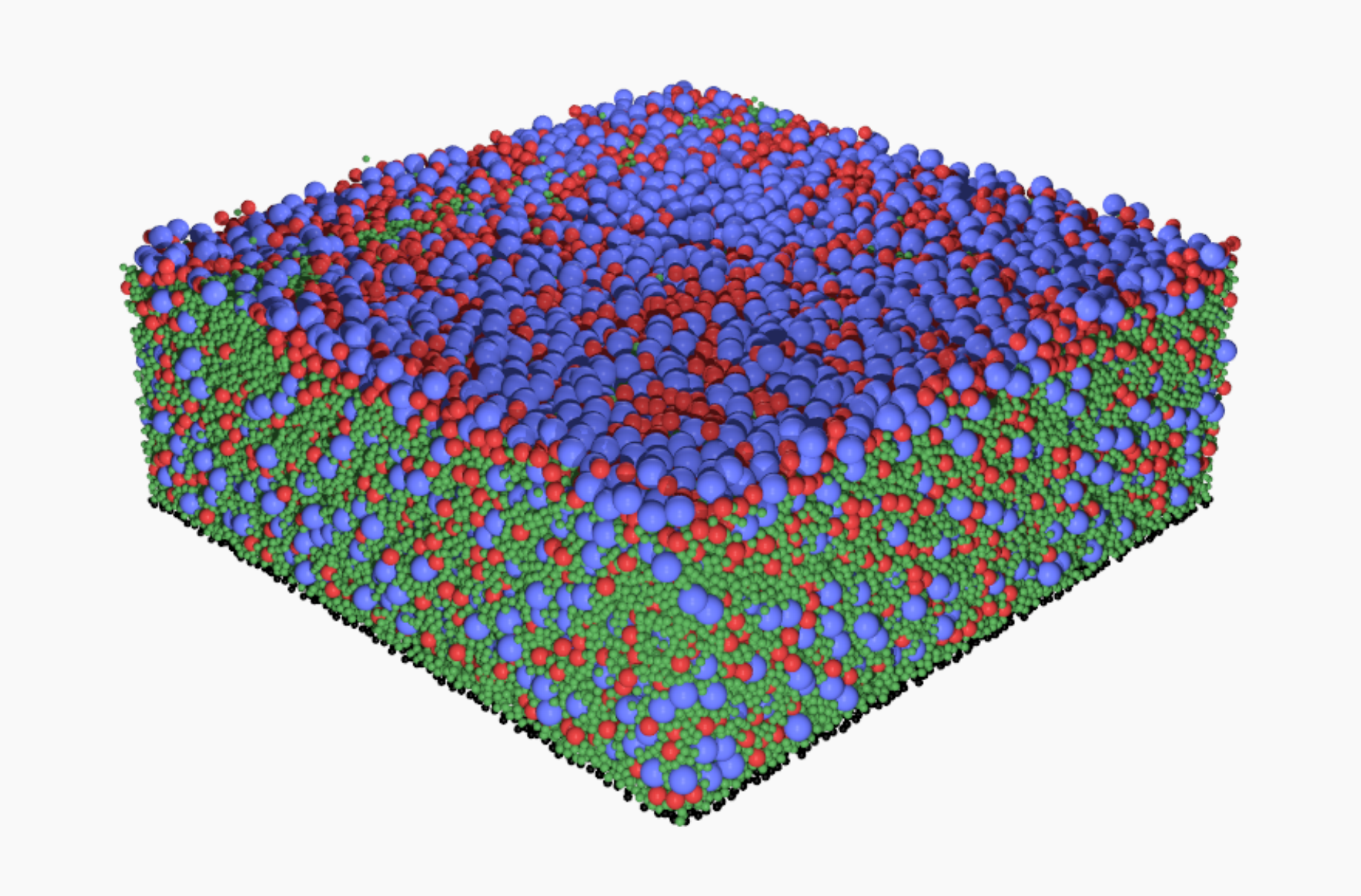}\put(-120,70){(f)}\put(-70,80){$t = 800$}\put(-65,30){$y$}\put(-36,13){$x$}\put(-90,12){$z$}\hfill
    \begin{tikzpicture}[overlay, remember picture]
          \draw[->, orange, very thick, line width=1.7pt, >=latex] (-2.0,1.05) -- (-0.8,1.7); 
         \draw[->, black, very thick, line width=0.5pt, >=latex] (-2.25,0.05) -- (-2.25,1.0); 
        \draw[->, black, very thick, line width=0.5pt, >=latex] (-2.25,0.05) -- (-1.4,0.55); 
        \draw[->, black, very thick, line width=0.5pt, >=latex] (-2.25,0.05) -- (-3.1,0.55);  
    \end{tikzpicture}
    \caption{DEM snapshots of an equal composition ternary mixture consisting of particles with size ratios $3.0 : 2.0 : 1.0$ flowing down an inclined plane at $\theta = 25^\circ$ at different times: (a) $t = 200$, (b) $t = 400$, (c) $t = 500$, (d) $t = 600$, (e) $t = 700$, (f) $t = 800$. The flow starts from a large-near-base configuration. Blue particles indicate larger particles, and green particles represent smaller ones, while red particles denote the intermediate size particles. The black particles are static particles that are used to form a rough, bumpy base to minimize slip at the chute base. \textcolor{black}{The orange arrow denotes the direction of the flow.}}
\label{fig:ternary_snap_LNB_100by100}
\end{figure}
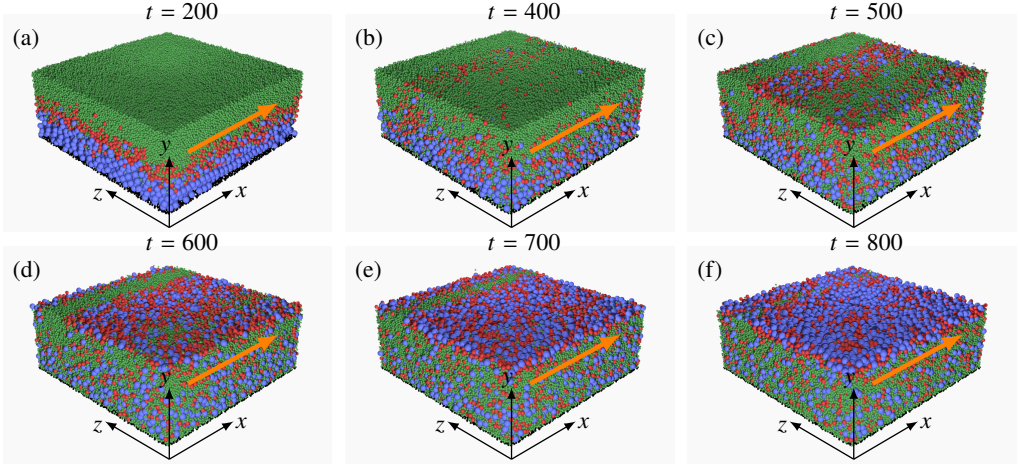
\begin{figure}
    \centering
    \includegraphics[scale=0.22, trim=750 350 650 350, clip]{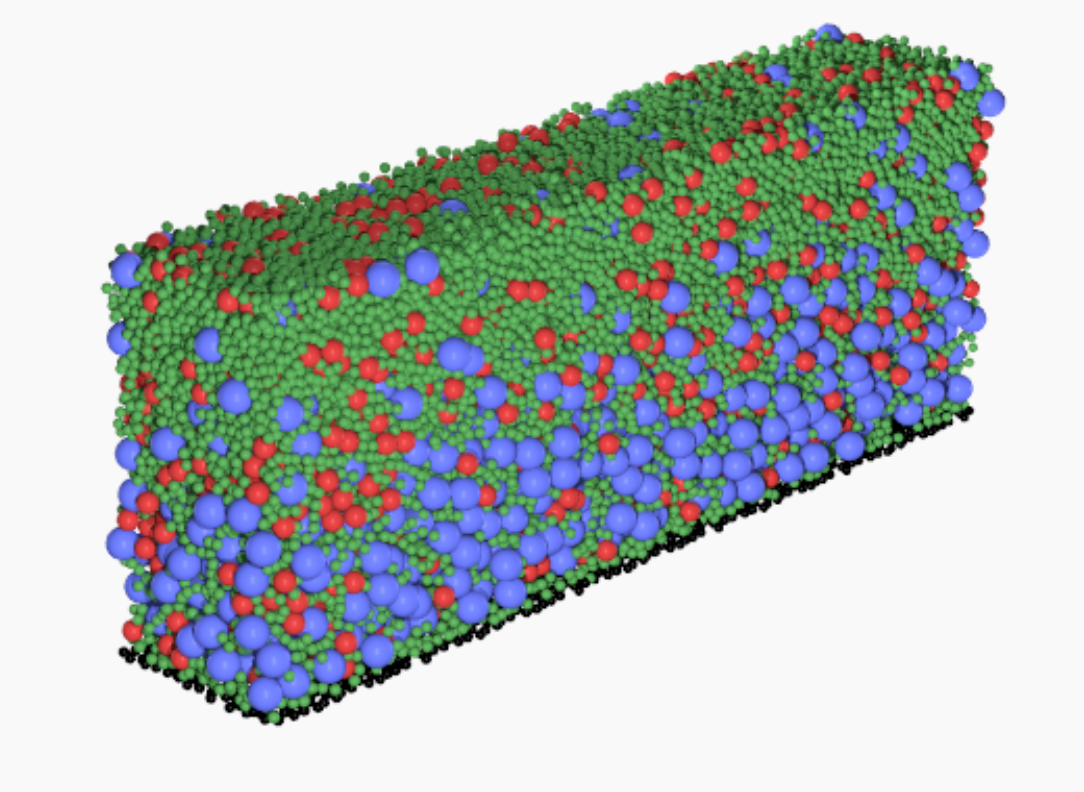}\put(-110,75){(a)}\put(-80,87){$ z = 30d_s$}\hfill
     \includegraphics[scale=0.23, trim=650 400 760 320, clip]{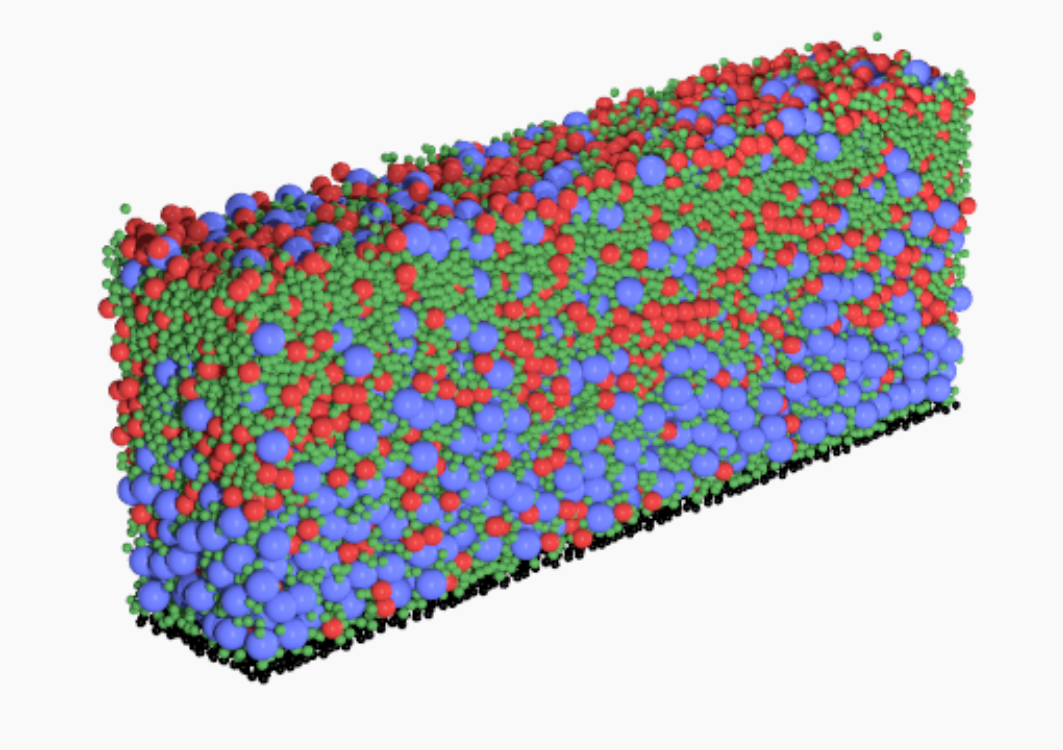}\put(-110,75){(b)}\put(-80,87){$ z = 60d_s$}\hfill
      \includegraphics[scale=0.25, trim=570 450 850 300, clip]{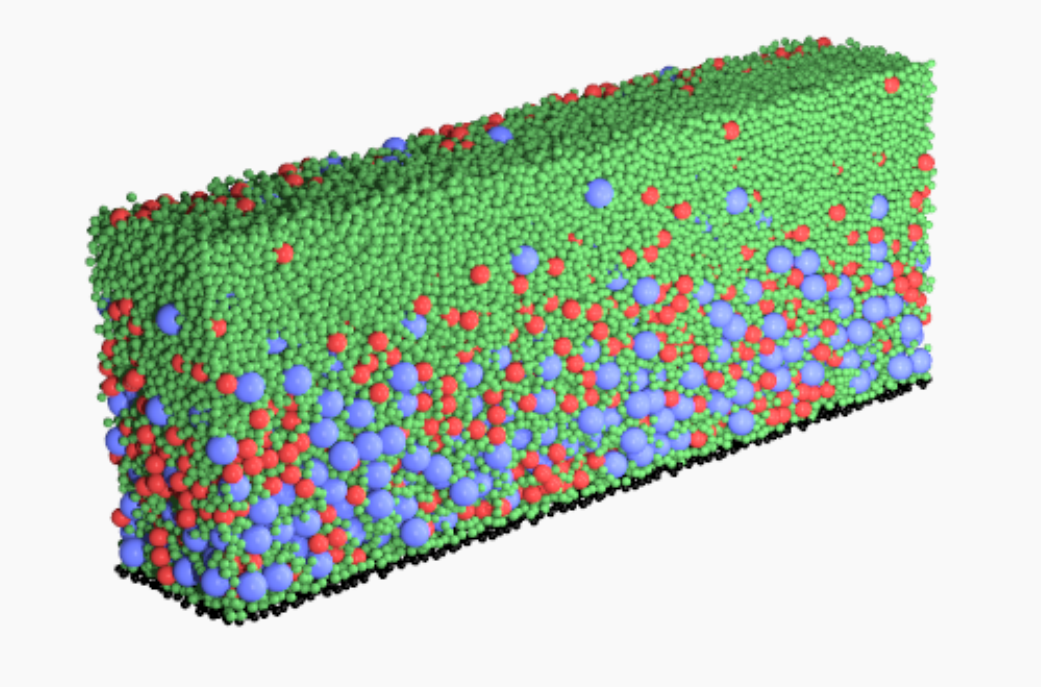}\put(-120,75){(c)}\put(-85,87){$ z = 90d_s$}\\
     \includegraphics[scale=0.38]{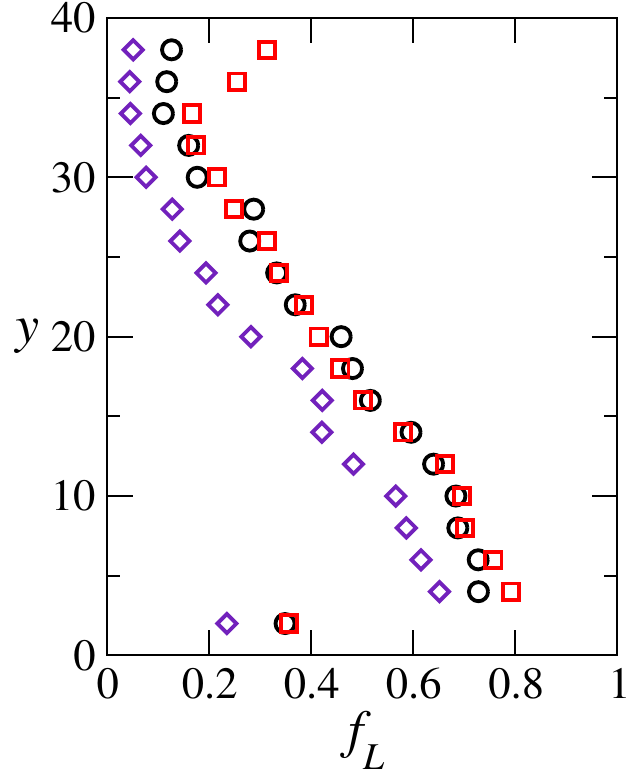}\put(-117,142){(d)} \quad 
    \includegraphics[scale=0.38]{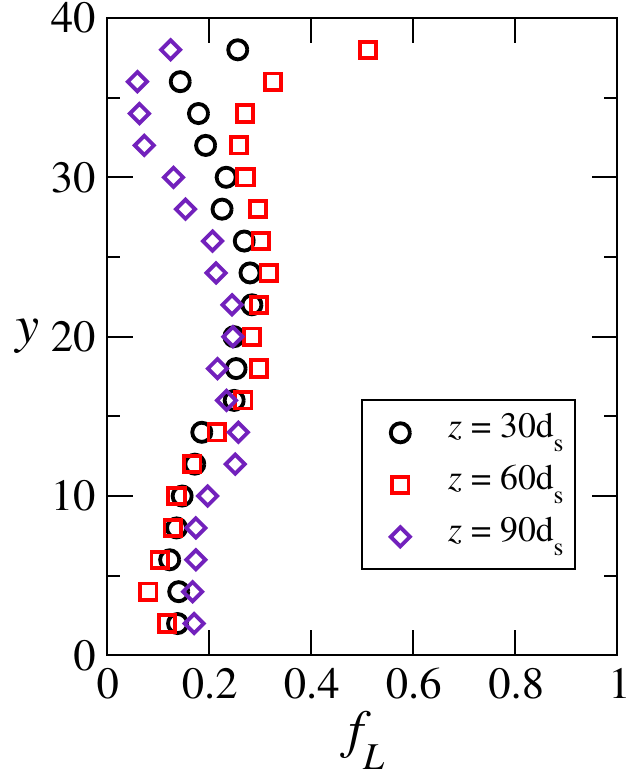}\put(-117,142){(e)} \quad 
    \includegraphics[scale=0.38]{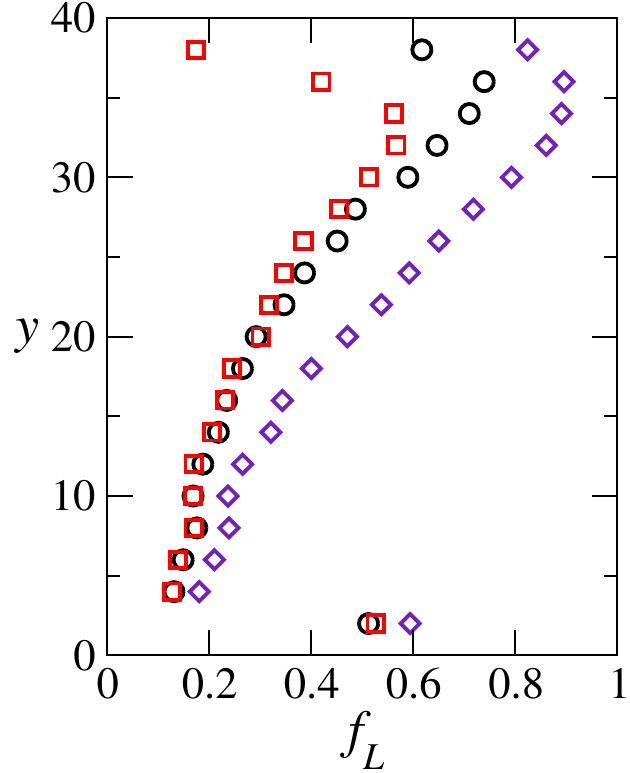}\put(-117,142){(f)} \quad \quad 
    \caption{DEM snapshot of different slices of width $20d_s$ at $t = 500$ as in figure~\ref{fig:ternary_snap_LNB_100by100}c centred at (a) $ z = 30d_s$, (b) $z = 60d_s$, (d) $z = 90d_s$. Variation of species concentration across the layer at different spanwise locations for (d) large particles, (e) medium particles, and (f) small particles.}
\label{fig:ternary_snap_LNB_500_slices}
\end{figure}
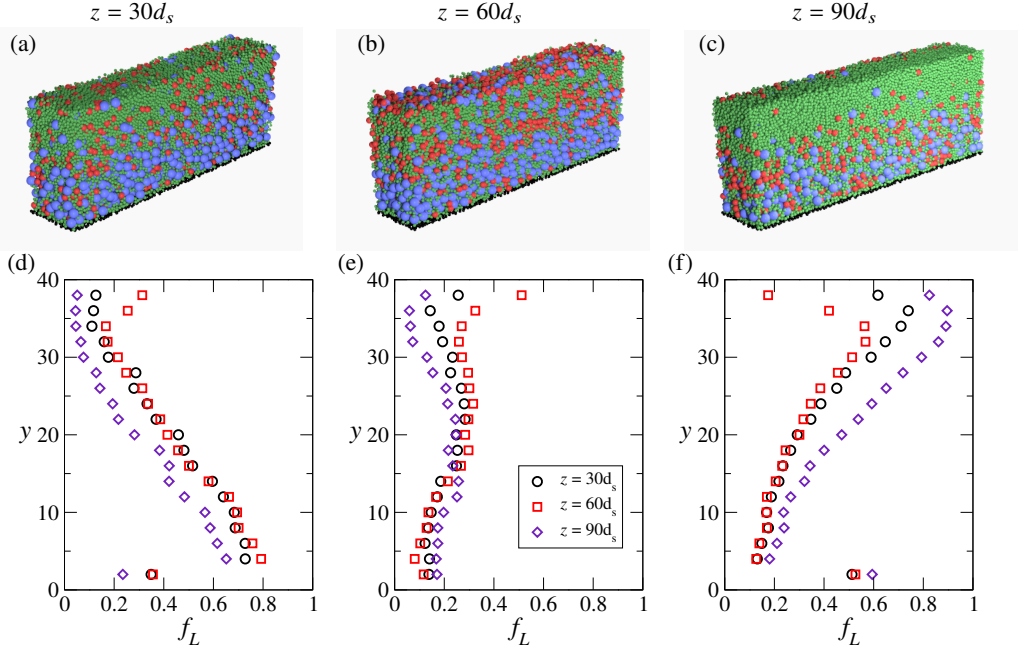

\textcolor{black}{We next turn our attention to large-near-base initial configuration for the large size ratio mixtures.} Figure~\ref{fig:ternary_snap_LNB_100by100} shows DEM snapshots of an equal composition ternary mixture with size ratio $3.0:2.0:1.0$ flowing down an inclined plane at $\theta = 25^\circ$ for this configuration. At $t = 200$, \textcolor{black}{the medium size (red) particles remain sandwiched between large (blue) particles that occupy the region near the base and small (green) particles at the free surface. However,} the interfaces \textcolor{black}{between the species} become diffused \textcolor{black}{due to the low shear at early times (figure~\ref{fig:ternary_snap_LNB_100by100}a).} As the flow evolves, the large and medium particles move upward, while the small particles move downward. Unlike the well-mixed case, figure~\ref{fig:ternary_snap_LNB_100by100}b shows \textcolor{black}{the emergence of a} streamwise band of small and medium particles \textcolor{black}{near the free surface} at $t = 400$. Figure~\ref{fig:ternary_snap_LNB_100by100}c shows that the large (blue) particles also \textcolor{black}{become} visible within the medium-sized particles rich bands at $0 \leq z \leq 20d_s$ and \textcolor{black}{more prominently at} $ 50d_s \leq z \leq 70d_s$ at the free surface. Bands of small (green) particles are also clearly visible \textcolor{black}{next to these bands of medium (red) and large (blue) particles}. This behavior can be seen more clearly in the slices at different spanwise ($z$) locations at $t = 500$ shown in figure~\ref{fig:ternary_snap_LNB_500_slices}. Figures~\ref{fig:ternary_snap_LNB_500_slices}a,~\ref{fig:ternary_snap_LNB_500_slices}b, and~\ref{fig:ternary_snap_LNB_500_slices}c show the slices \textcolor{black}{of $20d_s$ width centred at $z = 30d_s$, $z = 60d_s$, and $z = 90d_s$}, respectively. \textcolor{black}{It is evident that concentration of large (blue) and medium (red)} species particles \textcolor{black}{is higher} at $z = 60d_s$ in comparison to $z = 30d_s$ and $z = 90d_s$. The corresponding concentration profiles of large ($f_L$), medium ($f_M$), and small ($f_S$) species at these three location are shown in figures~\ref{fig:ternary_snap_LNB_500_slices}d,~\ref{fig:ternary_snap_LNB_500_slices}e, and~\ref{fig:ternary_snap_LNB_500_slices}f, respectively. \textcolor{black}{The concentration of large particles is noticeably smaller at $z = 90d_s$ all across the layer (figure~\ref{fig:ternary_snap_LNB_500_slices}d) while that for the medium species differs only near the free surface (figure~\ref{fig:ternary_snap_LNB_500_slices}e).} The profiles of $f_S$ also differ significantly from each other (figure~\ref{fig:ternary_snap_LNB_500_slices}f). \textcolor{black}{These variations in the concentration profiles at different $z$ locations in the spanwise direction confirm that the species concentration fields vary not only along the vertical $y$ direction, but also along the spanwise $z$ direction.} 

Coming back to figure~\ref{fig:ternary_snap_LNB_100by100}, at time $t = 600$, the concentration of large (blue) and medium (red) particles \textcolor{black}{increases} near the free surface, except in the region around $80d_s \leq z \leq 95d_s$ as shown in figure~\ref{fig:ternary_snap_LNB_100by100}d. With further evolution, these \textcolor{black}{small particle} bands \textcolor{black}{at the free surface} gradually \textcolor{black}{shrink due to diffusion of large and medium size particles} at $t = 700$ (figure~\ref{fig:ternary_snap_LNB_100by100}e) and disappear at $t = 800$ (figure~\ref{fig:ternary_snap_LNB_100by100}f). \textcolor{black}{These observations confirm the presence of a Rayleigh-Taylor instability which has been recently reported by~\cite{d2026rayleigh} for binary mixtures with different particle sizes in a large periodic box. While we do not clearly observe such an instability to be present for the well-mixed case, the instability appears \textcolor{black}{to be present} in this large-near-base initial configuration.} \textcolor{black}{The two dimensional evolution of the concentration field due to this Rayleigh-Taylor instability, observed at early times, \textcolor{black}{is not captured by} the one dimensional continuum model. The presence of this instability leads to a non-monotonic evolution of bulk solids fraction $\phi_{avg}$ and average velocity $v_{avg}$ (not shown here) which is not captured by the $1D$ continuum model.}

\section{Conclusion}
\label{chp4_sec:conclusion}
In this work, we have generalized the particle force-based theoretical framework proposed by~\citet{tripathi2021size} to model size-driven segregation in multicomponent granular mixtures. \textcolor{black}{We utilise the particle force expression given by~\cite{yennemadi2023drag} to compute the forces acting on a large size intruder particle in the dilute limit, and following~\cite{tripathi2021size}, generalise that to a binary mixture.} Given the lack of reliable data for forces acting on the small size particles in the assembly of large particles, we first employed a species segregation flux balance in binary mixtures to determine the expression of dimensionless measure of net upward force for small size particles. We then generalized this particle force-based segregation model for multicomponent mixtures by accounting for inter-species interactions on each component in the presence of other species. These inter-species interactions account for the size ratio of the species and local volume fractions, ensuring that the segregation flux for each species arises from net particle level forces. The model is found to be consistent with physical expectations across different limiting mixture compositions and captures the essential physics of segregation by explicitly computing the net force on the species, derived from pairwise interactions with all other species in the mixture. The segregation model shows that in a multicomponent mixture, the largest particle species will always experience a net upward motion, while the smallest species will always experience a net downward motion \textcolor{black}{in dense, free surface granular flows}. For intermediate-sized species, the direction of motion is dependent on the local composition of mixture. This feature, which is very well captured by our particle force-based segregation model, is confirmed from DEM simulations of ternary mixture having different combinations of mixture compositions.

We incorporated this segregation model along with the generalized form of binary mixture rheological model of~\citet{tripathi2011rheology} into the convection–diffusion–segregation equations and developed a continuum model that predicts the spatial and temporal evolution of species concentration and velocity fields. In contrast to other approaches that rely on flow kinematics obtained using experiments or DEM simulations, the present work solves the time-dependent momentum balance equations by directly coupling the generalized granular mixture rheology with this multicomponent size segregation model. We emphasize the importance of this coupling by pointing out that our recent work~\citep{kumawatpandg2025size} demonstrated that concentration fields can be accurately predicted only when their evolution is coupled with the time dependent velocity field for bidisperse mixtures with size ratios greater than $1.5$. The proposed continuum model is utilized to predict the segregation in ternary and quaternary mixtures having different sizes flowing over a periodic chute. The continuum model predictions show good agreement with DEM results obtained for the evolution of species center of mass positions as well as that for the evolution of average mixture velocity over a wide range of mixture compositions at different size ratios. \textcolor{black}{The model works well} for moderate particle size contrasts among the species \textcolor{black}{for both well-mixed and small-near-base initial configurations. In the large-near-base initial configuration, the continuum model is unable to predict the non-monotonic evolution of average mixture velocity and bulk layer solids fraction.} For higher size contrasts, the predictions deviate from the DEM results indicating \textcolor{black}{that the} effect of mixture composition on the solids fraction needs to be accounted explicitly \textcolor{black}{and can not be ignored}. \textcolor{black}{The role of percolation mechanism, where small particles percolate through the voids between larger particles may also contribute to these differences as shown by~\cite{kumawat2025sizetransient}.} 

For large-near-base configuration, the existence of Rayleigh-Taylor instability, which is discussed in detail in the recent work of~\cite{d2026rayleigh} is observed in our DEM simulations of large size contrasts as well. \textcolor{black}{For predicting these features,} the continuum model proposed in this work needs to account for the two dimensional variation of species concentration, which in turn leads to two dimensional variation of other mixture properties. An important direction of future work that emerges from this study is the crucial role of mixture composition on the solids fraction. For an accurate prediction of flow properties of such polydisperse mixtures, a generalized model relating the local species concentration to local mixture solids fraction should be sought for. \textcolor{black}{In future, it will be worth investigating if the continuum model is able to predict the segregation of polydisperse mixtures in long chutes and other flow geometries.}

\backsection[Acknowledgements]{SK acknowledges the help of Mr. Vishal Singh in performing the large scale DEM simulations.}

\backsection[Funding]{AT and SK gratefully acknowledge the funding support provided to SK by the Prime Minister's Research Fellowship (Government of India) grant.}

\backsection[Declaration of interests]{The authors report no conflict of interest.}

\backsection[Data availability statement]{The data that support the findings of this study are available from the corresponding author, AT, upon reasonable request.}

\backsection[Author ORCIDs]{
\newline
Soniya Kumawat  \url{https://orcid.org/0000-0003-3314-9875};\\
Anurag Tripathi \url{ https://orcid.org/0000-0001-9945-3197}.}

\bibliographystyle{jfm}
\bibliography{Denseg_v1}

@article{savage1988particle,
  title={Particle size segregation in inclined chute flow of dry cohesionless granular solids},
  author={Savage, S. B. and Lun, C. K. K.},
  journal={Journal of fluid mechanics},
  volume={189},
  pages={311--335},
  year={1988},
  publisher={Cambridge University Press}
}

@article{d2026rayleigh,
  title={Rayleigh-Taylor instability in size-bidisperse isodense granular flow down an incline},
  author={d'Ortona, Umberto and Lueptow, Richard M and Thomas, Nathalie},
  journal={Physical Review E},
  volume={113},
  number={2},
  pages={025404},
  year={2026},
  publisher={APS}
}

@article{asachi2018experimental,
  title={Experimental evaluation of the effect of particle properties on the segregation of ternary powder mixtures},
  author={Asachi, M. and Hassanpour, A. and Ghadiri, M. and Bayly, A.},
  journal={Powder Technology},
  volume={336},
  pages={240--254},
  year={2018},
  publisher={Elsevier}
}

@article{ayeni2015discrete,
  title={A discrete element method study of granular segregation in non-circular rotating drums},
  author={Ayeni, O. O. and Wu, C. L. and Joshi, J. B. and Nandakumar, K.},
  journal={Powder Technology},
  volume={283},
  pages={549--560},
  year={2015},
  publisher={Elsevier}
}

@article{BRANDAO20201,
title = {Experimental study and {DEM} analysis of granular segregation in a rotating drum},
journal = {Powder Technology},
volume = {364},
pages = {1-12},
year = {2020},
issn = {0032-5910},
doi = {https://doi.org/10.1016/j.powtec.2020.01.036},
url = {https://www.sciencedirect.com/science/article/pii/S0032591020300541},
author = {Rodolfo, J. B. and Rondinelli M. L. and Raphael L. S. and Claudio R. D. and Marcos A. S. B.},
}

@article{barker2021OpenFoam, 
    title={Coupling rheology and segregation in granular flows}, 
    volume={909}, 
    DOI={10.1017/jfm.2020.973}, 
    journal={Journal of Fluid Mechanics}, 
    publisher={Cambridge University Press}, 
    author={Barker, T. and Rauter, M. and Maguire, E. S. F. and Johnson, C. G. and Gray, J. M. N. T.}, 
    year={2021}, 
    pages={A22}
}

@article{barker2017wellposedrheology,
  title={Well-posed continuum equations for granular flow with compressibility and ${\it\mu}(\mathrm{I})$ rheology},
  author={Barker, T. and Schaeffer, D. G. and Shearer, M. and Gray, J. M. N. T.},
  journal={Proceedings of the Royal Society A: Mathematical, Physical and Engineering Sciences},
  volume={473},
  number={2201},
  pages={20160846},
  year={2017},
  publisher={The Royal Society Publishing}
}

@article{chand2012discrete,
  title={Discrete particle simulation of radial segregation in horizontally rotating drum: Effects of drum-length and non-rotating end-plates},
  author={Chand, R. and Khaskheli, M. A. and Qadir, A. and Ge, B. and Shi, Q.},
  journal={Physica A: Statistical Mechanics and its Applications},
  volume={391},
  number={20},
  pages={4590--4596},
  year={2012},
  publisher={Elsevier}
}

@article{combarros2014segregation,
  title={Segregation of particulate solids: Experiments and {DEM} simulations},
  author={Combarros, M. and Feise, H. J. and Zetzener, H. and Kwade, A.},
  journal={Particuology},
  volume={12},
  pages={25--32},
  year={2014},
  publisher={Elsevier}
}

@article{deng2018continuum,
  title={Continuum modelling of segregating tridisperse granular chute flow},
  author={Deng, Z. and Umbanhowar, P. B. and Ottino, J. M. and Lueptow, R. M.},
  journal={Proceedings of the Royal Society A: Mathematical, Physical and Engineering Sciences},
  volume={474},
  number={2211},
  pages={20170384},
  year={2018},
  publisher={The Royal Society Publishing}
}

@article{Duan2021, 
  title={Modelling segregation of bidisperse granular mixtures varying simultaneously in size and density for free surface flows}, 
  volume={918}, 
  DOI={10.1017/jfm.2021.342}, journal={Journal of Fluid Mechanics}, 
  publisher={Cambridge University Press}, 
  author={Duan, Y. and Umbanhowar, P. B. and Ottino, J. M. and Lueptow, R. M.},
  year={2021}, 
  pages={A20}
}

@article{fan2014modelling,
  title={Modelling size segregation of granular materials: the roles of segregation, advection and diffusion},
  author={Fan, Y. and Schlick, C. P. and Umbanhowar, P. B. and Ottino, J. M. and Lueptow, R. M.},
  journal={Journal of Fluid Mechanics},
  volume={741},
  pages={252-279},
  year={2014},
  publisher={Cambridge University Press}
}

@article{fry2019,
author = {Fry, A. M. and Umbanhowar, P. B. and Ottino, J. M. and Lueptow, R. M.},
title = {Diffusion, mixing, and segregation in confined granular flows},
journal = {AIChE Journal},
volume = {65},
number = {3},
pages = {875-881},
doi = {https://doi.org/10.1002/aic.16494},
url = {https://aiche.onlinelibrary.wiley.com/doi/abs/10.1002/aic.16494},
year = {2019}
}

@article{gray_ancey_2011,     
  title={Multi-component particle-size segregation in shallow granular avalanches},  
  volume={678}, 
  DOI={10.1017/jfm.2011.138}, 
  journal={Journal of Fluid Mechanics}, 
  publisher={Cambridge University Press}, 
  author={Gray, J. M. N. T. and Ancey, C.}, 
  year={2011}, 
  pages={535–588}
}

@article{grayandchugunov2006particle,
  title={Particle-size segregation and diffusive remixing in shallow granular avalanches},
  author={Gray, J. M. N. T. and Chugunov, V. A.},
  journal={Journal of Fluid Mechanics},
  volume={569},
  pages={365--398},
  year={2006},
  publisher={Cambridge University Press}
}

@article{gray1997pattern,
  title={Pattern formation in granular avalanches},
  author={Gray, J. M. N. T. and Hutter, K.},
  journal={Continuum Mechanics and Thermodynamics},
  volume={9},
  number={6},
  pages={341--345},
  year={1997},
  publisher={Springer}
}

@article{HSIAU2002,
  title = "Density effect of binary mixtures on the segregation process in a vertical shaker",
  journal = "Advanced Powder Technology",
  volume = "13",
  number = "3",
  pages = "301-315",
  year = "2002",
  issn = "0921-8831",
  doi = "https://doi.org/10.1163/156855202320252462",
  url = "http://www.sciencedirect.com/science/article/pii/S0921883108600692",
  author = "Hsiau, S. S. and Chen, W. C.",
}

@article{Jain2005,
  author={Jain, N. and Ottino, J. M. and Lueptow, R. M.},
  title={Regimes of segregation and mixing in combined size and density granular systems: an experimental study},
  journal={Granular Matter},
  year={2005},
  month={Jul},
  day={01},
  volume={7},
  number={2},
  pages={69-81},
  issn={1434-7636},
  doi={10.1007/s10035-005-0198-x},
  url={https://doi.org/10.1007/s10035-005-0198-x}
}

@article{jones2018asymmetric,
  title={Asymmetric concentration dependence of segregation fluxes in granular flows},
  author={Jones, R. P. and Isner, A. B. and Xiao, H. and Ottino, J. M. and Umbanhowar, P. B. and Lueptow, R. M.},
  journal={Physical Review Fluids},
  volume={3},
  number={9},
  pages={094304},
  year={2018},
  publisher={APS}
}

@article{jop2006constitutivenature,
  title={A constitutive law for dense granular flows},
  author={Jop, P. and Forterre, Y. and Pouliquen, O.},
  journal={Nature},
  volume={441},
  number={7094},
  pages={727-730},
  year={2006},
  publisher={Nature Publishing Group}
}

@article{liao2020behavior,
  title={Behavior of density-induced segregation with multi-density granular materials in a thin rotating drum},
  author={Liao, C. C. and Tsai, C. C.},
  journal={Granular Matter},
  volume={22},
  pages={1--9},
  year={2020},
  publisher={Springer}
}

@article{Pereira2017,
  author={Pereira, G. G. and Cleary, P. W.},
  title={Segregation due to particle shape of a granular mixture in a slowly rotating tumbler},
  journal={Granular Matter},
  year={2017},
  month={Mar},
  day={07},
  volume={19},
  number={2},
  pages={23},
  issn={1434-7636},
  doi={10.1007/s10035-017-0708-7},
  url={https://doi.org/10.1007/s10035-017-0708-7}
}

@article{pereira2011insights,
  title={Insights from simulations into mechanisms for density segregation of granular mixtures in rotating cylinders},
  author={Pereira, G. G. and Sinnott, M. D. and Cleary, P. W. and Liffman, K. and Metcalfe, G. and {\v{S}}utalo, I. D.},
  journal={Granular Matter},
  volume={13},
  number={1},
  pages={53--74},
  year={2011},
  publisher={Springer}
}

@article{pillitteri2020size,
  title={How size ratio and segregation affect the packing of binary granular mixtures},
  author={Pillitteri, S. and Opsomer, E. and Lumay, G. and Vandewalle, N.},
  journal={Soft Matter},
  volume={16},
  number={39},
  pages={9094--9100},
  year={2020},
  publisher={Royal Society of Chemistry}
}

@article{Qiao2021,
  title = {{DEM} study of segregation degree and velocity of binary granular mixtures subject to vibration},
  journal = {Powder Technology},
  volume = {382},
  pages = {107-117},
  year = {2021},
  issn = {0032-5910},
  doi = {https://doi.org/10.1016/j.powtec.2020.12.064},
  url = {https://www.sciencedirect.com/science/article/pii/S0032591020312419},
  author = {Qiao, J. and Duan, C. and Dong, K. and Wang, W. and Jiang, H. and Zhu, H. and Zhao, Y.},
}

@article{sahu_kumawat_agrawal_tripathi_2023, 
    title={Particle force-based density segregation theory for multi-component granular mixtures in a periodic chute flow}, 
    volume={956}, 
    DOI={10.1017/jfm.2023.7}, 
    journal={Journal of Fluid Mechanics}, 
    publisher={Cambridge University Press}, 
    author={Sahu, V. K. and Kumawat, S. and Agrawal, S. and Tripathi, A.}, 
    year={2023}, 
    pages={A8}
}

@article{schlick2016, 
  title={A continuum approach for predicting segregation in flowing polydisperse granular materials},
  volume={797}, 
  DOI={10.1017/jfm.2016.260}, 
  journal={Journal of Fluid Mechanics}, 
  publisher={Cambridge University Press}, 
  author={Schlick, C. P. and Isner, A. B. and Freireich, B. J. and Fan, Y. and Umbanhowar, P. B. and Ottino, J. M. and Lueptow, R. M.}, 
  year={2016}, 
  pages={95–109}
}

@article{trewhela2021experimental,
  title={An experimental scaling law for particle-size segregation in dense granular flows},
  author={Trewhela, T. and Ancey, C. and Gray, J. M. N. T.},
  journal={Journal of Fluid Mechanics},
  volume={916},
  year={2021},
  pages={A55},
  publisher={Cambridge University Press}
}

@article{trewhela2024segregation,
  title={Segregation--rheology feedback in bidisperse granular flows: A coupled {Stokes’} problem},
  author={Trewhela, T.},
  journal={Journal of Fluid Mechanics},
  volume={983},
  pages={A45},
  year={2024},
  publisher={Cambridge University Press}
}

@article{tripathi2021size,
  title = {Theory for size segregation in flowing granular mixtures based on computation of forces on a single large particle},
  author = {Tripathi, A. and Kumar, A. and Nema, M. and Khakhar, D. V.},
  journal = {Physical Review E},
  volume = {103},
  issue = {3},
  pages = {L031301},
  numpages = {5},
  year = {2021},
  month = {Mar},
  publisher = {American Physical Society},
  doi = {10.1103/PhysRevE.103.L031301},
  url = {https://link.aps.org/doi/10.1103/PhysRevE.103.L031301}
}

@article{tripathi2011rheology,
  title={Rheology of binary granular mixtures in the dense flow regime},
  author={Tripathi, A. and Khakhar, D. V.},
  journal={Physics of Fluids},
  volume={23},
  number={11},
  pages={113302},
  year={2011},
  publisher={AIP}
}

@article{tripathi2013density,
  title={Density difference-driven segregation in a dense granular flow},
  author={Tripathi, A. and Khakhar, D. V.},
  journal={Journal of Fluid Mechanics},
  volume={717},
  pages={643-669},
  year={2013},
  publisher={Cambridge University Press}
}

@article{yennemadi2023drag,
  title={Drag, lift, and buoyancy forces on a single large particle in dense granular flows},
  author={Yennemadi, A. V. and Khakhar, D. V.},
  journal={Physical Review Fluids},
  volume={8},
  number={7},
  pages={074301},
  year={2023},
  publisher={APS}
}

@article{maguire2024particle,
  title={Particle-size segregation patterns in a partially filled triangular rotating drum},
  author={Maguire, E. S. F. and Barker, T. and Rauter, M. and Johnson, C. G. and Gray, J. M. N. T.},
  journal={Journal of Fluid Mechanics},
  volume={979},
  pages={A40},
  year={2024},
  publisher={Cambridge University Press}
}

@article{liuandhennan2023coupled,
  title={Coupled continuum modelling of size segregation driven by shear-strain-rate gradients and flow in dense, bidisperse granular media},
  author={Liu, D. and Singh, H. and Henann, D. L.},
  journal={Journal of Fluid Mechanics},
  volume={976},
  pages={A16},
  year={2023},
  publisher={Cambridge University Press}
}

@article{singhandhennan2024continuum,
  title={Continuum modelling of size segregation and flow in dense, bidisperse granular media: accounting for segregation driven by both pressure gradients and shear-strain-rate gradients},
  author={Singh, H. and Liu, D. and Henann, D. L.},
  journal={Journal of Fluid Mechanics},
  volume={988},
  pages={A43},
  year={2024},
  publisher={Cambridge University Press}
}

@article{larcher2013segregation,
  title={Segregation and mixture profiles in dense, inclined flows of two types of spheres},
  author={Larcher, M. and Jenkins, J. T.},
  journal={Physics of Fluids},
  volume={25},
  number={11},
  year={2013},
  publisher={AIP Publishing}
}

@article{larcher2015evolution,
  title={The evolution of segregation in dense inclined flows of binary mixtures of spheres},
  author={Larcher, M. and Jenkins, J. T.},
  journal={Journal of Fluid Mechanics},
  volume={782},
  pages={405--429},
  year={2015},
  publisher={Cambridge University Press}
}

@article{kumaran2015kinetic,
  title={Kinetic theory for sheared granular flows},
  author={Kumaran, V.},
  journal={Comptes Rendus. Physique},
  volume={16},
  number={1},
  pages={51--61},
  year={2015}
}

@article{jenkins1985kinetic,
  title={Kinetic theory for plane flows of a dense gas of identical, rough, inelastic, circular disks},
  author={Jenkins, J. T. and Richman, M. W.},
  journal={The Physics of fluids},
  volume={28},
  number={12},
  pages={3485--3494},
  year={1985},
  publisher={American Institute of Physics}
}

@article{kumaran2008dense,
  title={Dense granular flow down an inclined plane: from kinetic theory to granular dynamics},
  author={Kumaran, V.},
  journal={Journal of Fluid Mechanics},
  volume={599},
  pages={121--168},
  year={2008},
  publisher={Cambridge University Press}
}

@article{jenkins2020segregation,
  title={Segregation in a dense, inclined, granular flow with basal layering},
  author={Jenkins, J. T. and Larcher, M.},
  journal={Granular Matter},
  volume={22},
  pages={1--8},
  year={2020},
  publisher={Springer}
}

@PREAMBLE{
 "\providecommand{\noopsort}[1]{}" 
 # "\providecommand{\singleletter}[1]{#1}%" 
}

@article{barker_2021_gray, 
title={Coupling rheology and segregation in granular flows},
volume={909}, DOI={10.1017/jfm.2020.973},
journal={J. Fluid Mech.},
publisher={Cambridge University Press},
author={Barker, T. and Rauter, M. and Maguire, E. S. F. and Johnson, C. G. and Gray, J. M. N. T.},
year={2021},
pages={A22}}

@article{khakhar1999mixing,
  title={Mixing and segregation of granular materials in chute flows},
  author={Khakhar, D. V. and McCarthy, J. J. and Ottino, J. M.},
  journal={Chaos: An Interdisciplinary Journal of Nonlinear Science},
  volume={9},
  number={3},
  pages={594--610},
  year={1999},
  publisher={American Institute of Physics}
}

@article{kumawat2025transient,
  title={Transient segregation of different density granular mixtures},
  author={Kumawat, S. and Sahu, V. K. and Tripathi, A.},
  journal={Journal of Fluid Mechanics},
  volume={1008},
  pages={A53},
  year={2025},
  publisher={Cambridge University Press}
}

@article{jing2020rising,
  title={Rising and sinking intruders in dense granular flows},
  author={Jing, L. and Ottino, Julio M and Lueptow, R. M. and Umbanhowar, Paul B},
  journal={Physical Review Research},
  volume={2},
  number={2},
  pages={022069},
  year={2020},
  publisher={APS}
}

@article{bhattacharya2014chute,
  title={Chute flow as a means of segregation characterization},
  author={Bhattacharya, T. and McCarthy, J. J.},
  journal={Powder technology},
  volume={256},
  pages={126--139},
  year={2014},
  publisher={Elsevier}
}

@article{guillard2016scaling,
  title={Scaling laws for segregation forces in dense sheared granular flows},
  author={Guillard, Fran{\c{c}}ois and Forterre, Yo{\"e}l and Pouliquen, Olivier},
  journal={Journal of Fluid Mechanics},
  volume={807},
  pages={R1},
  year={2016},
  publisher={Cambridge University Press}
}

@article{jing2021unified,
  title={A unified description of gravity-and kinematics-induced segregation forces in dense granular flows},
  author={Jing, L. and Ottino, Julio M and Lueptow, Richard M and Umbanhowar, Paul B},
  journal={Journal of Fluid Mechanics},
  volume={925},
  pages={A29},
  year={2021},
  publisher={Cambridge University Press}
}

@article{duan2022segregation,
  title={Segregation forces in dense granular flows: closing the gap between single intruders and mixtures},
  author={Duan, Y. and Jing, L. and Umbanhowar, P. B. and Ottino, J. M. and Lueptow, R. M.},
  journal={Journal of Fluid Mechanics},
  volume={935},
  pages={R1},
  year={2022},
  publisher={Cambridge University Press}
}

@article{golick2009mixing,
  title={Mixing and segregation rates in sheared granular materials},
  author={Golick, L. A. and Daniels, K. E.},
  journal={Physical Review E—Statistical, Nonlinear, and Soft Matter Physics},
  volume={80},
  number={4},
  pages={042301},
  year={2009},
  publisher={APS}
}

@article{van2015underlying,
  title={Underlying asymmetry within particle size segregation},
  author={van der Vaart, K. and Gajjar, P. and Epely-Chauvin, G. and Andreini, N. and Gray, J. M. N. T. and Ancey, C.},
  journal={Physical Review Letters},
  volume={114},
  number={23},
  pages={238001},
  year={2015},
  publisher={APS}
}

@article{kumawat2025sizetransient,
  title = {Transient segregation of bidisperse granular mixtures in a periodic chute flow},
  author = {Kumawat, S. and Sahu, V. K. and Tripathi, A.},
  journal = {Phys. Rev. Fluids},
  volume = {10},
  issue = {12},
  pages = {L122301},
  numpages = {9},
  year = {2025},
  month = {Dec},
  publisher = {American Physical Society},
  doi = {10.1103/3xfm-4x3r},
  url = {https://link.aps.org/doi/10.1103/3xfm-4x3r}
}

@article{kumawatpandg2025size,
	author = {Kumawat, S. and Tripathi, A.},
	title = {Transient size segregation of binary granular mixtures},
	DOI= "10.1051/epjconf/202534005001",
	url= "https://doi.org/10.1051/epjconf/202534005001",
	journal = {EPJ Web Conf.},
	year = 2025,
	volume = 340,
	pages = "05001",
}


\clearpage
\phantomsection 
\section*{Supplementary Material}
\addcontentsline{toc}{section}{Supplementary Material} 

\setcounter{section}{0}
\renewcommand{\thesection}{SM\arabic{section}}

\renewcommand{\theHsection}{SM\arabic{section}} 

\setcounter{figure}{0}
\renewcommand{\thefigure}{S\arabic{figure}}
\renewcommand{\theHfigure}{S\arabic{figure}}

\setcounter{equation}{0}
\renewcommand{\theequation}{SM\arabic{section}.\arabic{equation}}
\renewcommand{\theHequation}{SM\arabic{section}.\arabic{equation}} 
\makeatletter
\@addtoreset{equation}{section} 
\makeatother


\section{Comparison of predictions for species centre of mass by solving the species transport equation for large as well as small species}
\label{sec:comparison_small_large_v1}
\begin{figure}
    \centering
    \includegraphics[width=0.45\linewidth]{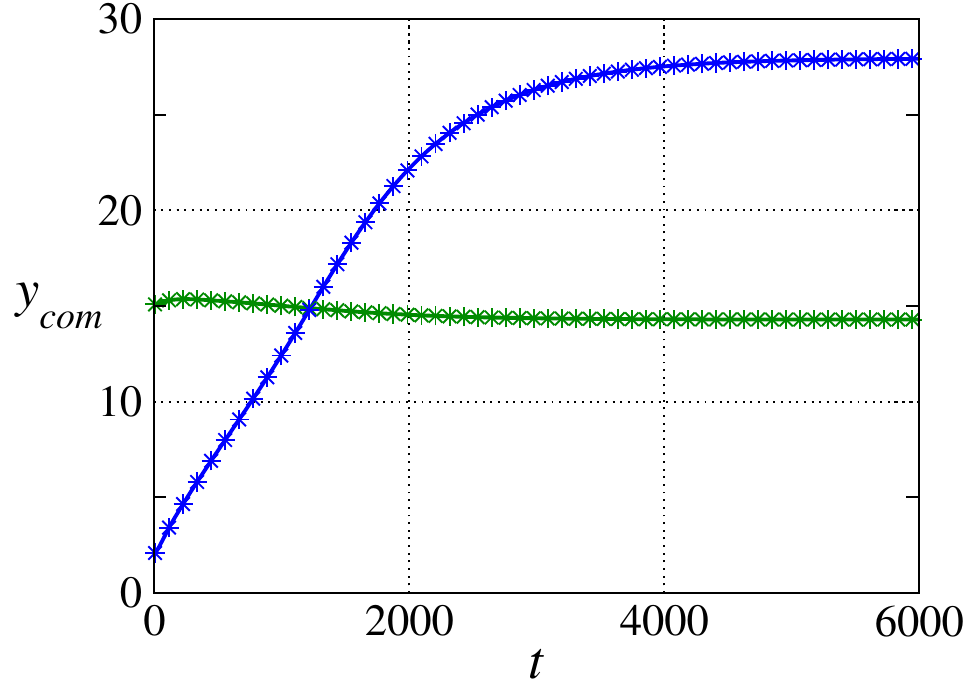} \put(-170,125){(a)}\put(-110,125){\small $r = 2.0, f^T_L = 0.05$} \put(-60,50){\textcolor{OliveGreen}{Small}}\put(-60,95){\textcolor{Blue}{Large}}
    \quad
    \vspace{0.5cm}
    \includegraphics[width=0.45\linewidth]{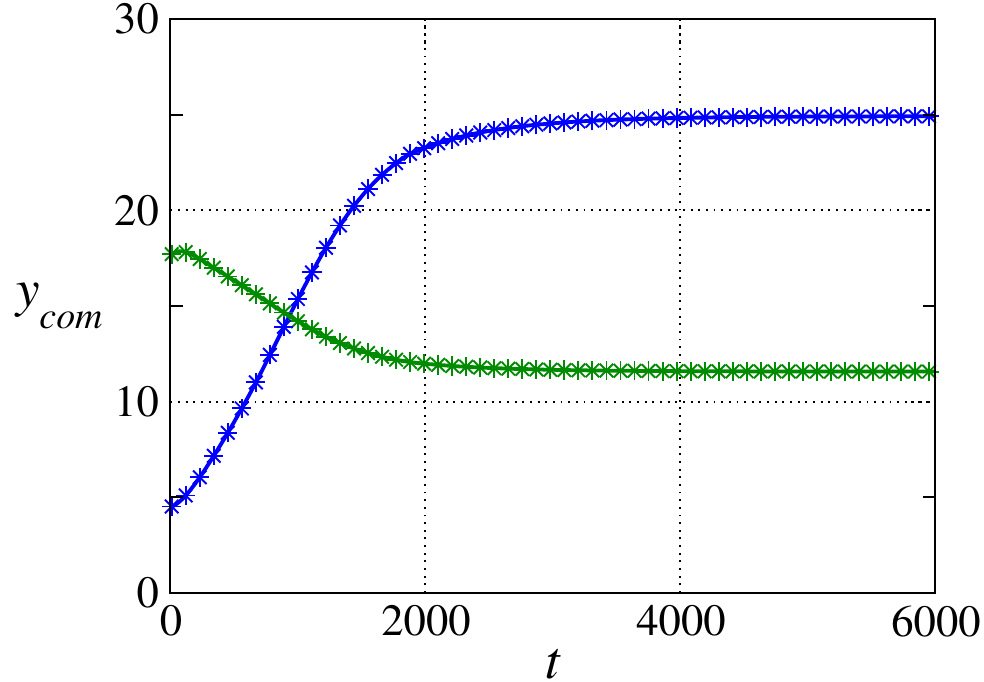} \put(-170,125){(b)}\put(-110,125){\small $r = 2.0, f^T_L = 0.25$}\put(-60,40){\textcolor{OliveGreen}{Small}}\put(-60,87){\textcolor{Blue}{Large}} \quad
    \vspace{0.5cm}
    \includegraphics[width=0.45\linewidth]{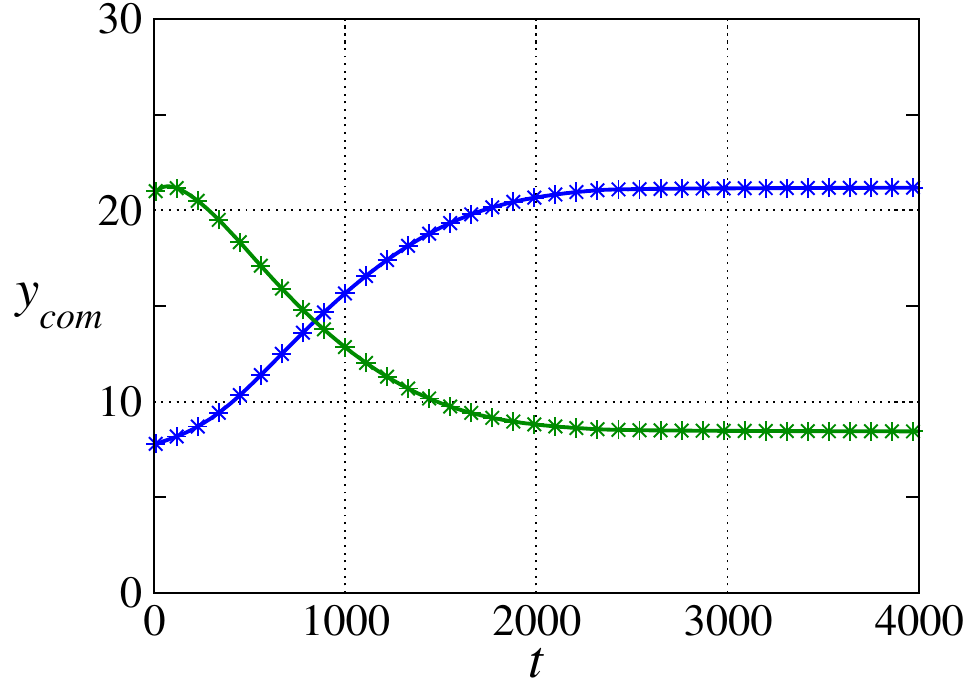}\put(-170,125){(c)}\put(-110,125){\small $r = 2.0, f^T_L = 0.50$}\put(-60,53){\textcolor{OliveGreen}{Small}}\put(-60,95){\textcolor{Blue}{Large}}\quad
\includegraphics[width=0.45\linewidth]{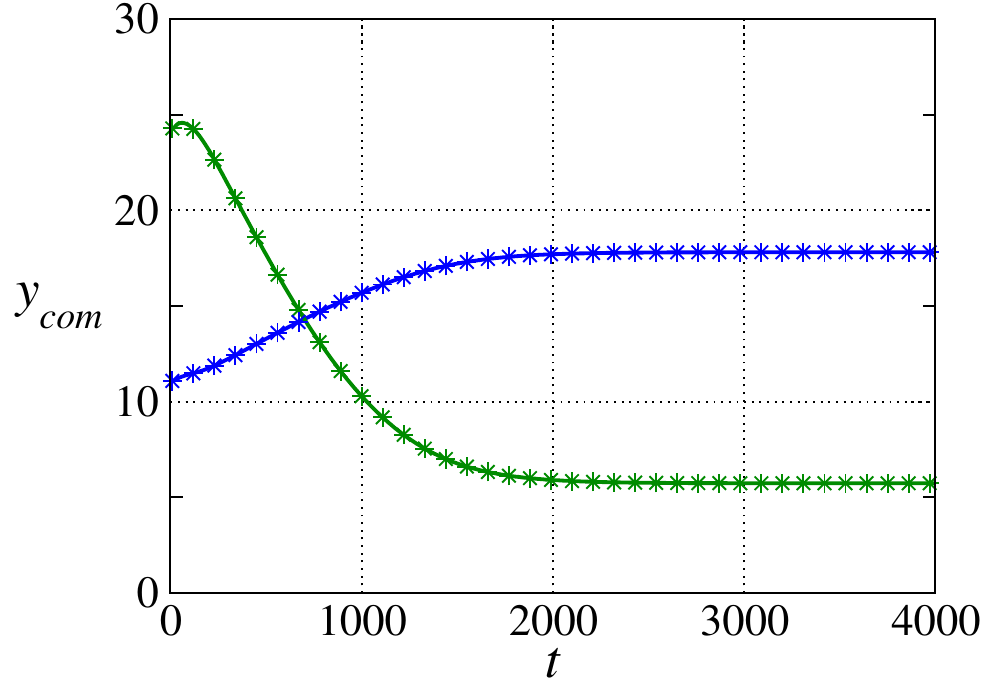} \put(-170,125){(d)}\put(-110,125){\small $r = 2.0, f^T_L = 0.75$} \put(-60,43){\textcolor{OliveGreen}{Small}}\put(-60,87){\textcolor{Blue}{Large}}\quad
  \vspace{0.5cm}
\includegraphics[width=0.45\linewidth]{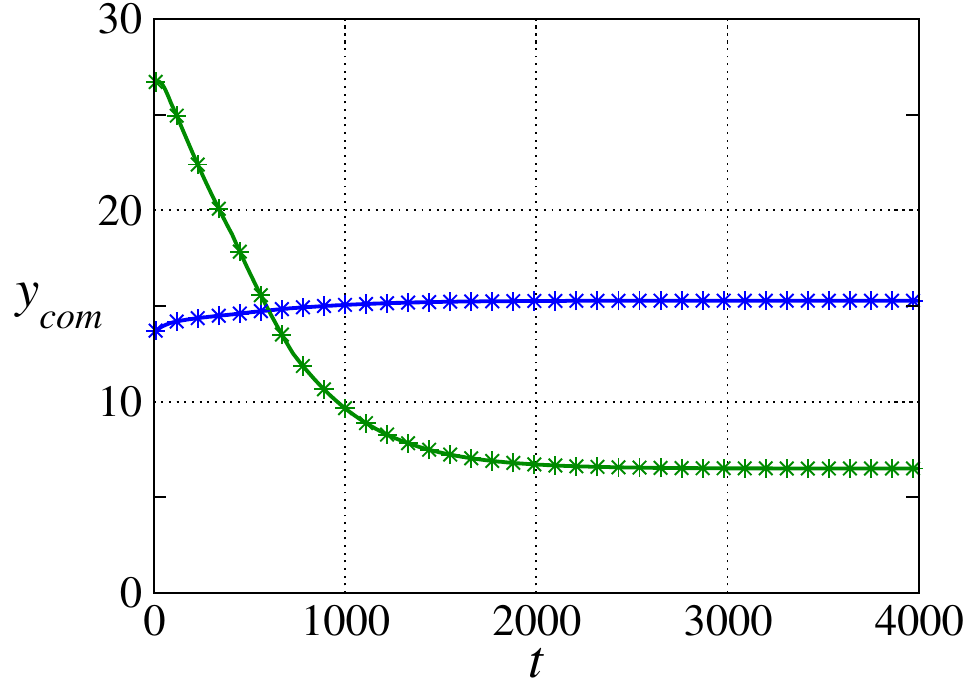}\put(-170,125){(e)}\put(-110,125){\small $r = 2.0, f^T_L = 0.95$}\put(-60,45){\textcolor{OliveGreen}{Small}}\put(-60,80){\textcolor{Blue}{Large}}
\caption{Comparison of time evolution of $y_{com}$ by solving the species transport equation for the small particles (star symbols) and by solving it for the large particles (solid lines) for a binary mixture having size ratio $r = 2.0$ flowing over inclination angle $\theta = 25^\circ$. \textcolor{black}{Both approaches give identical results, confirming the consistent formulation of the theory.}}
\label{fig:speciesflux_comparison}
\end{figure}

We discuss the consistency of particle force-based segregation for binary mixtures where we predict segregation by solving the segregation-diffusion transport equation for large species (Equation~\ref{eq:large_con_seg_diff_eq}) and also for small species (Equation~\ref{eq:small_con_seg_diff_eq}) following the approach of~\cite{kumawat2025sizetransient}. Figure~\ref{fig:speciesflux_comparison} shows the center of mass variation of large and small species in a binary mixture of size ratio $r = 2.0$ flowing over an inclined plane $\theta = 25^\circ$. Figures~\ref{fig:speciesflux_comparison}a -~\ref{fig:speciesflux_comparison}e show time variation of $y_{com}$ for total large particle compositions of $f^T_L = 0.05$, $0.25$, $0.50$, $0.75$, $0.95$, respectively. Symbols (star) represent the model predictions by solving the segregation diffusion equation for the small particles, while the solid lines correspond to the solution using the segregation flux expression for the large particles. In each case, both approaches give identical results, confirming that the model reliably captures segregation behavior irrespective of whether the transport equation is solved for the small or the large species.

\clearpage
\section{Ternary mixture segregation model to binary mixture segregation model}
\label{sec:appendix_ternary_to_binary}
In section~\ref{subsec:size_multicomponent}, discussion near equation~\ref{eq:chp4_mediumfluxternary}, we have shown that the ternary mixture segregation expression reduces to that for binary mixtures in the limit of one species concentration approaching zero. Alternatively, ternary mixture can approach a binary mixture if the sizes of any of the two species become identical. In this section, we show that our segregation flux formulation based on particle forces consistently reduces to the appropriate binary mixture expression from those of ternary mixtures.
The segregation flux of large species in a ternary mixture is given as 
\begin{equation}
    J_L^{seg}=\frac{m_Lg_y}{c_L \pi \eta d_L} \left[ \alpha_{LS}^0 (1+k_{LS}f_L) f_S + \alpha^0_{LM}(1+k_{LM}f_L)f_M \right]f_L
    \label{eq:Appendicx_large_seg_ternary_to_binary}
\end{equation}
As before the diameters of large, medium, and small species are $d_L$, $d_M$, and $d_S$, respectively. In order to reduce it to a binary mixture, we consider $d_M = d_S$ so that size ratio $r_{MS} = 1$. Since $d_M = d_S$, we have $k_{LS} = d_L/d_S$ becomes equal to $k_{LM} = d_L/d_M$, i.e., $k_{LS}=k_{LM}$. Further, since small and medium species are identical we have $\alpha^0_{LS}=\alpha^0_{LM}$. 
Using these, equation~\ref{eq:Appendicx_large_seg_ternary_to_binary} becomes 
\begin{equation}
    J_L^{seg}=\frac{m_Lg_y}{c_L \pi \eta d_S}\left[ \alpha^0_{LS}(1+k_{LS}f_L)(f_S+f_M) \right]f_L.
\end{equation}

\[\begin{aligned}
     1- f_L = f_S+f_M
\end{aligned}\]

\begin{equation}
    J_L^{seg}=\frac{m_Lg_y}{c_L \pi \eta d_L}\alpha^0_{LS}(1+k_{LS}f_L) (1 - f_L) f_L \hspace{0.5 cm} 
\end{equation}

Similarly, the segregation flux of small species in ternary mixture is written as
\begin{equation}
    J_S^{seg}=\frac{m_Sg_y}{c_S \pi \eta d_S} \left[ \alpha^0_{SL}[1+k_{SL}(1 - f_L)]f_L\ +  \alpha^0_{SM}[1+k_{SM}(1 - f_M)]f_M \right] f_S
\end{equation}

For $d_S=d_M$, $ \alpha^0_{SM}=\alpha^0_{SS}=0$

\begin{equation}
     J_S^{seg}=\frac{m_Lg_y}{c_L \pi \eta d_S}[ \alpha_{SL}^0 (1+k_{SL} (1 - f_L))]f_Lf_S
\end{equation}

Hence, the segregation flux of large and small species in the ternary mixture reduces as in the binary mixture.

\clearpage
\section{Sum of segregation flux in multicomponent mixtures}
\label{sec:sum_segregation_zero}
The generalized transport equation given by equation~\ref{eq:ch4_adv_diff_seg_vectoreqn} when written for all the $N$ components and summed over gives $ \nabla \cdot \textbf{v}  +  \nabla \cdot (\textbf{J}_{i}^{seg}) + \nabla \cdot (\textbf{J}_{i}^{diff})=0$ since $\sum f_i = 1$. For the incompressible flow $\nabla \cdot \textbf{v} = 0$ giving $\nabla \cdot (\textbf{J}_{i}^{seg} + \textbf{J}_{i}^{diff}) = 0$. In order to be valid for all multicomponent mixtures in variety of conditions at different positions, it is required that $\sum \textbf{J}_{i}^{seg} + \textbf{J}_{i}^{diff} = 0$. Since our formulation assumes the diffusion coefficient to be equal for all the species, $\sum \textbf{J}_{i}^{diff} = - D \nabla \sum f_i = 0$ as $\sum f_i = 1$. This requires that $ \sum \textbf{J}_{i}^{seg} = 0 $. In this section, we consider the case of a ternary mixture consisting of large, medium, and small species and show that the total segregation flux, $J_L^{\mathrm{seg}} + J_M^{\mathrm{seg}} + J_S^{\mathrm{seg}}$ is equal to zero. 
Segregation flux of large species is written as,
\begin{equation}
    J^{seg}_{L}=\frac{m_Lg_y}{c_L\pi\eta d_L}[\alpha_{LS}^{0} (1+k_{LS}~f_L) ~f_S+\alpha_{LM}^{0} (1+k_{LM}~f_L) ~f_M] f_L.
\end{equation}
Segregation flux of small species is written as,
\begin{align}
   J^{seg}_{S} = \frac{m_S g_y}{c_S\pi\eta d_S} \bigg(\alpha_{SL}^{0}  [1 + k_{SL}~(1-f_L)] ~f_L + \alpha_{SM}^{0}  [1 + k_{SM} ~(1 - f_M)]~f_M \bigg) f_S.
\end{align}
where $k_{SL} = -\frac{k_{LS}}{1 + k_{LS}}$, $k_{SM} = -\frac{k_{MS}}{1 + k_{MS}}$, $\alpha_{SL}^{0} = - \alpha_{LS}^{0}~\frac{c_S}{c_L} r^2_{LS} (1+k_{LS})$ and $\alpha_{SM}^{0} = - \alpha_{MS}^{0}~\frac{c_S}{c_M} r^2_{MS} (1+k_{MS})$. \\

Segregation flux of medium species is written as,
\begin{align}
    J^{seg}_{M}  =\frac{m_M g_y}{c_M\pi\eta d_M} \bigg(\alpha_{ML}^{0} [1 + k_{ML}~(1 - f_L) ]~f_L
    +\alpha_{MS}^{0} (1 + k_{MS}~f_M )~f_S \bigg) f_M.
\end{align}
where $k_{ML} = -\frac{k_{LM}}{1 + k_{LM}}$, and $\alpha_{ML}^{0} =  - \alpha_{LM}^{0}~\frac{c_M}{c_L} r^2_{LM} (1+k_{LM})$. \\

Adding $J^{seg}_L$, $J^{seg}_S$, and $J^{seg}_M$
\begin{align}
J_L^{seg} + J_M^{seg} + J_S^{seg}
&= \frac{m_L g_y}{c_L \pi \eta d_L}
\big[\alpha_{LS}^{0} (1+k_{LS}f_L)f_S + \alpha_{LM}^{0} (1+k_{LM}f_L)f_M\big] f_L
\nonumber \\[4pt]
&\quad + \frac{m_M g_y}{c_M \pi \eta d_M}
\big[\alpha_{ML}^{0} \big(1 + k_{ML}(1-f_L)\big)f_L
+ \alpha_{MS}^{0} (1+k_{MS}f_M)f_S\big] f_M
\nonumber \\[4pt]
&\quad + \frac{m_S g_y}{c_S \pi \eta d_S}
\big[\alpha_{SL}^{0} \big(1+k_{SL}(1-f_L)\big)f_L
+ \alpha_{SM}^{0} \big(1+k_{SM}(1-f_M)\big)f_M\big] f_S.
\end{align}

Let us assume the following: 
\begin{align}
A_L &= \frac{m_L g_y}{c_L \pi \eta d_L}
= \frac{\left(\frac{\pi}{6} d_L^3\right)\rho_p g_y}{c_L \pi \eta d_L}
= \left(\frac{\rho_p g_y}{6\eta}\right)\left(\frac{d_L^2}{c_L}\right),
\\[6pt]
A_M &= \frac{m_M g_y}{c_M \pi \eta d_M}
= \frac{\left(\frac{\pi}{6} d_M^3\right)\rho_p g_y}{c_M \pi \eta d_M}
= \left(\frac{\rho_p g_y}{6\eta}\right)\left(\frac{d_M^2}{c_M}\right),
\\[6pt]
A_S &= \frac{m_S g_y}{c_S \pi \eta d_S}
= \frac{\left(\frac{\pi}{6} d_S^3\right)\rho_p g_y}{c_S \pi \eta d_S}
= \left(\frac{\rho_p g_y}{6\eta}\right)\left(\frac{d_S^2}{c_S}\right).
\end{align}

\begin{align}
J_L^{\mathrm{seg}} + J_S^{\mathrm{seg}} + J_M^{\mathrm{seg}}
&=
\left(\frac{\rho_p g_y}{6\eta}\right)
\Bigg[
\frac{d_L^2}{c_L}
\left\{
\alpha_{LS}^{0}(1+k_{LS}f_L)f_S
+
\alpha_{LM}^{0}(1+k_{LM}f_L)f_M
\right\}f_L
\nonumber\\
&\quad+
\frac{d_M^2}{c_M}
\left\{
\alpha_{ML}^{0}\left[1+k_{ML}(1-f_L)\right]f_L
+
\alpha_{MS}^{0}(1+k_{MS}f_M)f_S
\right\}f_M
\nonumber\\
&\quad+
\frac{d_S^2}{c_S}
\left\{
\alpha_{SL}^{0}\left[1+k_{SL}(1-f_L)\right]f_L
+
\alpha_{SM}^{0}\left[1+k_{SM}(1-f_M)\right]f_M
\right\}f_S
\Bigg].
\end{align}

Now grouping the pairwise terms:

\begin{align}
\text{Term }(1)
&=
\frac{d_L^2}{c_L}\alpha_{LS}^{0}(1+k_{LS}f_L)f_Sf_L
+
\frac{d_S^2}{c_S}\alpha_{SL}^{0}\left[1+k_{SL}(1-f_L)\right]f_Lf_S.
\end{align}

\begin{align}
\text{Term }(2)
&=
\frac{d_L^2}{c_L}\alpha_{LM}^{0}(1+k_{LM}f_L)f_Mf_L
+
\frac{d_M^2}{c_M}\alpha_{ML}^{0}\left[1+k_{ML}(1-f_L)\right]f_Lf_M.
\end{align}

\begin{align}
\text{Term }(3)
&=
\frac{d_S^2}{c_S}\alpha_{SM}^{0}\left[1+k_{SM}(1-f_M)\right]f_Mf_S
+
\frac{d_M^2}{c_M}\alpha_{MS}^{0}(1+k_{MS}f_M)f_Sf_M.
\end{align}
Now, we simplify $\text{Term }(1)$, $\text{Term }(2)$, and $\text{Term }(3)$ separately.

\begin{align}
\text{Term }(1)
&=
\left\{
\frac{d_L^2}{c_L}\alpha_{LS}^{0}(1+k_{LS}f_L)
-
\frac{d_S^2}{c_S}\alpha_{LS}^{0}\left(\frac{c_S}{c_L}\right) r_{LS}^{2}(1+k_{LS})
\left[
1-\frac{k_{LS}}{1+k_{LS}}(1-f_L)
\right]
\right\}f_Lf_S
\nonumber\\
&=
\left\{
\frac{d_L^2}{c_L}\alpha_{LS}^{0}(1+k_{LS}f_L)
-
\frac{d_S^2}{c_S}\alpha_{LS}^{0}\left(\frac{c_S}{c_L}\right)
\left(\frac{d_L^2}{d_S^2}\right)(1+k_{LS})
\left[
1-\frac{k_{LS}}{1+k_{LS}}(1-f_L)
\right]
\right\}f_Lf_S
\nonumber\\
&=
\frac{d_L^2}{c_L}\alpha_{LS}^{0}
\left[
(1+k_{LS}f_L)
-
(1+k_{LS})
\left(
1-\frac{k_{LS}}{1+k_{LS}}(1-f_L)
\right)
\right]f_Lf_S
\nonumber\\
&=
\frac{d_L^2}{c_L}\alpha_{LS}^{0}
\left[
1+k_{LS}f_L - 1-k_{LS}+k_{LS}(1-f_L)
\right]f_Lf_S
\nonumber\\
&=
\frac{d_L^2}{c_L}\alpha_{LS}^{0}
\left[
k_{LS}f_L-k_{LS}+k_{LS}-k_{LS}f_L
\right]f_Lf_S
\nonumber\\
&=0.
\end{align}

\begin{align}
\text{Term }(2)
&=
\Bigg\{
\frac{d_L^2}{c_L}\alpha_{LM}^{0}(1+k_{LM}f_L)
\nonumber\\
&\qquad\qquad
-
\frac{d_M^2}{c_M}\alpha_{LM}^{0}\left(\frac{c_M}{c_L}\right) r_{LM}^{2}(1+k_{LM})
\left[
1-\frac{k_{LM}}{1+k_{LM}}(1-f_L)
\right]
\Bigg\}f_Lf_M
\nonumber\\
&=
\Bigg\{
\frac{d_L^2}{c_L}\alpha_{LM}^{0}(1+k_{LM}f_L)
\nonumber\\
&\qquad\qquad
-
\frac{d_M^2}{c_M}\alpha_{LM}^{0}\left(\frac{c_M}{c_L}\right)
\left(\frac{d_L^2}{d_M^2}\right)(1+k_{LM})
\left[
1-\frac{k_{LM}}{1+k_{LM}}(1-f_L)
\right]
\Bigg\}f_Lf_M
\nonumber\\
&=
\frac{d_L^2}{c_L}\alpha_{LM}^{0}
\left[
(1+k_{LM}f_L)
-
(1+k_{LM})
\left(
1-\frac{k_{LM}}{1+k_{LM}}(1-f_L)
\right)
\right]f_Lf_M
\nonumber\\
&=
\frac{d_L^2}{c_L}\alpha_{LM}^{0}
\left[
1+k_{LM}f_L - 1-k_{LM}+k_{LM}(1-f_L)
\right]f_Lf_M
\nonumber\\
&=
\frac{d_L^2}{c_L}\alpha_{LM}^{0}
\left[
k_{LM}f_L-k_{LM}+k_{LM}-k_{LM}f_L
\right]f_Lf_M
\nonumber\\
&=0.
\end{align}

\begin{align}
\text{Term }(3)
&=
\Bigg\{
-\frac{d_S^2}{c_S}\alpha_{MS}^{0}\left(\frac{c_S}{c_M}\right) r_{MS}^{2}(1+k_{MS})
\left[
1-\frac{k_{MS}}{1+k_{MS}}(1-f_M)
\right]
\nonumber\\
&\qquad\qquad
+
\frac{d_M^2}{c_M}\alpha_{MS}^{0}(1+k_{MS}f_M)
\Bigg\}f_Sf_M
\nonumber\\
&=
\Bigg\{
-\frac{d_S^2}{c_S}\alpha_{MS}^{0}\left(\frac{c_S}{c_M}\right)
\left(\frac{d_M^2}{d_S^2}\right)(1+k_{MS})
\left[
1-\frac{k_{MS}}{1+k_{MS}}(1-f_M)
\right]
\nonumber\\
&\qquad\qquad
+
\frac{d_M^2}{c_M}\alpha_{MS}^{0}(1+k_{MS}f_M)
\Bigg\}f_Sf_M
\nonumber\\
&=
\frac{d_M^2}{c_M}\alpha_{MS}^{0}
\left[
-(1+k_{MS})
\left(
1-\frac{k_{MS}}{1+k_{MS}}(1-f_M)
\right)
+(1+k_{MS}f_M)
\right]f_Sf_M
\nonumber\\
&=
\frac{d_M^2}{c_M}\alpha_{MS}^{0}
\left[
-1-k_{MS}+k_{MS}(1-f_M)+1+k_{MS}f_M
\right]f_Sf_M
\nonumber\\
&=
\frac{d_M^2}{c_M}\alpha_{MS}^{0}
\left[
-1-k_{MS}+k_{MS}-k_{MS}f_M+1+k_{MS}f_M
\right]f_Sf_M
\nonumber\\
&=0.
\end{align}

Therefore,
\begin{align}
J_L^{\mathrm{seg}}+J_S^{\mathrm{seg}}+J_M^{\mathrm{seg}}= \left(\frac{\rho_p g_y}{6\eta}\right) [\text{Term (1)} + \text{Term (2)} + \text{Term (3)}] = 0.
\end{align}
Thus, the sum of segregation fluxes of all three species in the ternary mixture becomes zero. Similarly, it can be verified that for a multicomponent mixture containing $N$ species, the total segregation flux satisfies  $\sum_{ j=1}^{N} J_i^{seg} = 0$.

\FloatBarrier
\section{Effect of lift force on the continuum model predictions}
\begin{figure}
    \centering
\includegraphics[width=0.45\linewidth, trim=0 0 0 20, clip]{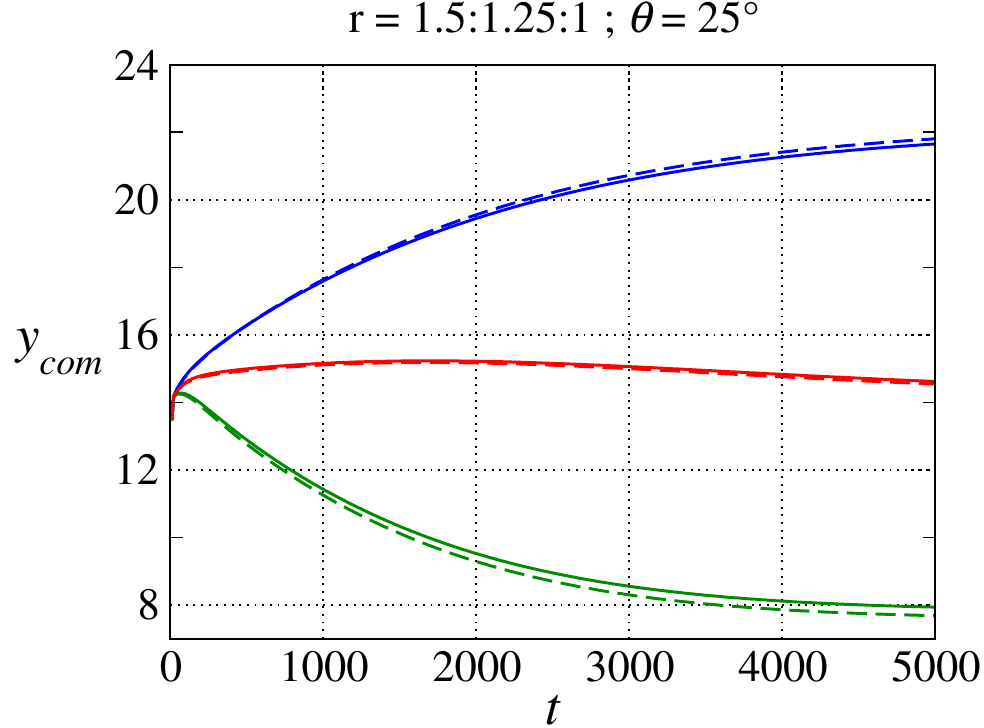} \put(-170,125){(a)} 
\includegraphics[width=0.45\linewidth, trim=0 0 0 20, clip]{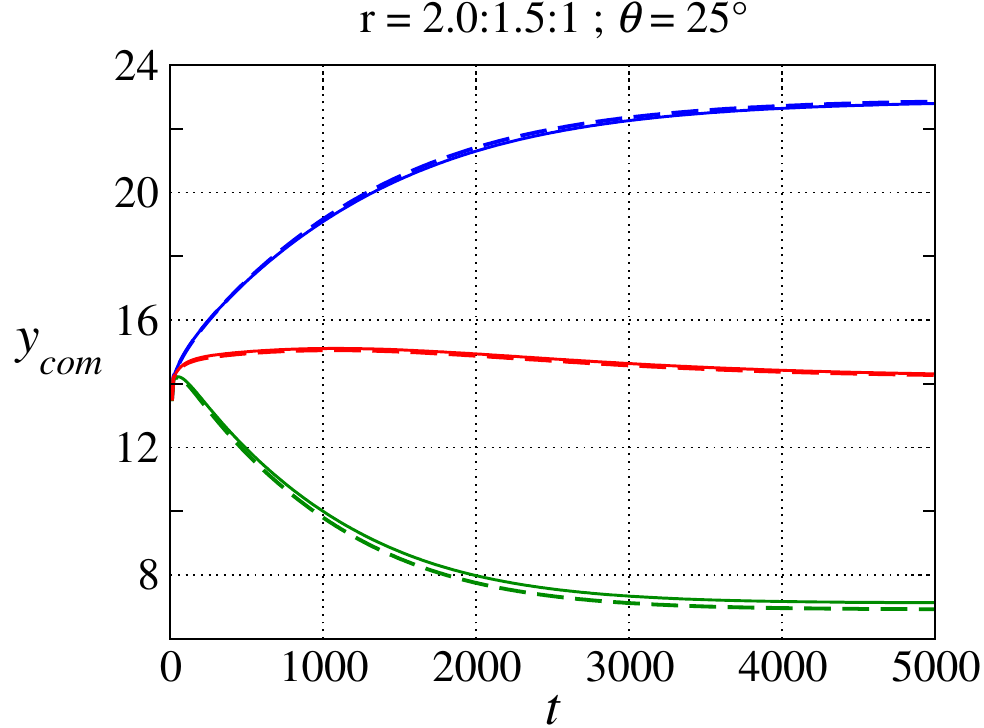} \put(-170,125){(b)} \\   
\caption{Comparison of evolution of centre of $y_{com}$ with or without accounting for the lift force contribution to the total upward force at two different mixtures having size ratios (a) $ 1.5:1.25:1.0$, and (b) $2.0:1.5:1.0$ at inclination $\theta = 25^\circ$. Solid lines correspond to the predictions after ignoring the contribution of the lift force while the dashed lines correspond to those obtained by accounting for the lift force in the dimensionless total upward force.}
\label{fig:lift_with_without}
\end{figure}

As mentioned in the text after equation~\ref{eq:generalised_segflux_multi}, the lift force accounts for a small fraction of the total upward force. In this section, we present the results of our segregation predictions by accounting and neglecting for the lift force contribution in the upward force in figure~\ref{fig:lift_with_without}.
Figure~\ref{fig:lift_with_without}a shows the time evolution of $y_{com}$ for equal composition ternary mixture having size ratios $1.5:1.25:1.0$. \ref{fig:lift_with_without}b shows results for size ratio $2.0:1.5:1.0$. The blue, red, and green colors represent the large, medium, and small species in the ternary mixture, respectively. In both cases, the predictions obtained by including the lift-force contribution (dashed lines) show no significant difference from those (solid lines) obtained without it. Therefore, we conclude that the lift-force contribution can be safely neglected. 

\section{Comparison of model predictions obtained by solving transport equations for any two of the three species in a ternary mixture}
\label{sec:app_LM_LS_comparison}
\begin{figure}
    \centering
\includegraphics[scale=0.4, trim=0 0 0 20, clip]{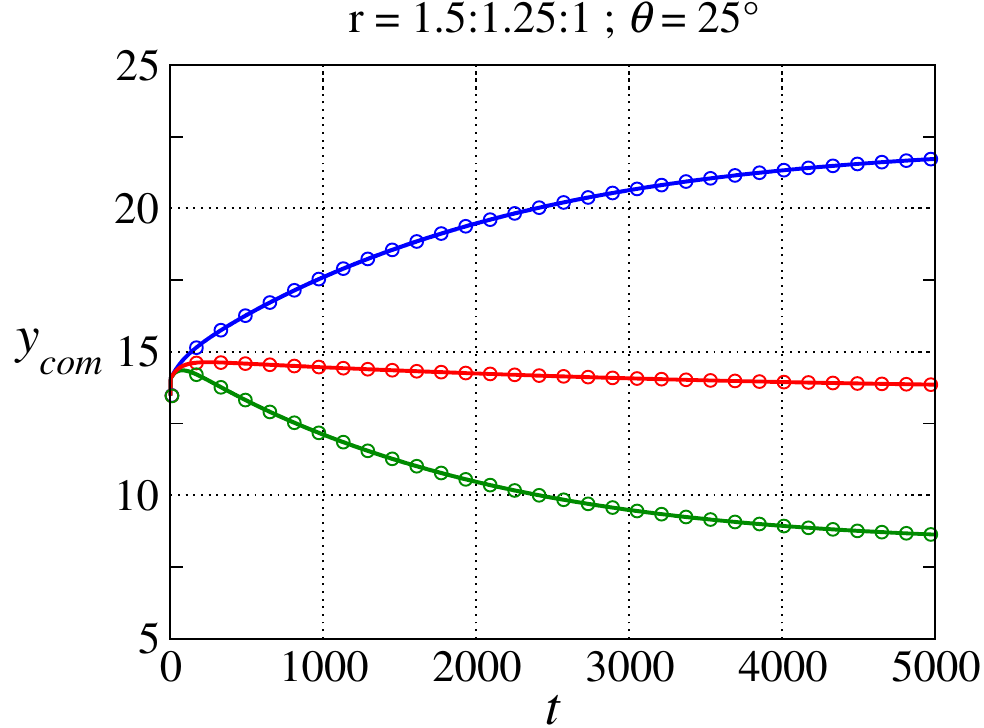}\put(-170,135){(a)} 
\includegraphics[scale=0.4, trim=0 0 0 20, clip]{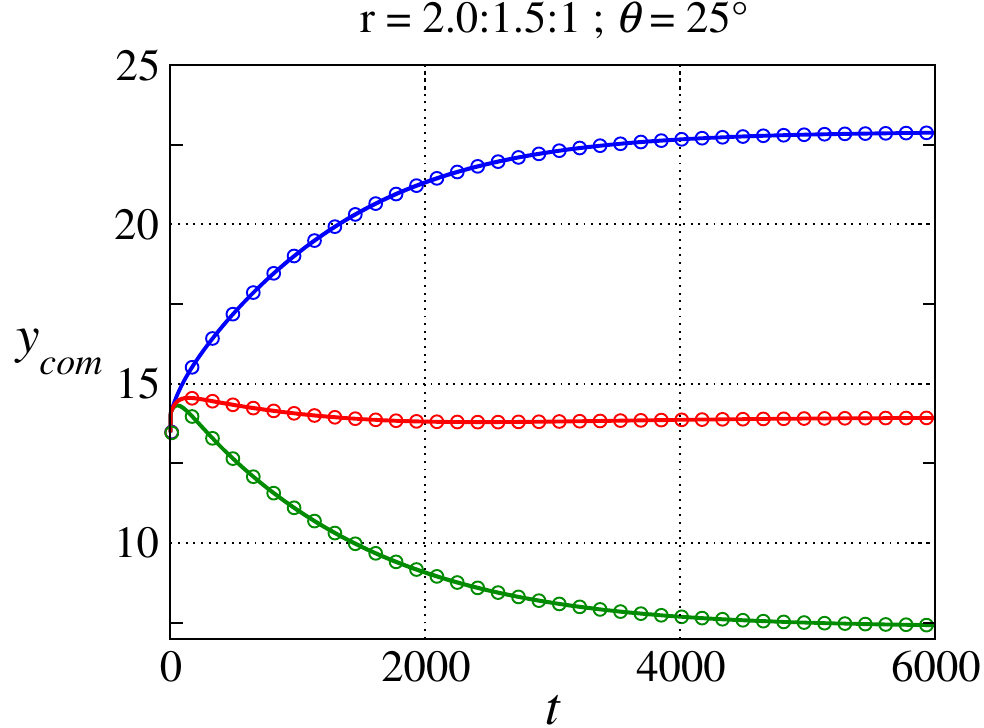}\put(-170,135){(b)} 
    \caption{Evolution of species centre of mass for equal composition ternary mixture having size ratio (a) $ 1.5:1.25:1$, and (b) $ 2.0:1.5:1$, flowing at inclination angle $\theta = 25^\circ$. Solid lines correspond to predictions by solving transport equations for large and small while symbols by solving large and medium transport equations.}
    \label{fig:LM_LS}
\end{figure}

For a $N$ component mixture, we only need to solve $N - 1$ equations to obtain the concentration of species $f_1$ to $f_{N-1}$. The concentration of the species $N^{th}$ is obtained as $f_N = 1 - \sum_{i=1}^{N-1} f_i$. In order to check the consistency of the generalized particle force-based segregation model, we predict the segregation by solving the transport equations first for large - small pair and get $f_L$ and $f_S$ from this solution and $f_M$ is then evaluated as $1 - (f_L + f_S)$. Next we solve the equations for large-medium pair in our continuum model and obtain the concentration of $f_L$ and $f_M$ from these. The value of $f_S$ is then obtained as $f_S = 1- (f_L + f_M)$. Figure~\ref{fig:LM_LS} shows the evolution of species centre of mass for equal composition ternary mixtures flowing at inclination angle $\theta = 25^\circ$ using these two different ways of calculating species concentration. Figure~\ref{fig:LM_LS}a shows $y_{com}$ for the size ratios $ 1.5:1.25:1.0$ while figure~\ref{fig:LM_LS}b corresponds to $2.0:1.5:1.0$. The blue, red, and green curves denote the large, medium, and small species, respectively. The model predictions for $y_{com}$ by solving large and medium transport equations (symbols) match very well with the predictions by solving large and small transport equations (solid lines) confirming the consistent formulation. 

\section{Comparison of results from the ternary model and the reduced quaternary model}
\label{sec:app_ternary_reduced_quat}
\begin{figure}
    \centering
\includegraphics[scale=0.4, trim=0 0 0 20, clip]{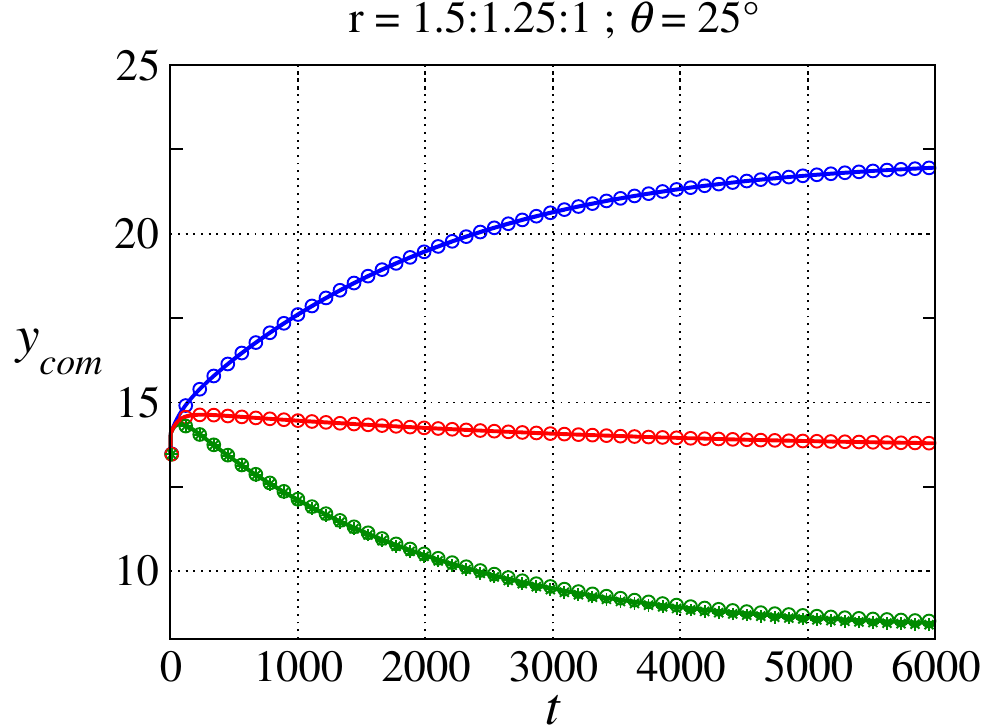}\put(-170,135){(a)} \put(-110,135){$1.5:1.25:1.0$}\put(-130,115){\tiny $f_L = 1/3; f_M = 1/3; f_S= 1/3$}
\includegraphics[scale=0.4, trim=0 0 0 20, clip]{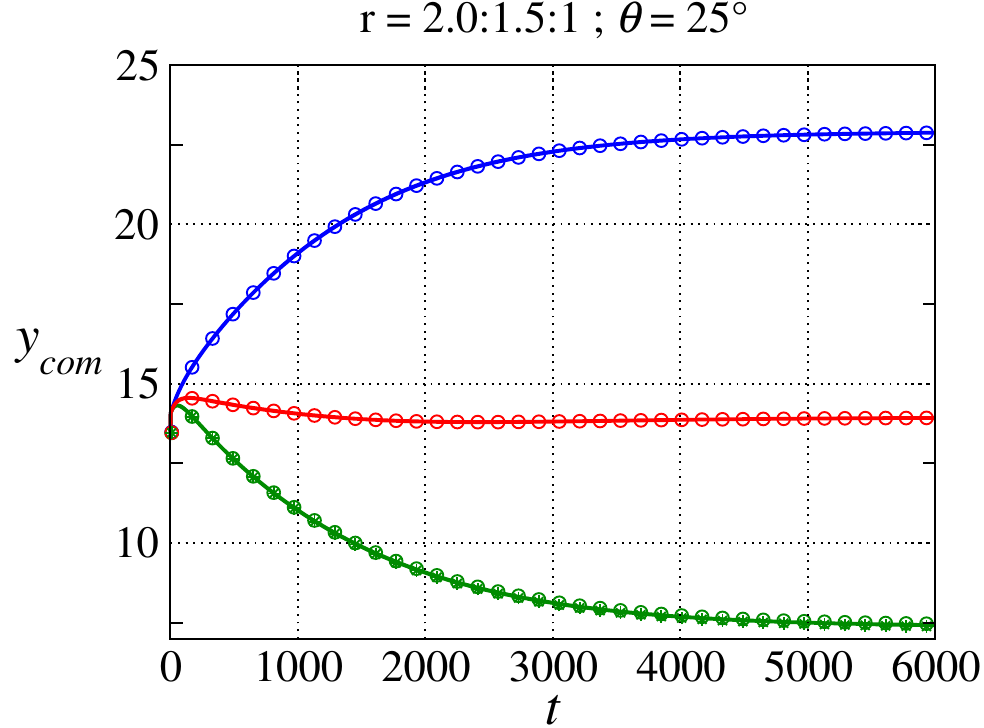}\put(-170,135){(b)} \put(-110,135){$2.0:1.5:1.0$}\put(-130,120){\tiny $f_L = 1/3; f_M = 1/3; f_S= 1/3$}\\
\includegraphics[scale=0.4, trim=0 0 0 20, clip]{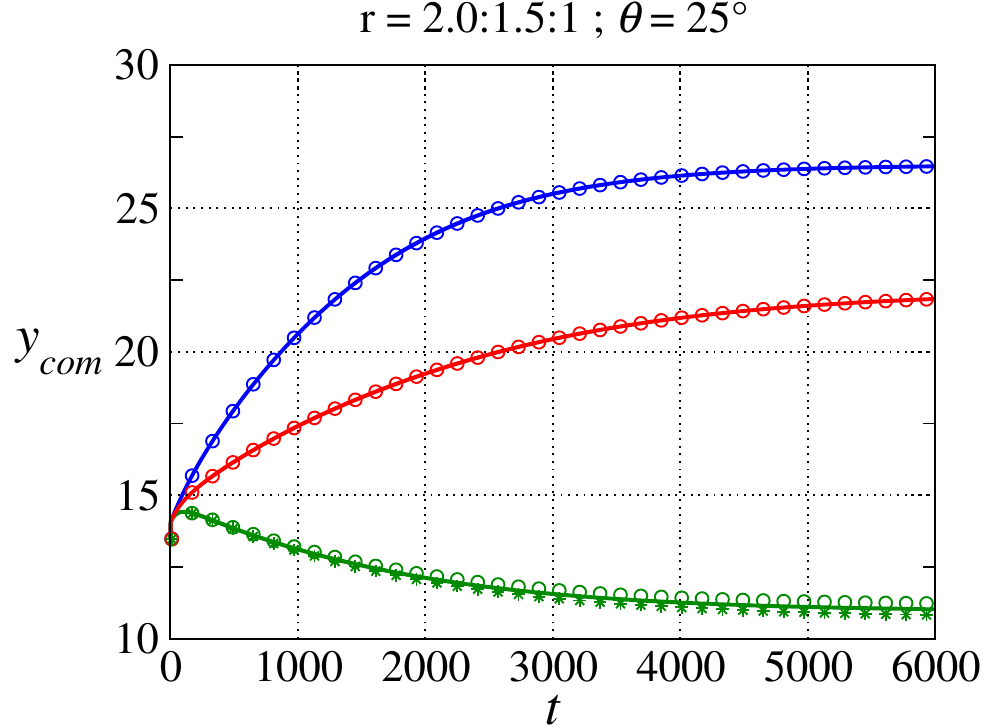}\put(-170,135){(c)} \put(-110,135){$2.0:1.5:1.0$}\put(-130,115){\tiny $f_L = 0.10; f_M = 0.20; f_S = 0.70$}
\includegraphics[scale=0.4, trim=0 0 0 20, clip]{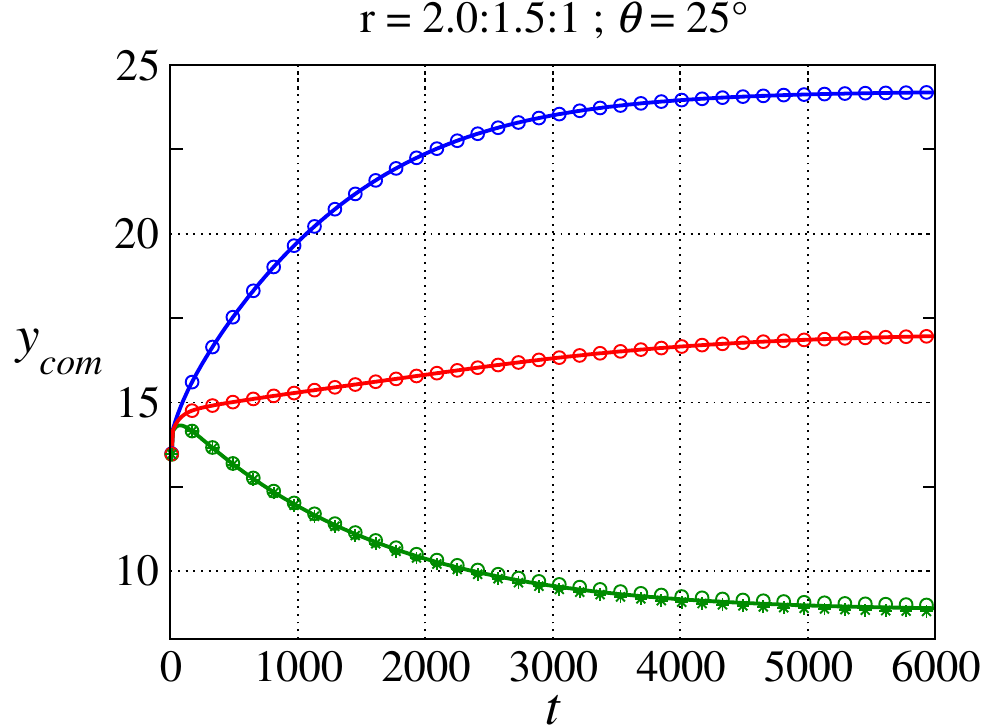}\put(-170,135){(d)} \put(-110,135){$2.0:1.5:1.0$}\put(-110,100){\tiny $f_L = 0.25; f_M = 0.25; f_S = 0.50$}
\caption{Evolution of centre of mass for different mixture flowing at inclination angle $\theta = 25^\circ$. Solid lines represent predictions obtained by solving the transport equations for the ternary system, while the symbols denote predictions obtained by reducing the quaternary system to a ternary system.}
    \label{fig:Quat_ternary}
\end{figure}

In this section, we check the consistency of continuum model for quaternary mixtures. To do this, we consider a quaternary mixture in which two species have the same particle size, thereby reducing the system effectively to a ternary mixture. We then predict the evolution of $y_{com}$ using both the reduced quaternary continuum model (symbols) and the ternary continuum model (solid lines). The blue, red, and green colors represent the large, medium, and small species in the ternary mixture, respectively. The predictions obtained from the two approaches are in very good agreement, as shown in figure~\ref{fig:Quat_ternary}.

\section{Discrete Element Method}
\label{sec:simulationMethod}
We numerically simulate spherical particles of different sizes and the same density ($\rho_p$) flowing down an inclined plane using the Discrete Element Method (DEM). The particles are assumed to be frictional (inter-particle friction coefficient $\mu_{pp} = 0.5$) and slightly inelastic (normal restitution coefficient $e = 0.88$). The simulation domain spans an area of $20d \times 20d$ in the streamwise ($x-$) and vorticity ($z-$) directions, with periodic boundary conditions imposed along both directions. We use $100d \times 100d$ base for mixtures of size ratios $r > 2$. The height of the simulation box in the $y$ direction varies with the size ratio; for ternary mixtures $H \approx 28d$ for $r < 2.0$ and $H \approx 50d$ for $r > 2.0$, whereas for quaternary mixtures $H \approx 40d$. At the start of the DEM simulation, particles are arranged in a cubic lattice to ensure no initial contact between adjacent particles. The particles are then allowed to settle under gravity at an inclination angle of $\theta = 0^\circ$. After settling, the inclination angle is increased to the desired value to initiate the flow. The linear spring and dashpot model is used to compute contact forces in both normal and tangential directions. The normal spring stiffness is $k_{n} = 2 \times 10^{5} mg/d$, while the tangential spring stiffness is $k_{t} = 2k_{n}/7$. More details on the contact force model and properties calculations are provided in \cite{tripathi2011rheology,sahu_kumawat_agrawal_tripathi_2023}.
\begin{figure}
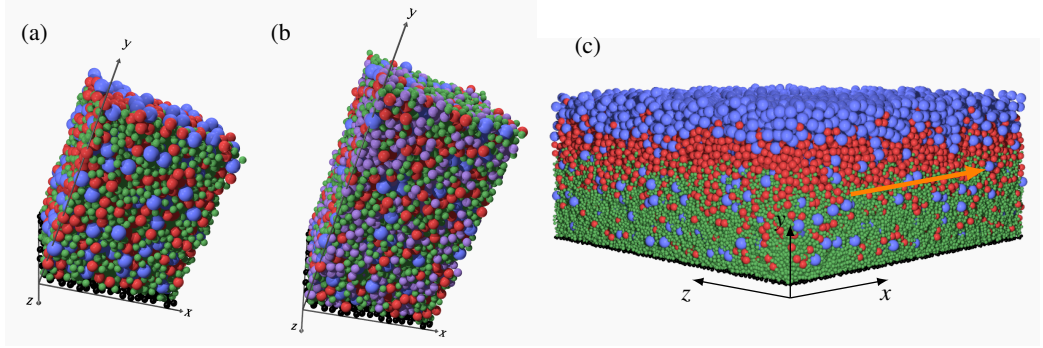

    \centering
    \includegraphics[scale=0.12, trim=520 0 520 0, clip]{Figures/Initial_2_1.5_1.pdf}\put(-100,115){(a)}
   \includegraphics[scale=0.12, trim=550 0 590 0, clip]{Figures/Fig8a.pdf}\put(-100,115){(b}
    \includegraphics[scale=0.23, trim=550 280 550 300, clip]{Figures/r_3_2_1_100by100_steady_mixed_v1.pdf} \put(-175,110){(c)} \put(-100,45){$y$}\put(-60,17){$x$}\put(-135,17){$z$}
      \begin{tikzpicture}[overlay, remember picture]
        \draw[->, orange, very thick, line width=1.7pt, >=latex] (-2.5,2.0) -- (-0.7,2.33); 
         \draw[->, black, very thick, line width=0.5pt, >=latex] (-3.3,0.63) -- (-3.3,1.6); 
        \draw[->, black, very thick, line width=0.5pt, >=latex] (-3.3,0.63) -- (-2.0,0.87); 
        \draw[->, black, very thick, line width=0.5pt, >=latex] (-3.3,0.63) -- (-4.6,0.87); 
    \end{tikzpicture}
    \caption{DEM snapshots (a) ternary ($ 2:1.5:1$), and (b) quaternary ($ 2:1.5:1.25:1$) mixtures having equal species volume fraction flowing down a plane inclined at $\theta = 25^\circ$ started from a uniformly almost mixed state. Panel (c) corresponds to DEM snapshot for larger size ratios $3:2:1$ with a larger system $100d_s \times 100d_s$. Blue particles indicate larger particles and green particles represent smaller ones while red and purple particles denote the intermediate size particles. The black particles are static particles that are used to form a rough bumpy base to minimize slip at the chute base.}
    \label{fig:DEM_snap_size_ratios}
\end{figure}

\end{document}